\documentclass[final,12pt]{clear2025} %

\newcommand{\Var}{\operatorname{Var}}

\title[Robustness of Median-based Splitting Rules for Causal Trees]{Robustness of Median-based Splitting Rules for Causal Trees}
\usepackage{times}
\usepackage{algorithm} 
\usepackage{dsfont} 
\usepackage{mathtools} 
\usepackage{booktabs} 
\usepackage{multirow} 
\usepackage{caption}
\usepackage{pdflscape}
\usepackage{array} 
\usepackage{amsmath} 
\usepackage{makecell} 
\usepackage{graphicx} 
\usepackage{ragged2e} 
\usepackage{rotating} 
\usepackage{bm} 
\usepackage{tikz}
\usepackage{physics} 
\usepackage{float}
\usepackage{placeins}
\usepackage{lscape}
\usepackage{subcaption}

\newenvironment{notes}
  {\begin{minipage}{0.9\linewidth}\smallskip\footnotesize\emph{Notes:}}
  {\end{minipage}}

\clearauthor{%
 \Name{Karolina Gliszczy\'nska-Schroeder} \Email{karolina.gliszczynska(at)vwl.uni-due.de}\\
 \addr Chair of Econometrics, Faculty of Business Administration and Economics, University of Duisburg-Essen, Universitätsstraße 12, 45117 Essen, Germany\\
 \Name{Jens Klenke} \Email{jens.klenke(at)vwl.uni-due.de}\\
 \addr Chair of Econometrics, Faculty of Business Administration and Economics, University of Duisburg-Essen, Universitätsstraße 12, 45117 Essen, Germany%
}

\begin{document}

\newtheorem{assump}{Assumption}
\newtheorem{rem}{Remark}[section]


\maketitle


\begin{abstract}%
Causal forests \citep{athey_recursive_2016} are a widely used method for the estimation of heterogeneous treatment effects. Standard implementations use mean-based splitting criteria, which perform well when the outcome distribution is well-behaved and free of contamination, such as outliers or heavy tails. However, real-world data are rarely clean. Outcome perturbations can affect the tree structure selected by mean-based splitting rules and lead to unreliable treatment effect estimates. In this study, we systematically examine the robustness of causal forests to irregularities in the outcome distribution.  Building on the median-based splitting criteria introduced by \citet{maßmann2026medianbasedsplittingrulescausal}, we empirically investigate three median-based splitting criteria: Median Absolute Deviation (MAD), Least Median of Squares (LMS), and Median Squared Deviation (MSD). We conduct Monte Carlo experiments and apply the methods to empirical benchmark datasets from the causal inference literature to compare the performance of the median-based approaches with standard causal forests. We additionally examine pointwise confidence interval coverage using the Bootstrap of Little Bags and the infinitesimal jackknife. The results indicate that median-based splitting criteria can reduce sensitivity to outcome irregularities in specific scenarios, particularly in the presence of extreme heavy-tailed noise and nonlinear treatment effect heterogeneity.  However, these benefits are scenario-dependent and do not consistently lead to more accurate conditional average treatment effect (CATE) estimates.

\end{abstract}

\begin{keywords}%
    Heterogeneous treatment effects, causal forests, Hodges--Lehmann estimator, robust estimation, heavy-tailed outcomes
\end{keywords}

\begin{jelclass}
    C14, C21, C52
\end{jelclass}


\section{Introduction}

Tree-based models, including regression trees and random forests \citep{cart84,Breiman2001}, traditionally use the mean squared error (MSE) for regression tasks or impurity measures such as the Gini index for classification tasks as splitting criteria. While these criteria are computationally attractive and perform well, they can be sensitive to extreme observations. In fact, contrary to the common perception that random forests are robust by construction, their standard regression implementation is not robust to outliers in the response variable. For example, in the empirical study of \citet{roy_robustness_2012}, the original random forest is frequently less stable under artificial contamination than robust variants based on median aggregation or least-absolute-deviation splitting. Further robust approaches have been investigated in the literature, including quantile-based \citep{Quantile_Regression_Forests,quantileregressiontree}, median-based \citep{rcate_Li}, $L_1$-based \citep{L1_splitting}, and weighted splitting rules \citep{ishwaran2015effect}. 

This paper studies the analogous robustness problem for causal trees and causal forests \citep{athey_recursive_2016,wager_estimation_2018}. We focus on the original causal tree-based formulation of causal forests, where recursive partitioning is explicitly used to uncover heterogeneity in conditional average treatment effects (CATEs).  In the causal tree framework considered here, causal forests extend causal trees by aggregating many trees, which typically rely on honesty, whereby the data used for tree construction are separated from the data used for treatment effect estimation. This sample-splitting structure reduces adaptive bias and supports valid treatment effect estimation \citep{wager_estimation_2018}. However, honesty and tree aggregation do not guarantee robustness against contaminated outcomes. Extreme outcome observations may still affect the selected partition and, consequently, the estimated treatment effect.

Research on alternative splitting rules for causal trees has become more prominent, but remains less developed than the corresponding literature for prediction problems. For example, \citet{athey_recursive_2016} compare several splitting rules within the causal tree framework, including transformed-outcome (TOT), fit-based (fit), and t-statistic (tstat) criteria. \citet{Lechner_Modified_CF} propose an extension of causal forests for multi-treatment settings using a criterion that combines MSE with the Mean Correlated Error (MCE). Evidence on the robustness of causal forests is also not uniform. While \citet{HTE_robust_gong} provide evidence for robustness properties of forest-based estimators of heterogeneous treatment effects, \citet{rcate_Li} find that the classical causal forest can perform poorly under contamination compared with more robust alternatives. Building on this literature, \citet{maßmann2026medianbasedsplittingrulescausal} introduce three median-based splitting rules for causal trees and derive their theoretical properties. Their main splitting criterion, the Median Squared Deviation (MSD), replaces the mean-based treatment effect estimator in the MSE criterion with the Hodges–Lehmann estimator \citep{hodges_estimates_1963}. Further, the authors introduced the Median Absolute Deviation (MAD), which evaluates the absolute difference between mean-based and Hodges–Lehmann leaf effects, and the Least Median of Squares (LMS) criterion, which replaces the average squared residual with its median. Building on this framework, we empirically investigate whether these median-based splitting rules improve the robustness of causal forests in practice.

Here, robustness refers to the stability of causal forest estimates under deviations from well-behaved outcome distributions. This definition of robustness differs from doubly robust machine learning \citep{robins1995semiparametric,Bang_Robins_2006,funk2011doubly} or double/debiased machine learning approaches \citep{chernozhukov_doubledebiased_2018}, where robustness usually refers to valid treatment effect estimation under misspecification of one or more nuisance components, which is not part of our study. We focus specifically on outcome-side irregularities, such as outliers and heavy-tailed outcome distributions. Importantly, the estimand of interest remains the CATE.  The only modification is the splitting step using the median-based criteria.  Across simulations and empirical noise and contamination designs, the results show that these modifications can improve stability, although the gains are not systematic and depend on the contamination setting and the underlying treatment effect structure.

In this context, we also examine the robustness of statistical inference by comparing confidence intervals constructed using the bootstrap of little bags (BLB) \citep{athey_generalized_2019} and the infinitesimal jackknife (IJ) approach \citep{wager_estimation_2018}. Both methods are designed to estimate the sampling variance of forest-based treatment effect estimates and thereby enable pointwise confidence intervals for the CATE. However, to the best of our knowledge, their behavior under outcome contamination and heavy-tailed noise has not yet been compared systematically in the causal forest literature.

The paper is structured as follows. Section~\ref{sec:PO_framework} introduces the potential-outcomes framework, the identifying assumptions, and the CATE as the main estimand. Section~\ref{sec:causal_trees} reviews honest causal trees and causal forests, including the standard MSE-based splitting criterion and the construction of leaf-level treatment effect estimates. Section~\ref{sec:splitting_criteria} presents the median-based splitting rules: the Median Squared Deviation, the Median Absolute Deviation, and the Least Median of Squares. Section \ref{sec:confidence_intervals} discusses pointwise inference and compares variance estimation using the Bootstrap of Little Bags and the infinitesimal jackknife.  Section~\ref{sec:robust_simulation_study} presents the simulation design and evaluates the performance of the splitting rules under clean data, outlier contamination, and heavy-tailed outcome distributions. Section~\ref{sec:empirical_contamination} provides an empirical robustness analysis using the Lalonde \citep{lalonde_evaluating_1986} and ACIC  \citep{grf_package} datasets, which are widely used benchmarks in the causal inference literature, alongside the Breast Cancer dataset  \citep{ems_package} as an additional empirical application.
Section~\ref{sec:Robust_discussion} concludes and discusses limitations and possible extensions.

\section{Potential Outcomes Framework}
\label{sec:PO_framework}
We use the potential outcome framework and notation of Rubin’s causal model \citep{rubin_estimating_1974, rubin_bayesian_1978, imbens_rubin_2010}. Consider a set of $N$ units, indexed by $i=1,\dots,N$.  For each unit $i$, let $D_i \in \{0,1\}$ be the binary treatment indicator and let $X_i \in \mathbb{R}^K$ be a $K$-dimensional covariate vector. The potential outcome for unit $i$ under treatment $d \in \{0,1\}$ is denoted by $Y_i(d)$, where $Y_i(0)$ represents the outcome under control, and $Y_i(1)$ the outcome under treatment. The individual treatment effect (ITE) for a given unit is then defined as
\begin{equation}\label{eq:ITE}
\tau_i \coloneqq Y_i(1) - Y_i(0). 
\end{equation}
Since each unit is observed either under treatment or control, but never both, $\tau_i$ is not directly observable.  
To identify valid causal treatment effects, the potential outcome framework relies on three key assumptions. The first is the Stable Unit Treatment Value Assumption (SUTVA), which rules out two potential complications. First, it excludes interference between units, implying that the potential outcome of unit $i$ depends only on its own treatment and not on the treatment received by other units. Second, it requires that the treatment is well defined, meaning that there are no hidden or alternative versions of the same treatment level.
\begin{assump}{Stable Unit Treatment Value Assumption (SUTVA).}
\label{assump:SUTVA}
\begin{align*}
\text{If } D_i = d \text{, then } Y_i(d) = Y_i^{obs} \text{ , } \forall d \in \{0, 1 \} \text{ , } \forall ~ i= 1, ..., N . 
\end{align*}
\end{assump}
Under SUTVA, the observed outcome can be written as
\[
Y_i^{\mathrm{obs}} = D_iY_i(1)+(1-D_i)Y_i(0).
\]
 Extensions allowing for interference or multiple treatment versions exist, but are not the focus of this paper \citep{hudgens_toward_2008,vanderweele_ignorability_2008,vanderweele_causal_2013}. Second, we also assume unconfoundedness, which requires that, conditional on $X_i$, the treatment is independent of the potential outcomes. 
\begin{assump}{Unconfoundedness.}
\label{assump:unconfoundedness}
\[
D_i \perp\!\!\!\perp \bigl(Y_i(1),Y_i(0)\bigr)\mid X_i.
\]
\end{assump}

This assumption is satisfied by design in randomized experiments. In observational studies, it is generally untestable and must be justified based on the institutional setting and the available covariates \citep{imbens_nonparametric_2004}.
Third, we also assume positivity, also referred to as overlap. Let the propensity score be the conditional probability of receiving treatment given covariates,
\begin{equation}
\label{eq:propensity}
p(x) \coloneqq Pr(D_i=1\mid X_i=x).
\end{equation}
Positivity requires that, for every covariate value $x$ in the support of $X_i$, both treatment states occur with positive probability.

\begin{assump}{Overlap.}
\label{assump:overlap}
There exists $\epsilon>0$ such that the conditional probability of receiving treatment given pre-treatment covariates $X$ is bounded away from zero and one,  
\begin{align*}
\epsilon < p(x) < 1-\epsilon \quad \text{for all } x \text{ in the support of } X_i.
\end{align*}
\end{assump}
Overlap ensures that treated and control observations are available throughout the relevant covariate space. Consequently, units with similar pre-treatment covariates must have a positive probability of being observed under either treatment status, ensuring that treated and control units are comparable on the basis of their observed characteristics. Weak overlap can lead to unstable treatment effect estimates, especially for methods that rely on local comparisons or inverse propensity weighting \citep{busso_new_2014,rothe_robust_2017,damour_overlap_2021}. Using Assumptions \ref{assump:SUTVA} to \ref{assump:overlap},  
we can identify two commonly used treatment effects. The average treatment effect (ATE), denoted as $\tau$, is defined as   
\begin{equation}\label{eq:ATE}
 \tau\coloneqq \mathbb{E}[Y_i(1)-Y_i(0)].  
\end{equation}
Second, the conditional average treatment effect (CATE), denoted by $\tau(X_i=x)$ and abbreviated as $\tau(x)$, is the conditional analog of the ATE: 
\begin{align}
\begin{split}
\label{eq:CATE}
\tau(X_i = x) &\coloneqq \mathbb{E}[Y_i(1) - Y_i(0) \mid X_i = x] \\
&= 
\underbrace{\mathbb{E}[Y_i \mid D_i = 1, X_i = x]}_{\coloneqq\mu_1(x)} 
- 
\underbrace{\mathbb{E}[Y_i \mid D_i = 0, X_i = x]}_{\coloneqq\mu_0(x)}, 
\end{split}
\end{align}
Thus, the CATE can be estimated by modeling the conditional mean outcomes for treated and control units, or directly by methods designed for heterogeneous treatment effect estimation. Throughout this paper, $\tau(X_i=x)$ is our main treatment effect of interest.  

\section{Causal Trees and Causal Forests}

\label{sec:causal_trees}
Tree-based methods divide the covariate space into smaller rectangles, or leaves, aiming to estimate the conditional expectation of the outcome variable given the regressors  \citep{hastie_09}. In the causal tree framework of \citet{athey_recursive_2016}, the partition is chosen to uncover heterogeneity in treatment effects. The goal is to form leaves in which treatment effects are relatively homogeneous, while treatment effects may differ across leaves. Formally, let $\Pi$ denote a tree or partition and let $l(x;\Pi)$ be the leaf $l \in \Pi$ containing a test point $x$. 
For treatment status $d\in\{0,1\}$, the leafwise conditional mean function is defined as
\begin{align}
\label{eq:conditional_mean_function}
    \mu_d(x; \Pi) = \mathbb{E}[Y_i|D_i=d, X_i \in l(x;\Pi) ]. 
\end{align}
An unbiased estimate of \eqref{eq:conditional_mean_function} is then given by 
\begin{equation}
\label{estim_cond_mean_func}
\hat\mu_d(x; \mathcal{S}, \Pi) \coloneqq \frac{1}{\# (i \in \mathcal{S}: D_i= d,  X_i \in l(x; \Pi))} \sum_{i\in \mathcal{S}:D_i=d, X_i \in l(x; \Pi)} Y_i.
\end{equation}
Then an estimator of the CATE in \eqref{eq:CATE} is the leafwise CATE estimator
\begin{equation} \label{eq:CATE_Def}
\hat{\tau}(x;\mathcal{S},\Pi)=
\hat{\mu}_1(x;\mathcal{S},\Pi)
-\hat{\mu}_0(x;\mathcal{S},\Pi).
\end{equation}
For simplicity, we often suppress the dependence on the partition $\Pi$ whenever it is clear from the context. Thus, $l(x)$ denotes the leaf containing $x$ under the candidate partition currently being evaluated. All comparisons of leafwise quantities are made within leaves of this same partition. 

Conventional CART algorithms \citep{cart84} use the same data both to construct the tree and to estimate the leaf-level responses. This approach is referred to as adaptive estimation. Generally, this adaptive use of the data can lead to overfitting, since additional splits may improve in-sample fit even when they do not reflect stable population structure. To reduce this source of bias, \citet{athey_recursive_2016} introduce the honesty approach for causal trees. 
A tree is called honest if the observations used to determine the splits are separated from those used to estimate treatment effects within the resulting leaves, which is automatically satisfied when the tree is grown
according to the Double-Sample Tree algorithm described in \cite{wager_estimation_2018} or Algorithm \ref{algo:double_sample} below. Formally, we distinguish between three disjoint samples. The training sample $\mathcal{S}^{tr}$ is used to construct the tree, the estimation sample $\mathcal{S}^{est}$ is used to estimate treatment effects, and the test sample $\mathcal{S}^{te}$ is used for out-of-sample evaluation. Their respective sample sizes are denoted by $N^{tr}$, $N^{est}$, and $N^{te}$. Within the training sample, $N^{tr}_1$ and $N^{tr}_0$ denote the number of treatment and control units, respectively.

Both approaches involve a trade-off. Adaptive estimation uses the available data more efficiently, but is more vulnerable to overfitting and adaptive bias. Honest estimation reduces this bias, but it can be costly in small samples because splitting the data leaves fewer observations for both partition construction and treatment effect estimation. A detailed discussion of this trade-off is beyond the scope of this study, but we refer the reader to \citet{hou2025honesty}. Throughout this paper, we use honest leaf-level treatment effect estimation. The resulting double-sample tree construction is summarized in Algorithm~\ref{algo:double_sample}.
\FloatBarrier
\begin{algorithm}[ht]
    \SetAlgoLined
    \KwIn{Training data $(X_i,Y_i^{\mathrm{obs}},D_i)_{i=1}^{N}$, a minimum leaf size $k$, splitting criterion $Q$.}

    \textbf{Step 1:} Draw a random subsample $\mathcal{S}\subset\{1,\dots,N\}$ without replacement. Split $\mathcal{S}$ into two disjoint samples, a splitting sample $\mathcal{S}^{tr}$ and an estimation sample $\mathcal{S}^{est}$, with
    \[
        \mathcal{S}^{tr}\cap\mathcal{S}^{est}=\emptyset,
        \qquad
        \mathcal{S}^{tr}\cup\mathcal{S}^{est}=\mathcal{S}.
    \]
    
    \textbf{Step 2:} Grow a tree by recursive partitioning using the splitting sample $\mathcal{S}^{tr}$. Candidate splits are selected according to the splitting criterion $Q$. In the benchmark case, $Q$ is the MSE-based causal tree criterion. In the robust variants, $Q$ is replaced by one of the median-based criteria introduced in Section~\ref{sec:splitting_criteria}. Each terminal leaf is restricted to contain at least $k$ observations from $\mathcal{S}^{est}$ in each treatment group. \\[0.3cm]

    \textbf{Step 3:} Holding the partition $\Pi$ fixed, estimate leafwise treatment effects using only the estimation sample $\mathcal{S}^{est}$. For a test point $x$, define
    \[
        \hat{\tau}(x;\mathcal{S}^{est})
        =
        \hat{\mu}_1(x;\mathcal{S}^{est})
        -
        \hat{\mu}_0(x;\mathcal{S}^{est}),
    \]
    where
    \[
        \hat{\mu}_d(x;\mathcal{S}^{est})
        =
        \frac{1}{
        |\{i\in\mathcal{S}^{est}:D_i=d,\;X_i\in l(x)\}|
        } \sum \limits_{i\in\mathcal{S}^{est}:D_i=d,\;X_i\in l(x)}
        Y_i^{\mathrm{obs}},
        \qquad d\in\{0,1\}.
    \]

    \caption{Honest Double-Sample Causal Tree}
    \label{algo:double_sample}
\end{algorithm}
\FloatBarrier
After constructing $B$ honest trees via Algorithm \ref{algo:double_sample}, the causal forest aggregates their predictions. Each tree $b=1,\ldots,B$ produces a treatment effect estimate $\hat{\tau}_b(x)$ and the final forest estimate is then obtained by averaging across trees,
\begin{equation} \label{eq:robust_aggragated_CATE}
\hat{\tau}(x)
=
\frac{1}{B}
\sum_{b=1}^{B}
\hat{\tau}_b(x).
\end{equation}

We now turn to Step~2 of Algorithm~\ref{algo:double_sample}, where the recursive partition is constructed. Following \citet{athey_recursive_2016}, candidate splits are evaluated using a criterion derived from an infeasible treatment effect mean squared error, which averages over a test sample $\mathcal{S}^{te}$ while the CATE is estimated on the estimation sample $\mathcal{S}^{est}$
\begin{equation}
\label{eq:MSE_treatment_effect}
\operatorname{MSE}_{\tau}(\mathcal{S}^{te},\mathcal{S}^{est})
=
\frac{1}{N^{te}}
\sum_{i\in\mathcal{S}^{te}}
\left[
\left(
\tau_i-\hat{\tau}(X_i;\mathcal{S}^{est})
\right)^2
-\tau_i^2
\right].
\end{equation}
The subtraction of $\tau_i^2$ does not affect the ranking of candidate splits since it is independent of $\Pi$. Furthermore, $\tau_i$ is unobserved, so \eqref{eq:MSE_treatment_effect} cannot be evaluated directly. \citet{athey_recursive_2016} derive a feasible honest splitting criterion by estimating the negative expected MSE, $-\mathrm{EMSE}_\tau(\Pi)$, using the training sample and accounting for the sampling variance of the leafwise treatment effect estimates.  The resulting criterion is given by
\begin{align}
\label{eq:robust_MSE_CT_H}
  \widehat{Q}_\tau^{\,\text{MSE, H}}(\mathcal{S}^{tr}, \Pi)
  = \sum_{l \in \Pi} \frac{N_l^{tr}}{N^{tr}}
    \left[\hat{\tau}(l;\mathcal{S}^{tr})^2
    - \left(\frac{1}{N_l^{tr}} + \frac{1}{N_l^{est}}\right)
      \!\left(\frac{\hat{S}^2_{1}(l;\mathcal{S}^{tr})}{p}
      + \frac{\hat{S}^2_{0}(l;\mathcal{S}^{tr})}{1-p}\right)\right].
\end{align}
Here, $\hat{\tau}(l;\mathcal{S}^{tr})$ denotes the treatment effect estimate in leaf $l$, $\hat{S}^2_d(l;\mathcal{S}^{tr})$ is the within-leaf outcome variance for treatment group $d$, $p$ denotes the treatment probability, and $N_l^{est}$ is the expected number of estimation-sample observations in leaf $l$. Candidate splits are chosen to maximize this criterion, which rewards treatment effect heterogeneity while penalizing imprecise leaf estimates. 

Throughout the rest of the paper, we refer to \eqref{eq:robust_MSE_CT_H} as the standard causal tree (CT) splitting rule. An adaptive version of \eqref{eq:robust_MSE_CT_H} can be found in Appendix \ref{appendix:additional_theory}. The median-based rules introduced in the next section modify only the auxiliary splitting criterion $Q$ in Step~2 of Algorithm~\ref{algo:double_sample}. The estimand remains the CATE, and treatment effects are still estimated by treated-control differences within leaves. 

\begin{figure}[H]
\centering
\caption{CATE estimation error for CT and MSD splitting rules under a single increasing outlier.}
\label{fig:motivation_robust_splitting}
\includegraphics[width=0.6\textwidth]{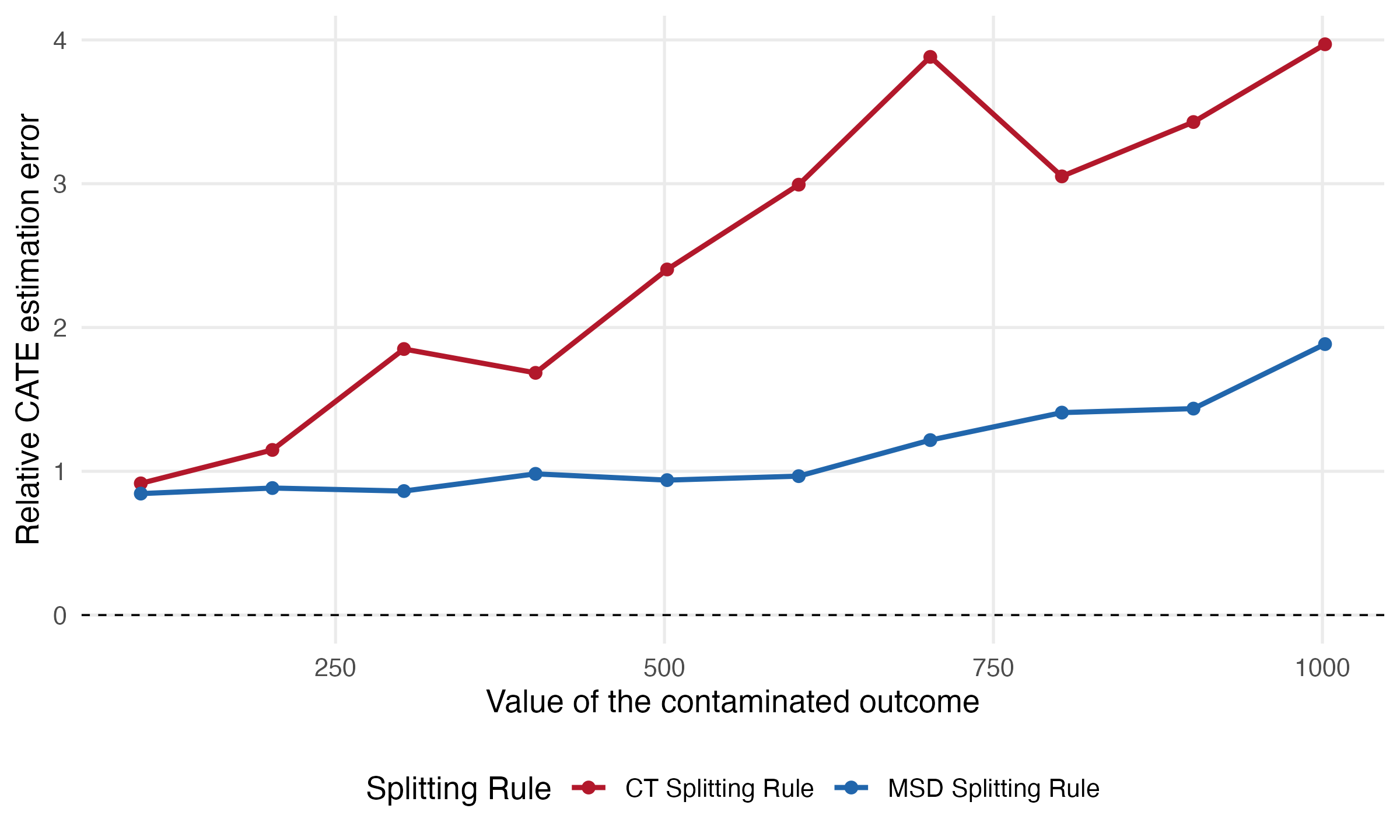}
\begin{notes} Estimation error is evaluated on uncontaminated observations as the magnitude of a single outlier in the training sample increases. In this illustrative design, the MSD-based splitting rule is less affected than the standard CT rule.
\end{notes}
\end{figure}

To motivate the need for robust split selection, we present a simple example. Consider a setting in which each unit $i$ is characterized by covariates $X_i = (X_{i1}, X_{i2})$, where both components are independently drawn from a standard normal distribution. Treatment assignment is random, $D_i \sim \operatorname{Bernoulli}(0.5)$, and the treatment effect  depends on the second covariate
\begin{align*}
\tau(X_i)=
\begin{cases}
8, & X_{i2}>0,\\
0, & X_{i2}\leq 0.
\end{cases}
\end{align*}
The observed outcome variable is generated according to
\[ Y_i=\tau(X_i)D_i+\varepsilon_i, \qquad \varepsilon_i\sim\mathcal{N}(0,1).\]
Beginning with an uncontaminated training sample, we introduce a single outlier to the outcome variable $Y_i$ by replacing its value with $Y_i^{c} \in \{102, 202, \ldots, 1002\}$, while holding the covariates $X_i$ and treatment assignment $D_i$ fixed. Estimation accuracy is then evaluated on an independent, uncontaminated test sample. We compare the CT splitting rule with the Median Squared Deviation (MSD) criterion introduced in the next section. Figure~\ref{fig:motivation_robust_splitting} shows the evolution of the estimation error as the magnitude of the outlier increases.
 The MSD criterion is substantially less sensitive to the increasing contamination. Up to an outlier magnitude of approximately 600, its estimation error remains close to its uncontaminated level, whereas the estimation error of the standard CT has already increased by roughly a factor of three. Over the full range of outlier magnitudes considered, the estimation error approximately doubles for MSD but increases by about a factor of four for CT. Although this example is purely illustrative and such an extreme observation would typically be investigated during data preprocessing, it nevertheless demonstrates that the standard causal tree is not robust to outcome contamination by default. Even a single extreme outcome can substantially alter the selected partition and thereby degrade CATE estimation accuracy.
 
\section{Median-based Splitting Criteria for the Causal Tree}
\label{sec:splitting_criteria}
\label{ch:Median_Splits}
This section briefly reviews the three median-based splitting rules introduced by \citet{maßmann2026medianbasedsplittingrulescausal}, as alternatives to the MSE-based splitting criterion in Equation~\eqref{eq:robust_MSE_CT_H}. The first two criteria, MSD and the MAD, use the Hodges--Lehmann estimator \citep{hodges_estimates_1963, hoyland_HL} as a robust splitting ``anchor'', that is, the location estimator around which the within-node dispersion is measured. The third criterion, LMS, instead robustifies an outcome-fit-based splitting criterion by replacing the average squared residual with the median squared residual. Of the three median-based criteria considered here, only MSD has a theoretically derived honest analog, while MAD and LMS are formulated as adaptive splitting criteria. In all implementations considered below, however, final leaf-level treatment effects are estimated honestly using the separate estimation sample $\mathcal{S}^{est}$. Before we discuss each splitting rule in detail, let us introduce the Hodges--Lehmann estimator. 

Let $\mathcal{S}_{1} \subset \mathcal{S}$ and $\mathcal{S}_{0}\subset \mathcal{S}$ denote the treatment and control group of a sample $\mathcal{S}$, with corresponding sample sizes $N_{1}$ and $N_{0}$. On a general level, the Hodges–Lehmann estimator is defined as the median of all pairwise treated-minus-control differences,
\begin{equation} \label{eq:Robust_HL_general}
\hat{\Delta}_{HL}:=\operatorname{med}\left\{Y_j-Y_m:j\in\mathcal S_1,m\in\mathcal S_0\right\}.
\end{equation}
Therefore, the Hodges–Lehmann estimator summarizes the relative location of the two outcome distributions using all $N_1\cdot N_0$ pairwise differences. Its natural interpretation is as an estimator of a location-shift parameter. In particular, if the treatment and control outcome distributions, $F_1$ and $F_0$, have the same shape and differ only by a shift $\Delta$, so that
\begin{equation}
F_1(z)=F_0(z-\Delta)
\qquad
\text{for all }z\in\mathbb R,
\end{equation}
then the population median of the pairwise treatment-minus-control differences equals $\Delta$, and $\hat{\Delta}_{\mathrm{HL}}$ is a consistent estimator of $\Delta$.
Since we use the Hodges–Lehmann estimator as an anchor for split selection, we also need a leafwise version of \eqref{eq:Robust_HL_general}. Given a tree $\Pi$ with a leaf $l$, we define the leafwise Hodges--Lehmann estimator as
\[
\hat{\tau}_{\mathrm{HL}}(l;\mathcal{S})
=
\operatorname{med}
\left\{
Y_j-Y_m
:
j\in\mathcal{S}_1,\;
m\in\mathcal{S}_0,\;
X_j\in l,\;
X_m\in l
\right\}.
\]
Because it is based on a median of pairwise differences rather than an average outcome difference, the Hodges--Lehmann estimator is less sensitive to extreme observations than the standard difference-in-means estimator \citep{hoyland_HL}.

\subsection{Median Squared Deviation (MSD)}
\label{sec:MSD_short}

The Median Squared Deviation (MSD) criterion is motivated by \eqref{eq:MSE_treatment_effect}, but replaces the mean-based leafwise treatment effect estimate with the Hodges--Lehmann estimator. As before, we start from the infeasible treatment effect criterion, 
\begin{equation}
\label{eq:mse_tau_med_short}
\operatorname{MSD}_{\tau}(\mathcal{S}^{te},\mathcal{S}^{est})
=
\frac{1}{N^{te}}
\sum_{i\in\mathcal{S}^{te}}
\left[
\left(
\tau_i-\hat{\tau}_{\mathrm{HL}}(X_i;\mathcal{S}^{est})
\right)^2
-\tau_i^2
\right].
\end{equation}

As in the mean-based criterion, the subtraction of $\tau_i^2$ does not affect the ranking of candidate partitions, since it is independent of $\Pi$. The formal honest MSD criterion replaces the mean-based leaf effect and its variance correction with the Hodges–Lehmann location estimator and the corresponding asymptotic variance term, which is discussed in detail in \cite{maßmann2026medianbasedsplittingrulescausal}. For completeness, the resulting honest MSD criterion is reported in Appendix~\ref{appendix:additional_theory}. Because evaluating this variance requires repeated density estimation within every candidate leaf, the authors instead use the computationally simpler form
\begin{equation}
\label{eq:msd_feasible_short}
\widehat Q^{\operatorname{MSD}}_{\tau}(\mathcal{S}^{tr},\Pi)
=\sum_{l\in\Pi}\frac{N_{l}^{tr}}{N^{tr}}
\left[
\hat{\tau}_{\mathrm{HL}}^2(l;\mathcal{S}^{tr})
-
\left(
\hat{\tau}_{\mathrm{HL}}(l;\mathcal{S}^{tr})
-
\hat{\tau}(l;\mathcal{S}^{tr})
\right)^2
\right].
\end{equation}
Candidate splits are chosen to maximize  $\widehat Q^{\operatorname{MSD}}_{\tau}(\mathcal{S}^{tr},\Pi)$. The first term favors partitions with larger Hodges--Lehmann leaf-level treatment effect estimates, while the second term penalizes the squared difference between the Hodges--Lehmann and mean-based leaf estimates.
\subsection{Median Absolute Deviation (MAD)}
\label{sec:MAD_short}

The Median Absolute Deviation (MAD) criterion is based on the least-absolute-deviation (LAD) splitting rules in robust regression forests \citep{roy_robustness_2012}. Compared with squared-error criteria, absolute deviations place less weight on large residuals and are therefore less sensitive to extreme outcome observations. 

Starting with the infeasible absolute deviation criterion,
\begin{equation}\label{eq:infeasible_MAD}
    \mathrm{MAD}_\tau(\mathcal{S}^{te})
    \;:=\;\frac{1}{N^{te}}\sum_{i\in\mathcal{S}^{te}}
    \bigl|\tau_i - \tau_{\mathrm{med}}(X_i)\bigr|,
\end{equation}

where $\tau_{\mathrm{med}}(X_i)$ denotes the conditional median treatment effect (CMTE) 
\begin{align} \label{eq:CMTE}
        \tau_{\operatorname{med}}(x) &\coloneq \operatorname{med}\bigl(Y_i(1) - Y_i(0)\mid X_i=x \bigr).  
\end{align}
As in the MSE criterion, \eqref{eq:infeasible_MAD} cannot be used directly for split selection because $\tau_i$ is unobserved. Moreover, $\tau_{\mathrm{med}}(X_i)$ is not point-identified without
restrictions on the joint distribution of potential outcomes
\citep{Addanki2024LimitsOA}. We obtain a practical and feasible rule by replacing $\tau_i$ by
the mean-based plug-in $\hat\tau(X_i;\mathcal{S}^{tr},\Pi)$ and
$\tau_{\mathrm{med}}(X_i)$ by the leafwise HL estimator
$\hat\tau_{\mathrm{HL}}(X_i;\mathcal{S}^{tr},\Pi)$, resulting in 
\begin{equation}\label{eq:MAD_split}
    \widehat Q^{\operatorname{MAD}}_{\tau}(\mathcal{S}^{tr},\Pi)
    \;:=\;\frac{1}{N^{tr}}\sum_{i\in\mathcal{S}^{tr}}
    \bigl|\hat\tau(X_i;\mathcal{S}^{tr},\Pi)
    -\hat\tau_{\mathrm{HL}}(X_i;\mathcal{S}^{tr},\Pi)\bigr|.
\end{equation}
Splits are chosen by minimizing \eqref{eq:MAD_split}. So the criterion
favors partitions on which the mean-based and Hodges--Lehmann leaf estimates agree. 

\subsection{Least Median of Squares (LMS)}
\label{sec:LMS_short}

The Least Median of Squares (LMS) criterion is motivated by the robust regression estimator of \citet{rousseeuw_least_1984}. Instead of minimizing the average squared residual as in least squares, LMS minimizes the median of squared residuals. This gives the estimator a high breakdown point and makes it less sensitive to extreme outcome observations, but can be computationally demanding, especially in high-dimensional settings \citep{mount_practical_2007,berenguer-rico_models_2019}.  Starting with the adaptive fit-based tree splitting criterion \citep{ZeileisModelBasedRP}, which is defined as 
\begin{equation}\label{eq:MSE_fit_A}
    \mathrm{MSE}^{\mathrm{fit\text{-}A}}_\mu(\mathcal{S}^{tr},\mathcal{S}^{tr},\Pi)
    \;:=\;\frac{1}{N^{tr}}\sum_{i\in\mathcal{S}^{tr}}
    \bigl[(Y_i-\hat\mu_{D_i}(X_i;\mathcal{S}^{tr},\Pi))^2-Y_i^2\bigr],
\end{equation}
where, $\hat{\mu}_{D_i}(X_i;\mathcal{S}^{tr},\Pi)$ denotes the fitted leaf mean for the observed treatment group of unit $i$. The LMS criterion robustifies this fit-based rule by replacing the mean of squared residuals with their median, leading to

\begin{equation}\label{eq:LMS_split_short}
\widehat Q_{\mu}^{\mathrm{LMS}}
(\mathcal{S}^{tr},\Pi)
:=
\sum_{l\in\Pi}
\frac{N_l^{tr}}{N^{tr}}
\underset{i\in\mathcal{S}^{tr}_l}{\operatorname{med}}
\left[
\left(
Y_i-\hat\mu_{D_i}(X_i;\mathcal{S}^{tr},\Pi)
\right)^2
\right].
\end{equation}

Candidate splits are chosen to minimize
$\widehat Q^{\operatorname{LMS}}_{\mu}(\mathcal{S}^{tr},\Pi)$. Unlike MSD and MAD, LMS is based on outcome fit rather than on a treatment effect deviation criterion. Compared with the classical fit-based criterion, LMS reduces the influence of large residuals and is therefore designed to produce partitions that are less driven by outlying outcomes.

\section{Confidence Interval Construction}
\label{sec:confidence_intervals}
In this section, we compare two methods for constructing pointwise confidence intervals, the Bootstrap of Little Bags (BLB) and the Infinitesimal Jackknife (IJ) approach. Let $\hat{\tau}(x)$ denote an honest forest-based estimator constructed from $B$  causal trees that targets the CATE defined in Equation \eqref{eq:CATE}. To quantify the uncertainty around each estimate $\hat{\tau}(x)$, we aim to estimate its asymptotic variance. For $D_i \in \{0, 1\}$, \citet{athey_generalized_2019} have shown that the asymptotic variance satisfies 
\begin{align} \label{eq:robust_variancedecomposition}
    \Var\left(\hat\tau_{\mathrm{GRF}}(x)\right) = V(x)^{-2} H_N^{GRF}(x).
\end{align}
The term $V(x)$ reflects the local identification strength and is defined as  $$V(x) =\mathbb{E}\left[ \left(D_i - p(X_i)\right)^2 \,\big|\, X_i = x \right],$$ while $\hat\tau_{\mathrm{GRF}}(x)$ denotes the CATE estimator obtained within the generalized random forest framework, and $H_N^{\mathrm{GRF}}(x)$ denotes the corresponding variance component associated with the treatment effect.  In contrast, our implementation follows the causal forest algorithm introduced by \citet{athey_recursive_2016} and applies both BLB and IJ directly to the tree-level CATE predictions $\hat{\tau}_b(x)$ produced by the individual trees.
 In the following, $\widehat H_N(x)$ denotes a generic estimator of the forest variance, while $\widehat H_N^{\mathrm{BLB}}(x)$ and $\widehat H_N^{\mathrm{IJ}}(x)$ refer to its BLB and IJ implementations, respectively.
The relationship between $\widehat H_N(x)$ and $H_N^{\mathrm{GRF}}(x)$ is discussed in Appendix~\ref{appendix:Robust_variance_discussion}. In the simulation study in Section \ref{sec:robust_simulation_study}, we consider both the unadjusted variance estimates $\widehat H_N^{\mathrm{BLB}}(x)$ and $\widehat H_N^{\mathrm{IJ}}(x)$ and curvature-adjusted versions obtained by multiplying these estimates by $\widehat V(x)^{-2}$.

We apply the variance estimation methods considered below to all splitting rules under the same forest-growing procedure. Since the median-based rules modify only the split-selection criterion, they do not change the subsampling scheme, the honesty construction, the random covariate selection, or the leaf-level CATE estimator. Thus, BLB and IJ estimate the uncertainty of the resulting forest predictions in the same way for CT, MSD, MAD, and LMS. The formal asymptotic guarantees in \citet{athey_generalized_2019} and \citet{wager_estimation_2018} are derived under their respective honesty and regularity conditions. For the median-based splitting rules, we therefore apply the same variance-estimation principles to the modified forest predictions rather than claim a separate theoretical validity result for each splitting criterion.
\subsection{Bootstrap of Little Bags (BLB) Algorithm} 

The first we consider is the BLB approach. Specifically, we use it to operationalize the infeasible half-sampling variance estimator  $H_N^{\text{HS}}(x)$ as suggested in Section 4.1 of \citet{athey_generalized_2019}. 

Let $l \ge 2$ be a fixed integer specifying the number of trees per bag, and set the total number of bags to $G = \lfloor B / l \rfloor$. For each bag $g = 1, \dots, G$, we independently draw a half-sample
\begin{align*}
    \mathcal{S}_{\text{half},g} \subset \{1,\dots,N\}, \qquad |\mathcal{S}_{\text{half},g}| = \lfloor N/2 \rfloor,
\end{align*}
uniformly without replacement from the full dataset. Given $\mathcal{S}_{\text{half},g}$, we then draw $l$ subsamples $I_{g,b} \subseteq \mathcal{S}_{\text{half},g}$, each of size
\begin{align*}
    |I_{g,b}| = \left\lfloor |\mathcal{S}_{\text{half},g}| \cdot \rho \right\rfloor, \qquad \rho \in (0,1),
\end{align*}
and grow one honest causal tree on each subsample $I_{g,b}$.

Within each bag $g$, let $\hat{\tau}_{g,b}(x)$ denote the CATE prediction of tree $b$. Trees belonging to the same bag share a common half-sample and are therefore dependent, whereas trees across different bags are approximately independent. We define the group-level mean prediction as
\begin{align*}
    \bar{\tau}_g(x) = \frac{1}{l} \sum_{b=1}^{l} \hat{\tau}_{g,b}(x),
\qquad g = 1,\dots,G,
\end{align*}
and the overall mean across all $G$ bags as
\begin{align*}
    \bar{\tau}(x) = \frac{1}{G} \sum_{g=1}^{G} \bar{\tau}_g(x).
\end{align*}

The BLB estimator of the forest variance is obtained from an ANOVA decomposition that separates between- and within-group variation of tree predictions:
\begin{align*}
\widehat{V}_{\text{between}}(x)
&=
\frac{1}{G-1}
\sum_{g=1}^{G}
\left( \bar{\tau}_g(x) - \bar{\tau}(x) \right)^2, \\
\widehat{V}_{\text{within}}(x)
&=
\frac{1}{G}
\sum_{g=1}^{G}
\left[
\frac{1}{l}
\sum_{b=1}^{l}
\left( \hat{\tau}_{g,b}(x) - \bar{\tau}_g(x) \right)^2
\right].
\end{align*}
The classical BLB estimator is then given by
\begin{align*}
\widehat{H}_N^{\text{BLB,raw}}(x)
=
\widehat{V}_{\text{between}}(x)
-
\dfrac{1}{l - 1}\widehat{V}_{\text{within}}(x),
\end{align*}
which can take negative values in finite samples. To guarantee non-negativity, we follow \citet{athey_generalized_2019} and adopt a Bayesian ANOVA adjustment corresponding to an improper uniform prior on $[0,\infty)$ for $H_N(x)$, yielding 
\begin{align*}
\widehat{H}_N^{\text{BLB}}(x)
=
\left( \widehat{V}_{\text{between}}(x)
-
\frac{1}{l-1}\widehat{V}_{\text{within}}(x)
\right)_+,
\end{align*}
where $\left(a\right)_+ = \max\{a,0\}$. As the number of trees $B \to \infty$, this estimator converges to the infeasible half-sampling variance. The BLB estimate of the forest variance is then used as the variance estimate for the forest CATE prediction, yielding 
\begin{align}
    \label{eq:BLB}
    \widehat{\mathrm{Var}}_N^{\text{BLB}}(\hat\tau(x)) =  \, \widehat{H}_N^{\text{BLB}}(x), \qquad \widehat{\mathrm{SE}}^{\mathrm{BLB}}(\hat\tau(x))
=
\sqrt{
\widehat{\mathrm{Var}}_N^{\mathrm{BLB}}(\hat\tau(x))
}.
\end{align}

\subsection{Infinitesimal Jackknife (IJ)}

Following \citet{wager_estimation_2018}, the infinitesimal jackknife variance estimator for the forest variance $H_N(x)$ is defined as

\begin{align*}
    \widehat{H}_N^{\text{IJ}}(x) = \frac{N-1}{N} \left(\frac{N}{N-s}\right)^2 \sum_{i=1}^{N} \widehat{\mathrm{Cov}}_b\left[ \hat{\tau}_b(x), I_{ib}^{*} \right]^2,
\end{align*}

where $I_{ib}^* \in \{0,1\}$ indicates whether observation $i$ was included in the subsample used to grow tree $b$, and the covariance is taken across all $B$ trees. The term $\frac{N-1}{N}\left(\frac{N}{N-s}\right)^2$ is a finite-sample correction for forests grown by subsampling without replacement of size $s$.

The estimated covariance for each observation $i$ is computed as
\begin{align*}
    \widehat{\mathrm{Cov}}_b\left[I_{ib}^*, \hat{\tau}_b(x)\right] = \frac{1}{B} \sum_{b=1}^{B} \left(I_{ib}^* - \bar{I}_{i}\right)\left(\hat{\tau}_b(x) - \bar{\tau}(x)\right),
\end{align*}

where $\bar{I}_i = B^{-1}\sum_{b=1}^B I_{ib}^*$ and $\bar{\tau}(x) = B^{-1}\sum_{b=1}^B \hat{\tau}_b(x)$ denote the empirical means across trees.
The IJ estimator measures the sensitivity of the forest prediction to the inclusion of each training observation by summing squared empirical influence estimates across all training observations. \citet{wager_estimation_2018} show in Theorem 3.1 that, under honesty and regularity conditions,

\[
\widehat H_N^{\mathrm{IJ}}(x)/\sigma_N^2(x)\rightarrow_p1,
\]

establishing consistency of the IJ estimator for the forest variance. The IJ estimate of the forest variance is then used as the variance estimate for the forest CATE prediction, yielding

\begin{align}
    \label{eq:IJ}
    \widehat{\mathrm{Var}}_N^{\mathrm{IJ}}(\hat\tau(x))
=
\widehat H_N^{\mathrm{IJ}}(x),
\qquad
\widehat{\mathrm{SE}}^{\mathrm{IJ}}(\hat\tau(x))
=
\sqrt{\widehat{\mathrm{Var}}_N^{\mathrm{IJ}}(\hat\tau(x))}.
\end{align}

\section{Simulation Study}
\label{sec:robust_simulation_study}

To investigate the performance of the median-based methods in the presence of outcome irregularities, we conduct a simulation study with three parts. First, we study estimation performance when the outcome errors are generated from heavy-tailed distributions. Second, within the same heavy-tailed designs, we evaluate pointwise confidence intervals and compare the Bootstrap of Little Bags procedure with the Infinitesimal Jackknife. Third, we examine robustness to isolated outcome contamination, considering both mean-shift and variance contamination. For each part, we use four simulation designs. The first three simulation designs are based on the heterogeneous treatment effect settings of  \citet{kunzel_metalearners_2019} and vary in the complexity of the underlying CATE function, ranging from linear treatment effects to nonlinear and zero-effect settings. The fourth simulation design is based on \citet{rcate_Li} and contains covariate interaction terms. 
\subsection{Setup}

For each design, we generate independent training and test samples. The training sample is used to estimate the CATE function, while the test sample is used only for evaluation. In the heavy-tailed outcome noise experiment, both training and test samples are generated from the same error distribution. In the contamination experiment, contamination is introduced only into the training sample, while the test sample remains uncontaminated. Across all designs, we compare the standard causal tree splitting rule from Equation~\eqref{eq:robust_MSE_CT_H}, which serves as the benchmark, with the median-based alternatives introduced in Section~\ref{ch:Median_Splits}. 

For each design, we use $K=10$ covariates and simulate $MC=1000$ Monte Carlo replications, with $N^{\text{training}} =N^{{te}}=1000$ in each replication\footnote{Since the honest sample splitting is done internally by the causal forest implementation, we report only the total training sample size. Formally, we have $\mathcal{S}^{\text{training}}=\mathcal{S}^{tr}\cup\mathcal{S}^{est}$ and $\mathcal{S}^{tr}\cap\mathcal{S}^{est}=\emptyset$.}. 
For each observation $i$, $X_{i1},\dots,X_{i10}$ are generated independently from a standard normal distribution,
\[
X_{ij} \sim \mathcal{N}(0, 1), \quad \text{for } j = 1, \dots, 10.
\]
The potential outcomes are generated as
\begin{align*}
Y_i(1)&=\mu_1(X_i)+\varepsilon_i(1),\\Y_i(0)&=\mu_0(X_i)+\varepsilon_i(0),
\end{align*}
where $\varepsilon_i(1),\varepsilon_i(0)\overset{iid}{\sim}\mathcal{N}(0,1)$. We simulate the treatment according to
 \[
D_i \sim \operatorname{Bernoulli}(0.5).
\]
Finally, we set \[Y_i = D_iY_i(1)+(1-D_i)Y_i(0),\] such that
the observed data consist of $(X_i,D_i,Y_i)$.
\subsection{Data-generating processes} \label{sec:simulation_dgps}
\paragraph{Linear treatment effect (\textbf{S1}).}

The first design considers a heterogeneous but linear treatment effect. Both potential outcome functions are linear in the covariates, but they are parameterized by different coefficient vectors:
\begin{align*}
\mu_0(x) &= x^\top \beta_0,
\qquad
\beta_0 \sim \operatorname{Unif}[1,30]^K, \\
\mu_1(x) &= x^\top \beta_1,
\qquad
\beta_1 \sim \operatorname{Unif}[1,30]^K.
\intertext{The resulting CATE is}
\tau(x)
&=
\mu_1(x)-\mu_0(x)
=
x^\top(\beta_1-\beta_0).
\end{align*}
This setting provides a baseline case with heterogeneous treatment effects but a simple linear structure.

\paragraph{Complex nonlinear treatment effect (\textbf{S2}).}

The second design introduces nonlinear treatment effect heterogeneity through a sigmoid transformation of the first two covariates. The potential outcome functions are defined as
\begin{align*}
\mu_0(x) &= -\frac{1}{2}\zeta(x_1)\zeta(x_2), \\
\mu_1(x) &= \phantom{-}\frac{1}{2}\zeta(x_1)\zeta(x_2),
\intertext{where}
\zeta(x)
&=
\frac{2}{1+\exp\{-12(x-0.5)\}}.
\intertext{This implies}
\tau(x)
&=
\mu_1(x)-\mu_0(x)
=
\zeta(x_1)\zeta(x_2).
\end{align*}
This design is more challenging than the linear case, because the treatment effect changes nonlinearly and sharply with the covariates.

\paragraph{Global linear case with no treatment effect (\textbf{S3}).}

The third design considers a setting in which the two potential outcome functions are identical. Hence, there is no treatment effect heterogeneity and the CATE is zero for all observations
\begin{align*}
\mu_0(x) &= x^\top\beta,
\qquad
\beta \sim \operatorname{Unif}[1,30]^K, \\
\mu_1(x) &= \mu_0(x).
\end{align*}
Therefore,
\[
\tau(x)
=
\mu_1(x)-\mu_0(x)
=
0
\quad
\text{for all } x.
\]
This design is included to examine whether the methods spuriously detect treatment effect heterogeneity when none is present.

\paragraph{Complex nonlinear CATE with covariate interactions (\textbf{S4}).}

The fourth design considers a more complex treatment effect function with nonlinear terms and covariate interactions. The baseline outcome function is given by
\begin{align*}
\mu_0(x)
&=
100 + 4x_1 + x_2 - 3x_3 - \frac{1}{2}\tau(x),
\intertext{and the treated outcome is defined as}
\mu_1(x)&=\mu_0(x) + \tau(x).
\end{align*}
The CATE is
\begin{align*}
\tau(x)
&=
6\sin(2x_1)
+
3(x_2+3)x_3
+
9\tanh(0.5x_4)
+
3x_5\left(2\mathds{1}\{x_4>0\}-1\right) \\
&\quad
+
3x_6
+
2x_7
+
x_8
-
2x_9
-
4x_{10}.
\end{align*}
This design is included because the treatment effect is both nonlinear and non-additive. In particular, the term $3(x_2+3)x_3$ introduces an interaction between covariates, while the sine, hyperbolic tangent, and threshold terms generate nonlinear treatment effect heterogeneity.
\subsection{Outcome irregularities} \label{sec:contamination}

We consider two forms of outcome irregularity. The first is heavy-tailed outcome noise, where extreme observations arise from the outcome-generating process itself. The second is artificial outcome contamination adapted from \citet{roy_robustness_2012}, where an increasing proportion of training observations is contaminated by adding an additional error component to the observed outcome.

\paragraph{Heavy-tailed outcome noise.}
To study robustness to heavy-tailed outcome distributions, we vary the distribution of the error terms in the potential outcomes. In the baseline simulation design, the potential outcomes are generated with
\[
\varepsilon_i(1),\varepsilon_i(0)\overset{iid}{\sim}\mathcal{N}(0,1).
\]
In the heavy-tailed outcome-noise designs, we replace the normal error terms with alternative error distributions. Specifically, we consider Student-$t$, Cauchy, and Laplace errors.  For the Student-$t$ distribution, we rescale the errors by their standard deviation so that the resulting errors have unit variance, consistent with the baseline design. This leads to the following error distributions
\[
\varepsilon_i(1),\varepsilon_i(0)\overset{iid}{\sim} F_{\varepsilon},
\qquad
F_{\varepsilon}\in\left\{\dfrac{\operatorname{t(3)}}{\sqrt{3}},\operatorname{Cauchy}(0,1),\operatorname{Laplace}(0,1)\right\}.
\]

Note that, under Cauchy errors, the conditional expectations defining the CATE do not exist. However, the functions $\mu_1(x)$ and $\mu_0(x)$ remain well-defined as location parameters of the potential outcome distributions. We therefore use $\mu_1(x)-\mu_0(x)$ as the reference function for evaluating estimation accuracy in the Cauchy design.

\paragraph{Outcome contamination.}
Next, we consider two artificial contamination strategies adapted from \citet{roy_robustness_2012}, namely variance contamination and mean-shift contamination. 
In contrast to the heavy-tailed outcome-noise experiment, the idea is to introduce a prespecified proportion of outliers into the training sample, while the test sample remains uncontaminated. 
This allows us to evaluate how well each method learns the underlying treatment effect structure when the training sample contains irregular observations. Contaminating the test sample as well would reflect a sample-level characteristic and would therefore no longer isolate sensitivity to contamination in the training data. Let $p_c$ denote the proportion of contaminated training observations. We set
\[p_c \in \{0,0.05,0.10,0.15,0.20,0.25,0.30,0.35\},\]
where $p_c=0$ corresponds to the uncontaminated baseline.
Furthermore, let $ \sigma_Y^j $ denote the sample standard deviation of the observed outcomes $\{Y_i\}_{i=1}^{N^{\mathrm{training}}}$ in simulation design $j$. Under variance contamination, for a randomly selected fraction $p_c$ of the training observations, we add mean-zero noise with standard deviation $5\sigma_Y^j$:
\[ Y_i^c = Y_i + e_i, \qquad e_i \sim \mathcal{N}\!\left(0,(5\sigma_Y^j)^2\right).
\]

Under mean-shift contamination, for a randomly selected fraction $p_c$ of the training observations, we add shifted noise with mean $5\sigma_Y^j$ and standard deviation $\sigma_Y^j$:
\[
Y_i^c = Y_i + e_i,
\qquad
e_i \sim \mathcal{N}\!\left(5\sigma_Y^j,(\sigma_Y^j)^2\right).
\]

\subsection{Evaluation Criteria}
\label{sec:simulation_eval}

We evaluate performance along three dimensions. First, we assess estimation accuracy. Second, we report pointwise confidence interval coverage and the width of the confidence intervals. Third, we measure robustness by comparing performance under contamination to the clean baseline. The root mean squared error (RMSE) and the mean absolute error (MAE) are defined as

\begin{align}
\operatorname{RMSE}
&=
\sqrt{
\frac{1}{N^{te}}
\sum_{i=1}^{N^{te}}
\left(
\tau(X_i)-\hat{\tau}(X_i)
\right)^2
}, \\
\operatorname{MAE}
&=
\frac{1}{N^{te}}
\sum_{i=1}^{N^{te}}
\left|
\tau(X_i)-\hat{\tau}(X_i)
\right|.
\end{align}

In the heavy-tailed outcome-noise experiments, we additionally evaluate the quality of the pointwise $95\%$ confidence intervals. For observation $i$, the confidence interval is defined as
\[
\mathrm{CI}_{i}
=
\left[
\hat{\tau}(X_{i})
-
1.96\,\widehat{\mathrm{SE}}\!\left(\hat{\tau}(X_{i})\right),
\;
\hat{\tau}(X_{i})
+
1.96\,\widehat{\mathrm{SE}}\!\left(\hat{\tau}(X_{i})\right)
\right],
\]
where $\widehat{\mathrm{SE}}(\hat{\tau}(X_i))$ is computed using either the BLB or IJ variance estimator. For each simulation replication, empirical coverage and the average interval width are defined as
\begin{align}
\operatorname{Coverage}
&=
\frac{1}{N^{te}}
\sum_{i=1}^{N^{te}}
\mathds{1}
\left\{
\tau(X_i)
\in
\mathrm{CI}_i
\right\},\\
\operatorname{Width}
&=
\frac{1}{N^{te}}
\sum_{i=1}^{N^{te}}
\left(
\mathrm{CI}^{\mathrm{upper}}_i
-
\mathrm{CI}^{\mathrm{lower}}_i
\right).
\end{align}

For mean-shift and variance contamination, CATE estimation accuracy is additionally summarized by a stabilized mean absolute percentage error (MAPE). Since Scenario S3 has $\tau(x)=0$ for all observations, the usual percentage error with denominator $|\tau(X_i)|$ would be undefined. We therefore stabilize the denominator by using $\max\{|\tau(X_i)|,1\}$:
\[
\operatorname{MAPE}
=
\frac{100}{N^{te}}
\sum_{i=1}^{N^{te}}
\frac{
\left|\hat{\tau}(X_i)-\tau(X_i)\right|
}{
\max\{|\tau(X_i)|,1\}
}.
\]
This measure is a percentage error when $|\tau(X_i)|>1$, and equals the absolute error on the original scale multiplied by 100 when $|\tau(X_i)|\leq 1$.

To quantify robustness in the contamination design, we report percentage changes in the error measures relative to the clean baseline $p_c=0$. For each contamination level $p_c$, we define
\begin{align}
\operatorname{PRMSE}(p_c)
&=
100\cdot
\frac{
\operatorname{RMSE}(p_c)
-
\operatorname{RMSE}(0)
}{
\operatorname{RMSE}(0)
}, \\
\operatorname{PMAE}(p_c)
&=
100\cdot
\frac{
\operatorname{MAE}(p_c)
-
\operatorname{MAE}(0)
}{
\operatorname{MAE}(0)
},\\
\operatorname{PMAPE}(p_c)
&=
100\cdot
\frac{\operatorname{MAPE}(p_c)-\operatorname{MAPE}(0)}{\operatorname{MAPE}(0)}.
\end{align}
Positive values indicate a deterioration relative to the clean baseline, whereas values close to zero indicate that performance is largely unaffected by contamination. Negative values can occur when the contaminated training sample yields a lower test error than the uncontaminated baseline. The $\operatorname{RMSE}$, $\operatorname{MAE}$, and $\operatorname{MAPE}$ measure CATE estimation accuracy, so we refer to these quantities as CATE estimation accuracy metrics. By contrast, $\operatorname{PRMSE}$, $\operatorname{PMAE}$, and $\operatorname{PMAPE}$ measure percentage changes relative to the clean baseline and are therefore referred to as robustness metrics. All reported values are averaged across simulation replications, which is described in Appendix~\ref{appendix:Robust_ct_monte_additional_results}.
\subsection{Results}
Following the structure of the simulation study, we present the results in three parts. First, we evaluate estimation performance under heavy-tailed outcome noise. Second, we examine the corresponding pointwise confidence intervals. Third, we analyze robustness under mean-shift and variance contamination.

\subsubsection{Outcome Noise Analysis}

Table~\ref{tab:cate_metric} summarizes CATE estimation performance under the alternative outcome-noise distributions. Performance is measured by $\operatorname{RMSE}$ and $\operatorname{MAE}$, with standard deviations across simulation runs reported in parentheses.  The normal-error design serves as the benchmark, while the Student-$t$, Cauchy, and Laplace designs are the alternative outcome-noise distributions.

Overall, the results show that median-based splitting rules do not uniformly improve CATE estimation accuracy under heavy-tailed outcome noise. 
Under Student-$t$ and Laplace errors, the results are very similar to the normal error benchmark. In S1, CT has the lowest $\operatorname{RMSE}$ and $\operatorname{MAE}$ under all four noise distributions. For example, under Student-$t$ errors, CT has $\operatorname{RMSE}=28.93$ and $\operatorname{MAE}=22.80$, while the corresponding values are $30.04$ and $23.74$ for LMS, $29.86$ and $23.56$ for MAD, and $33.13$ and $26.31$ for MSD. A similar pattern appears under Laplace errors. In S4, CT also performs best under normal, Student-$t$, and Laplace errors, while the median-based rules produce slightly larger errors. These results suggest that moderate deviations from normality do not by themselves cause clear advantages for median-based split selection.

The biggest differences arise under Cauchy noise. Because the Cauchy distribution has extremely heavy tails, both estimation errors and variability across simulation runs increase substantially. In this setting, the median-based criteria improve performance in some scenarios. For example, in S2, MSD achieves the lowest $\operatorname{RMSE}$, with $10.55$, compared to $11.36$ for CT, $11.16$ for LMS, and $11.32$ for MAD. In S3, MSD again performs best, with $\operatorname{RMSE}=9.04$, compared with $9.85$ for CT, $9.62$ for LMS, and $9.84$ for MAD. The same pattern is visible for $\operatorname{MAE}$ in S3, where MSD has the smallest value among all methods. Since the Hodges--Lehmann estimator is constructed from the  median of the pairwise treated-control outcome differences, it is less sensitive to extreme observations than a mean-based estimator. Hence, we expect these rules to be most effective when the outcome distribution has extreme observations that would otherwise distort split selection. However, the Cauchy results do not imply uniform superiority of the median-based criteria. In S1, CT still has the lowest $\operatorname{RMSE}$ and $\operatorname{MAE}$. In S4, the methods have very similar $\operatorname{RMSE}$ values, with MAD only marginally below CT, while CT retains the lowest $\operatorname{MAE}$. 

\begin{table}[!h]
\centering
\caption{\label{tab:cate_metric} CATE estimation accuracy under different outcome-noise distributions across the four simulation scenarios (S1–S4).}
\centering
\resizebox{\ifdim\width>\linewidth\linewidth\else\width\fi}{!}{
\fontsize{10}{12}\selectfont
\begin{tabular}[t]{ll>{\raggedright\arraybackslash}p{0.25cm}rr>{\raggedleft\arraybackslash}p{0.25cm}rr>{\raggedleft\arraybackslash}p{0.25cm}rr>{\raggedleft\arraybackslash}p{0.25cm}rr}
\toprule
\multicolumn{1}{c}{ } & \multicolumn{1}{c}{ } & \multicolumn{1}{c}{ } & \multicolumn{2}{c}{CT} & \multicolumn{1}{c}{ } & \multicolumn{2}{c}{LMS} & \multicolumn{1}{c}{ } & \multicolumn{2}{c}{MAD} & \multicolumn{1}{c}{ } & \multicolumn{2}{c}{MSD} \\
\cmidrule(l{3pt}r{3pt}){4-5} \cmidrule(l{3pt}r{3pt}){7-8} \cmidrule(l{3pt}r{3pt}){10-11} \cmidrule(l{3pt}r{3pt}){13-14}
Scenario & Noise &  & RMSE & MAE &  & RMSE & MAE &  & RMSE & MAE &  & RMSE & MAE\\
\midrule
 &  &  & \phantom{(}37.53\phantom{)} & \phantom{(}29.59\phantom{)} &  & \phantom{(}38.28\phantom{)} & \phantom{(}30.43\phantom{)} &  & \phantom{(}38.11\phantom{)} & \phantom{(}30.18\phantom{)} &  & \phantom{(}40.36\phantom{)} & \phantom{(}32.61\phantom{)}\\

 & \multirow{-2}{*}{\raggedright\arraybackslash Cauchy} &  & (101.07) & (79.00) &  & (96.78) & (78.71) &  & (101.19) & (80.69) &  & (84.32) & (74.85)\\
\addlinespace[4pt]

 &  &  & \phantom{(}28.86\phantom{)} & \phantom{(}22.75\phantom{)} &  & \phantom{(}30.00\phantom{)} & \phantom{(}23.72\phantom{)} &  & \phantom{(}29.80\phantom{)} & \phantom{(}23.51\phantom{)} &  & \phantom{(}33.10\phantom{)} & \phantom{(}26.29\phantom{)}\\

 & \multirow{-2}{*}{\raggedright\arraybackslash Laplace} &  & (5.91) & (4.72) &  & (6.08) & (4.85) &  & (5.99) & (4.79) &  & (6.46) & (5.16)\\
\addlinespace[4pt]

 &  &  & \phantom{(}28.61\phantom{)} & \phantom{(}22.55\phantom{)} &  & \phantom{(}29.72\phantom{)} & \phantom{(}23.49\phantom{)} &  & \phantom{(}29.53\phantom{)} & \phantom{(}23.29\phantom{)} &  & \phantom{(}32.79\phantom{)} & \phantom{(}26.04\phantom{)}\\

 & \multirow{-2}{*}{\raggedright\arraybackslash Normal} &  & (6.21) & (4.94) &  & (6.41) & (5.11) &  & (6.30) & (5.02) &  & (6.82) & (5.44)\\
\addlinespace[4pt]

 &  &  & \phantom{(}28.93\phantom{)} & \phantom{(}22.80\phantom{)} &  & \phantom{(}30.04\phantom{)} & \phantom{(}23.74\phantom{)} &  & \phantom{(}29.86\phantom{)} & \phantom{(}23.56\phantom{)} &  & \phantom{(}33.13\phantom{)} & \phantom{(}26.31\phantom{)}\\

\multirow{-8}{*}{\raggedright\arraybackslash S1} & \multirow{-2}{*}{\raggedright\arraybackslash Student-$t$} &  & (6.25) & (4.98) &  & (6.44) & (5.13) &  & (6.33) & (5.05) &  & (6.82) & (5.44)\\
\addlinespace[4pt]

\cmidrule{1-14}
 &  &  & \phantom{(}11.36\phantom{)} & \phantom{(}9.01\phantom{)} &  & \phantom{(}11.16\phantom{)} & \phantom{(}9.02\phantom{)} &  & \phantom{(}11.32\phantom{)} & \phantom{(}9.05\phantom{)} &  & \phantom{(}10.55\phantom{)} & \phantom{(}8.97\phantom{)}\\

 & \multirow{-2}{*}{\raggedright\arraybackslash Cauchy} &  & (46.90) & (36.05) &  & (46.08) & (36.43) &  & (46.99) & (36.65) &  & (44.82) & (37.67)\\
\addlinespace[4pt]

 &  &  & \phantom{(}0.95\phantom{)} & \phantom{(}0.57\phantom{)} &  & \phantom{(}1.00\phantom{)} & \phantom{(}0.60\phantom{)} &  & \phantom{(}0.98\phantom{)} & \phantom{(}0.59\phantom{)} &  & \phantom{(}1.01\phantom{)} & \phantom{(}0.61\phantom{)}\\

 & \multirow{-2}{*}{\raggedright\arraybackslash Laplace} &  & (0.03) & (0.06) &  & (0.03) & (0.06) &  & (0.03) & (0.06) &  & (0.03) & (0.06)\\
\addlinespace[4pt]

 &  &  & \phantom{(}0.93\phantom{)} & \phantom{(}0.55\phantom{)} &  & \phantom{(}0.99\phantom{)} & \phantom{(}0.60\phantom{)} &  & \phantom{(}0.97\phantom{)} & \phantom{(}0.58\phantom{)} &  & \phantom{(}0.99\phantom{)} & \phantom{(}0.60\phantom{)}\\

 & \multirow{-2}{*}{\raggedright\arraybackslash Normal} &  & (0.02) & (0.04) &  & (0.03) & (0.05) &  & (0.03) & (0.05) &  & (0.03) & (0.05)\\
\addlinespace[4pt]

 &  &  & \phantom{(}0.92\phantom{)} & \phantom{(}0.55\phantom{)} &  & \phantom{(}1.00\phantom{)} & \phantom{(}0.61\phantom{)} &  & \phantom{(}0.97\phantom{)} & \phantom{(}0.58\phantom{)} &  & \phantom{(}0.99\phantom{)} & \phantom{(}0.60\phantom{)}\\

\multirow{-8}{*}{\raggedright\arraybackslash S2} & \multirow{-2}{*}{\raggedright\arraybackslash Student-$t$} &  & (0.02) & (0.04) &  & (0.03) & (0.05) &  & (0.03) & (0.05) &  & (0.03) & (0.04)\\
\addlinespace[4pt]

\cmidrule{1-14}
 &  &  & \phantom{(}9.85\phantom{)} & \phantom{(}7.88\phantom{)} &  & \phantom{(}9.62\phantom{)} & \phantom{(}7.83\phantom{)} &  & \phantom{(}9.84\phantom{)} & \phantom{(}7.91\phantom{)} &  & \phantom{(}9.04\phantom{)} & \phantom{(}7.75\phantom{)}\\

 & \multirow{-2}{*}{\raggedright\arraybackslash Cauchy} &  & (43.09) & (33.35) &  & (41.73) & (33.06) &  & (42.33) & (33.07) &  & (40.00) & (33.36)\\
\addlinespace[4pt]

 &  &  & \phantom{(}0.08\phantom{)} & \phantom{(}0.07\phantom{)} &  & \phantom{(}0.08\phantom{)} & \phantom{(}0.07\phantom{)} &  & \phantom{(}0.08\phantom{)} & \phantom{(}0.07\phantom{)} &  & \phantom{(}0.07\phantom{)} & \phantom{(}0.06\phantom{)}\\

 & \multirow{-2}{*}{\raggedright\arraybackslash Laplace} &  & (0.03) & (0.03) &  & (0.03) & (0.03) &  & (0.03) & (0.03) &  & (0.03) & (0.03)\\
\addlinespace[4pt]

 &  &  & \phantom{(}0.06\phantom{)} & \phantom{(}0.05\phantom{)} &  & \phantom{(}0.06\phantom{)} & \phantom{(}0.05\phantom{)} &  & \phantom{(}0.06\phantom{)} & \phantom{(}0.05\phantom{)} &  & \phantom{(}0.05\phantom{)} & \phantom{(}0.04\phantom{)}\\

 & \multirow{-2}{*}{\raggedright\arraybackslash Normal} &  & (0.02) & (0.02) &  & (0.02) & (0.02) &  & (0.02) & (0.02) &  & (0.02) & (0.02)\\
\addlinespace[4pt]

 &  &  & \phantom{(}0.05\phantom{)} & \phantom{(}0.05\phantom{)} &  & \phantom{(}0.05\phantom{)} & \phantom{(}0.04\phantom{)} &  & \phantom{(}0.05\phantom{)} & \phantom{(}0.05\phantom{)} &  & \phantom{(}0.05\phantom{)} & \phantom{(}0.04\phantom{)}\\

\multirow{-8}{*}{\raggedright\arraybackslash S3} & \multirow{-2}{*}{\raggedright\arraybackslash Student-$t$} &  & (0.02) & (0.02) &  & (0.02) & (0.02) &  & (0.02) & (0.02) &  & (0.02) & (0.02)\\
\addlinespace[4pt]

\cmidrule{1-14}
 &  &  & \phantom{(}17.09\phantom{)} & \phantom{(}13.46\phantom{)} &  & \phantom{(}17.12\phantom{)} & \phantom{(}13.61\phantom{)} &  & \phantom{(}17.08\phantom{)} & \phantom{(}13.48\phantom{)} &  & \phantom{(}17.52\phantom{)} & \phantom{(}14.20\phantom{)}\\

 & \multirow{-2}{*}{\raggedright\arraybackslash Cauchy} &  & (70.34) & (55.12) &  & (67.84) & (55.02) &  & (69.08) & (54.46) &  & (63.74) & (54.71)\\
\addlinespace[4pt]

 &  &  & \phantom{(}10.18\phantom{)} & \phantom{(}7.91\phantom{)} &  & \phantom{(}10.34\phantom{)} & \phantom{(}8.06\phantom{)} &  & \phantom{(}10.45\phantom{)} & \phantom{(}8.12\phantom{)} &  & \phantom{(}11.53\phantom{)} & \phantom{(}9.03\phantom{)}\\

 & \multirow{-2}{*}{\raggedright\arraybackslash Laplace} &  & (0.14) & (0.11) &  & (0.15) & (0.11) &  & (0.15) & (0.11) &  & (0.19) & (0.15)\\
\addlinespace[4pt]

 &  &  & \phantom{(}10.17\phantom{)} & \phantom{(}7.91\phantom{)} &  & \phantom{(}10.30\phantom{)} & \phantom{(}8.04\phantom{)} &  & \phantom{(}10.45\phantom{)} & \phantom{(}8.12\phantom{)} &  & \phantom{(}11.52\phantom{)} & \phantom{(}9.03\phantom{)}\\

 & \multirow{-2}{*}{\raggedright\arraybackslash Normal} &  & (0.15) & (0.11) &  & (0.16) & (0.12) &  & (0.16) & (0.12) &  & (0.19) & (0.15)\\
\addlinespace[4pt]

 &  &  & \phantom{(}10.17\phantom{)} & \phantom{(}7.91\phantom{)} &  & \phantom{(}10.29\phantom{)} & \phantom{(}8.03\phantom{)} &  & \phantom{(}10.45\phantom{)} & \phantom{(}8.12\phantom{)} &  & \phantom{(}11.51\phantom{)} & \phantom{(}9.02\phantom{)}\\

\multirow{-8}{*}{\raggedright\arraybackslash S4} & \multirow{-2}{*}{\raggedright\arraybackslash Student-$t$} &  & (0.15) & (0.11) &  & (0.16) & (0.11) &  & (0.16) & (0.12) &  & (0.19) & (0.15)\\
\addlinespace[4pt]
\bottomrule
\end{tabular}}%
\vspace{0.5em}
\begin{minipage}{0.95\linewidth}
\footnotesize
\textit{Notes:} Entries report the Monte Carlo averages of the RMSE and MAE across $MC=1000$ replications, with the corresponding standard deviations in parentheses. Results are shown for the standard causal tree splitting rule (CT) and the three median-based alternatives (LMS, MAD, and MSD) under Normal, Student-$t$, Cauchy, and Laplace outcome errors.
\end{minipage}

\end{table}

Taken together, the heavy-tailed outcome-noise experiment implies that median-based splitting rules do not provide systematic gains under moderately heavy-tailed errors such as Student-$t$ or Laplace noise. Their benefits appear mainly in the most extreme Cauchy setting and only in some data-generating processes. 
\subsubsection{Confidence Intervals}

Tables~\ref{tab:ci_metric_ct_lms} and \ref{tab:ci_metric_mad_msd}, together with  Figures~\ref{fig:ci_s1}-\ref{fig:ci_s4} in Appendix \ref{appendix:Robust_ct_monte_additional_results}
 report the empirical coverage rates and average widths of the nominal $95\%$ pointwise confidence intervals across the four simulation scenarios and outcome-noise distributions. Results are shown for the unadjusted BLB and IJ variance estimators and for their curvature-adjusted counterparts, denoted by $\mathrm{BLB}^{V}$ and $\mathrm{IJ}^{V}$. As discussed in Appendix~\ref{appendix:Robust_variance_discussion}, the curvature adjustment based on $\widehat V(x)^{-2}$ is motivated by the generalized random forest framework, whereas in our implementation BLB and IJ are applied directly to completed tree-level CATE predictions. We report the adjusted intervals primarily as a sensitivity comparison. 

The findings show variations among the four data-generating processes. In S1, both BLB and IJ show severe undercoverage under Normal, Student–$t$, and Laplace noise.  For CT, for example, the unadjusted coverage rates are approximately $0.06$ for BLB and $0.09$ for IJ, while the corresponding curvature-adjusted intervals reach coverage rates of about $0.23$ and $0.34$. The same pattern holds for LMS, MAD, and MSD. Thus, although the curvature adjustment substantially increases coverage, the resulting intervals remain far below the nominal $95\%$ level. Under Cauchy noise, coverage is generally higher, but this is accompanied by substantially wider confidence intervals. 

In S2, a different pattern is observed. Here, the unadjusted BLB and IJ intervals again tend to undercover, but the curvature adjustment produces much larger improvements in empirical coverage. Under Normal, Student-$t$, and Laplace noise, the adjusted IJ reaches coverage rates of approximately $0.80$--$0.89$ across the four splitting rules, compared with approximately $0.02$--$0.12$ for the corresponding unadjusted IJ intervals. Since $\widehat{V}(x)^{-2}$ increases the estimated variance, the adjusted intervals are necessarily wider. In S2, however, this additional width appears useful, as it moves empirical coverage substantially closer to the nominal $95\%$ level.

Scenario S3 is particularly interesting since it involves a null-heterogeneity design with $\tau(x)=0$ for every observation. Because there is no heterogeneous treatment effect to estimate, the problem is closer to a standard regression setting. Here, the unadjusted IJ performs much better than in the other scenarios. Whether the noise is Normal, Student-$t$, or Laplace, its coverage is about $0.95$ to $0.96$ for all four splitting rules, so very close to the nominal level. With the curvature adjustment, coverage rises to almost $1.00$, indicating overcoverage. Thus, in this setting, the additional width is unnecessary to achieve adequate coverage. 

The results for S4 again resemble those for S1. Both BLB and IJ exhibit pronounced undercoverage under Normal, Student-$t$, and Laplace noise. The curvature adjustment increases coverage, but the resulting intervals remain well below the nominal level. 

Overall, the IJ generally achieves slightly higher empirical coverage than the BLB, although neither variance estimator is uniformly better or well calibrated across all simulation designs. Performance varies substantially with the underlying treatment effect structure. In particular, the unadjusted IJ performs well in S3, while substantial undercoverage is still observed in the situations with nonzero treatment effect heterogeneity. The differences between CT, LMS, MAD and MSD are relatively small, indicating that the simulation design and the variance-estimation method drive inference more strongly than the specific splitting rule.

\subsubsection{Contamination Analysis}

\begin{figure}[!htbp]
  \caption{Percentage results for Scenario S1}
  \label{fig:sim-percentage-s1}
  \centering
  \includegraphics[width=0.85\textwidth]{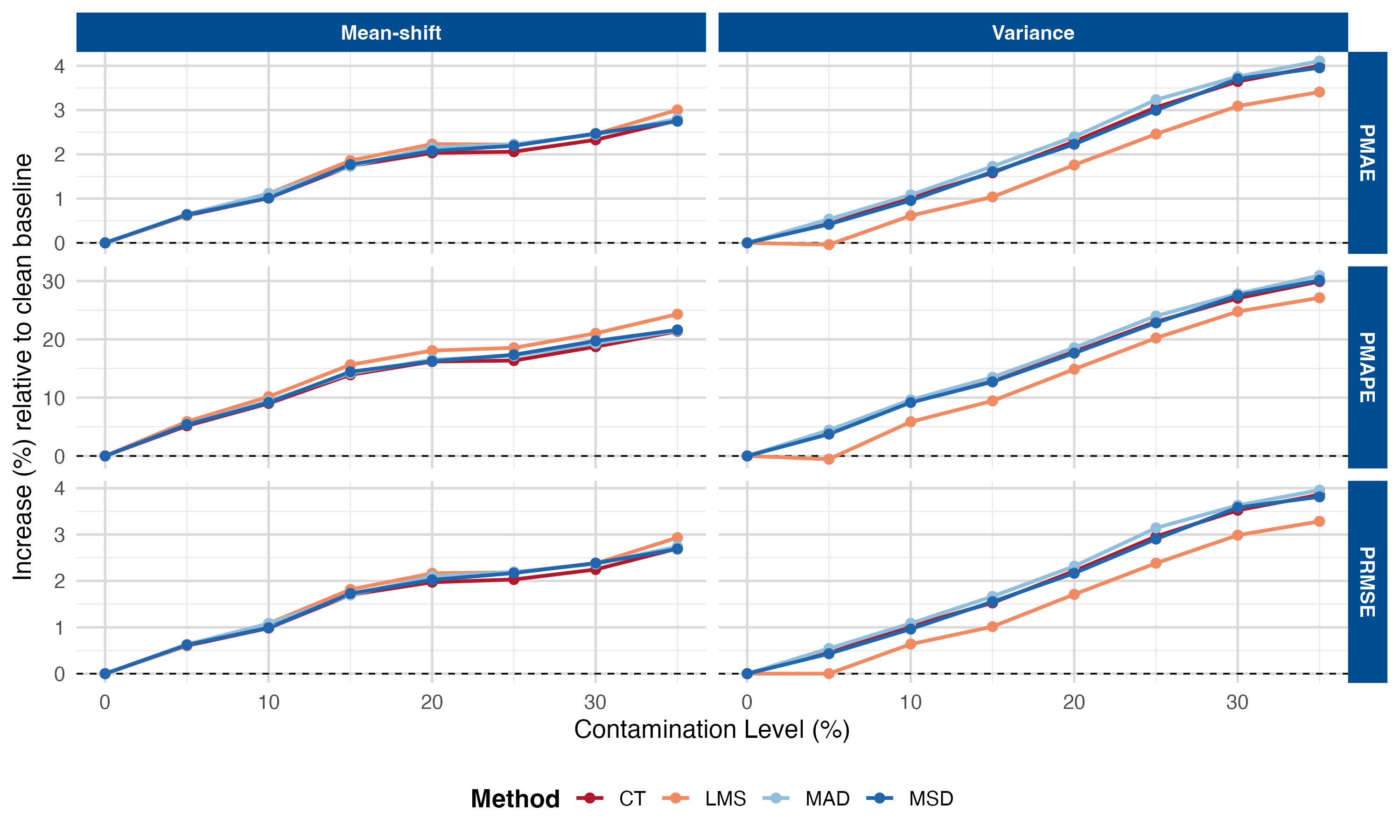}
  \caption*{\scriptsize\textit{Notes:} Percentage changes in PMAE, PMAPE, and PRMSE for each method when compared with the non-contaminated case under Scenario S1, as a function of the proportion of contamination. Columns distinguish mean-shift and variance contamination, while the lines correspond to CT, LMS, MAD, and MSD.}
\end{figure}

\begin{figure}[!htbp]
  \caption{Percentage results for Scenario S2}
  \label{fig:sim-percentage-s2}
  \centering
  \includegraphics[width=0.85\textwidth]{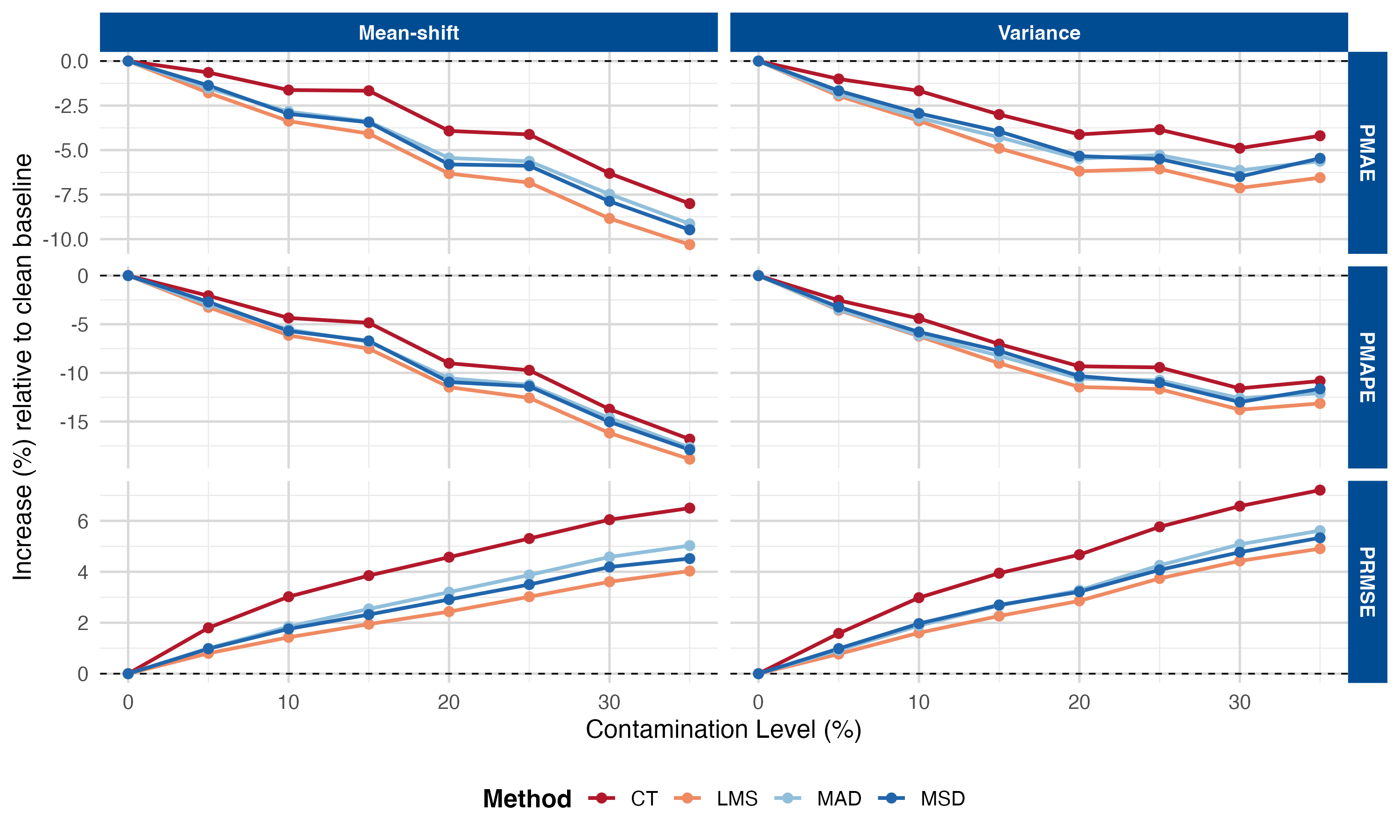}
  \caption*{\scriptsize\textit{Notes:} Percentage changes in PMAE, PMAPE, and PRMSE for each method when compared with the non-contaminated case under Scenario S2, as a function of the proportion of contamination. Columns distinguish mean-shift and variance contamination, while the lines correspond to CT, LMS, MAD, and MSD.}
\end{figure}

\begin{figure}[!htbp]
  \caption{Percentage results for Scenario S3}
  \label{fig:sim-percentage-s3}
  \centering
  \includegraphics[width=0.85\textwidth]{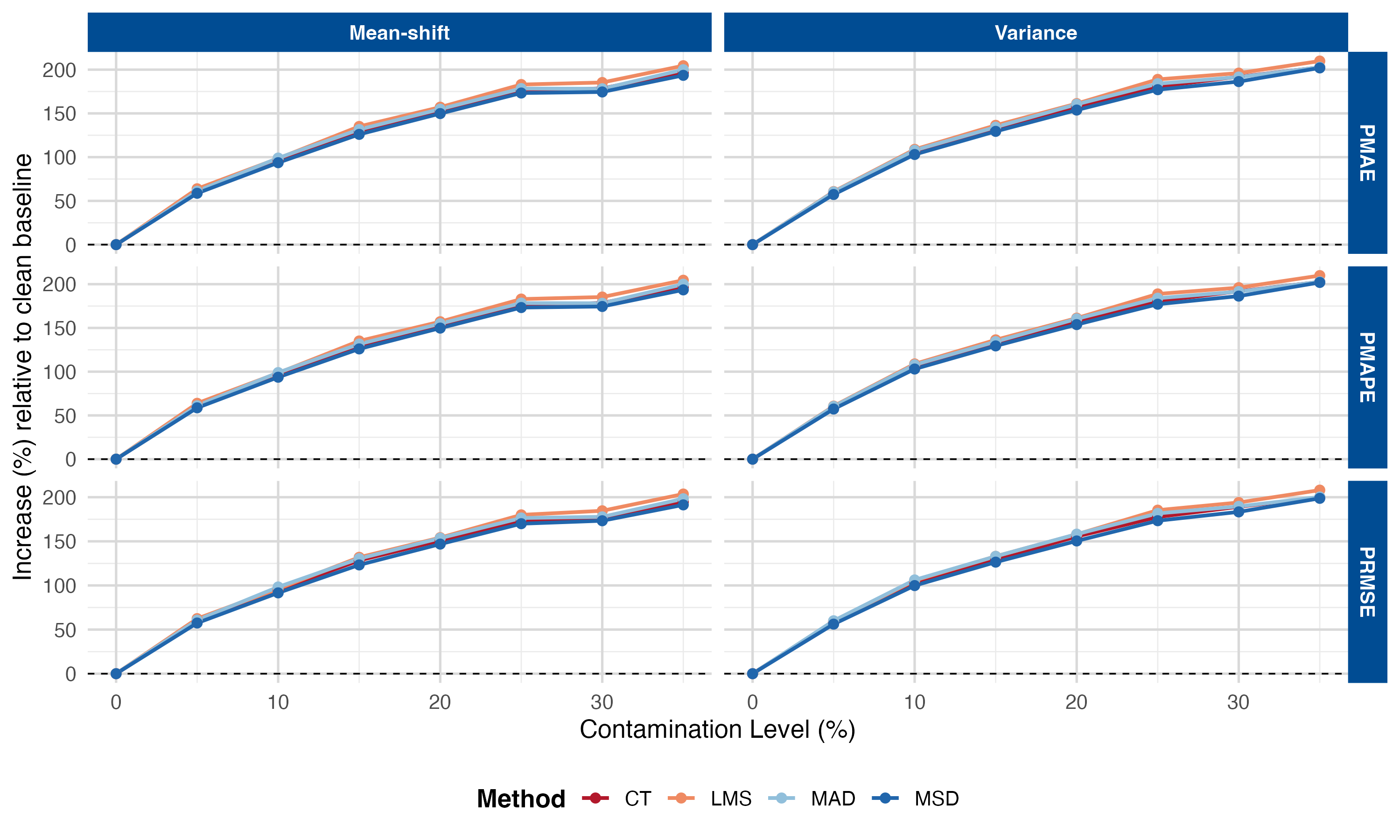}
  \caption*{\scriptsize\textit{Notes:} Percentage changes in PMAE, PMAPE, and PRMSE for each method when compared with the non-contaminated case under Scenario S3, as a function of the proportion of contamination. Columns distinguish mean-shift and variance contamination, while the lines correspond to CT, LMS, MAD, and MSD.}
\end{figure}

\begin{figure}[!htbp]
  \caption{Percentage results for Scenario S4}
  \label{fig:sim-percentage-s4}
  \centering
  \includegraphics[width=0.85\textwidth]{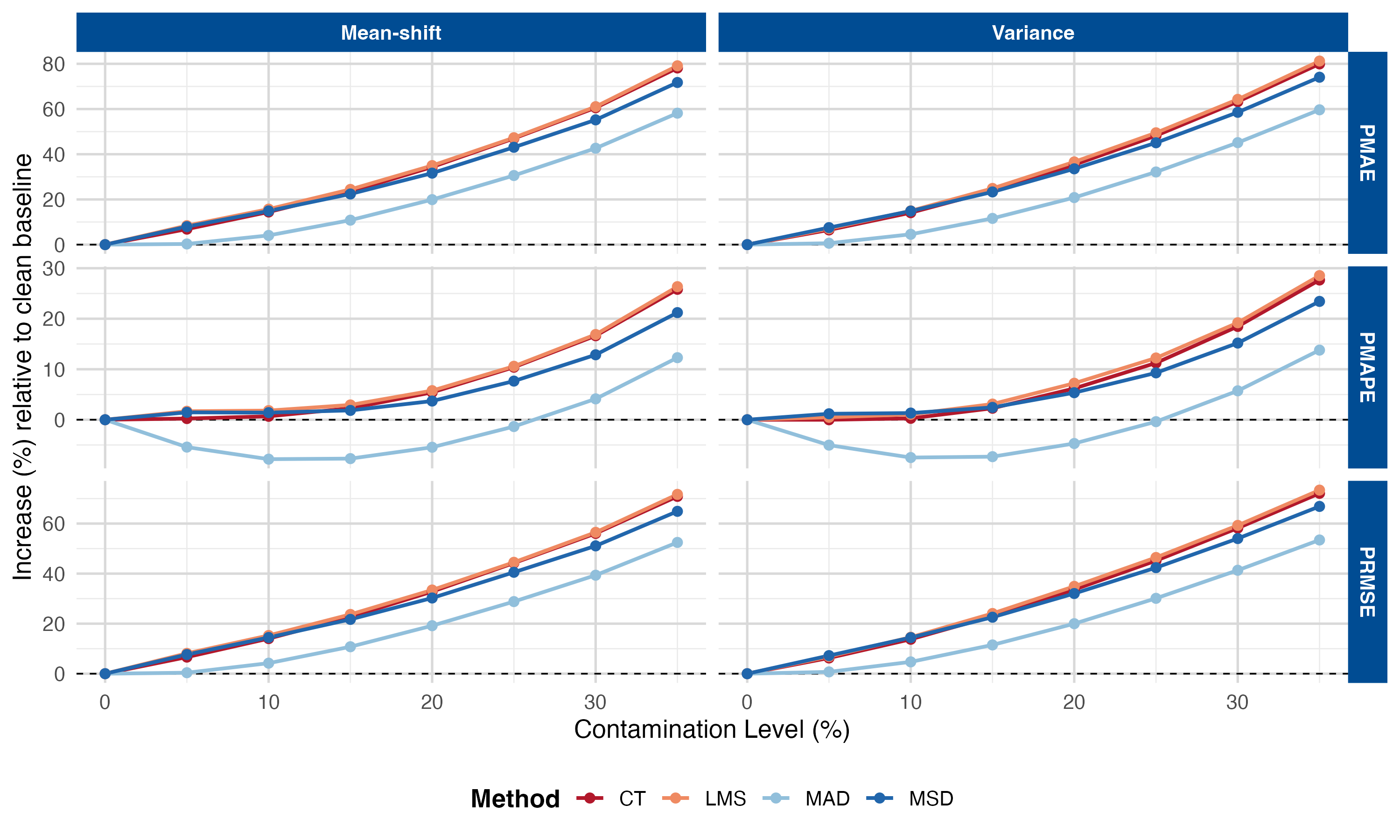}
  \caption*{\scriptsize\textit{Notes:} Percentage changes in PMAE, PMAPE, and PRMSE for each method when compared with the non-contaminated case under Scenario S4, as a function of the proportion of contamination. Columns distinguish mean-shift and variance contamination, while the lines correspond to CT, LMS, MAD, and MSD.}
\end{figure}

Figures~\ref{fig:sim-percentage-s1}--\ref{fig:sim-percentage-s4} summarize the robustness metrics under mean-shift and variance contamination. The CATE estimation accuracy measures, $\operatorname{RMSE}$, $\operatorname{MAE}$, and $\operatorname{MAPE}$, are reported in Tables~\ref{tab:sim_raw_pct_s1}--\ref{tab:sim_raw_pct_s4} in Appendix~\ref{appendix:Robust_ct_monte_additional_results}. Overall, we observe that robustness gains depend heavily on the underlying data-generating process. Furthermore, the results indicate a trade-off between robustness and CATE estimation accuracy. Splitting rules that are less affected by contamination are not necessarily the best-performing rules in the clean baseline.

In the linear treatment effect design S1, contamination has only a limited effect on $\operatorname{RMSE}$ and $\operatorname{MAE}$. Even at $p_c=0.35$, the increase in $\operatorname{RMSE}$ remains below $3\%$ under mean-shift contamination for all four splitting rules. CT increases by $2.70\%$, while MSD is almost identical at $2.69\%$. The same pattern appears for $\operatorname{MAE}$, where CT and MSD increase by $2.77\%$ and $2.75\%$, respectively. Thus, in the mean-shift design, the median-based rules do not substantially change the behavior of the forest, relative to the CT benchmark. Under variance contamination, the deterioration is somewhat larger, especially for $\operatorname{MAPE}$. Here, LMS is the least affected rule at the highest contamination level, with increases of $3.28\%$ in $\operatorname{RMSE}$, $ 3.41\%$ in $\operatorname{MAE}$, and $27.12\%$ in $\operatorname{MAPE}$. The corresponding values for CT are $3.86\%, 4.00\%$, and $29.92\%$. Hence, S1 provides only weak evidence for robustness gains from median-based splitting and all rules are fairly stable in this simple linear design.

\FloatBarrier
\FloatBarrier

Scenario S2 shows a different pattern. The treatment effect in this design is bounded, nonlinear, and characterized by sharp transition regions. Here, contamination leads to an increase in $\operatorname{RMSE}$, whereas $\operatorname{MAE}$ and $\operatorname{MAPE}$ decrease relative to the uncontaminated baseline. These decreases should not be interpreted as evidence that contamination generally improves CATE estimation accuracy. Rather, they indicate that the contaminated training samples produce smaller average absolute deviations on the uncontaminated test set than the clean baseline. The ranking in $\operatorname{RMSE}$ is clear. At $p_c=0.35$, CT shows the largest increase under both mean-shift and variance contamination, with increases of $6.50\%$ and $7.21\%$, respectively. LMS is least affected, with corresponding increases of $4.03\%$ and $4.91\%$. MSD and MAD lie between these two methods. This suggests that the median-based criteria, especially LMS, reduce squared-error sensitivity in S2, although the negative changes in $\operatorname{MAE}$ and $\operatorname{MAPE}$ show that the robustness pattern is metric-dependent.

\FloatBarrier
\FloatBarrier
The strongest deterioration occurs in Scenario S3.  Since $\tau(x)=0$ for all observations, the clean baseline errors are small in this design and even moderate increases in the raw error measures translate into large percentage changes. In this setting, $\operatorname{MAPE}$ reduces to $100\cdot\operatorname{MAE}$, so $\operatorname{PMAPE}$ and $\operatorname{PMAE}$ coincide up to rounding. At $p_c=0.35$, all methods show very large increases in PRMSE, PMAE, and PMAPE, often close to or above $200\%$. MSD performs best under both contamination mechanisms. Under mean-shift contamination, MSD yields a PRMSE of $191.3\%$, compared to $194.8\%$ for CT, $198.2\%$ for MAD, and $203.6\%$ for LMS. Under variance contamination, MSD again gives the lowest PRMSE and PMAE. However, the differences between CT, MAD, and MSD are relatively small compared to the overall magnitude of the deterioration.

\FloatBarrier
\FloatBarrier

Scenario S4 differs from the previous settings because the clean baseline performances are no longer close across all splitting rules. In particular, MAD and MSD start from higher errors than CT and LMS, especially in $\operatorname{RMSE}$ and $\operatorname{MAE}$. Therefore, the percentage increases must be interpreted together with the CATE estimation accuracy. Under mean-shift contamination, MAD has the smallest relative increase at $p_c=0.35$, with $\operatorname{PRMSE}=52.47\%$  and $\operatorname{PMAE}=58.17\%$. MSD follows with $\operatorname{PRMSE}=64.90\%$ and $\operatorname{PMAE}=71.75\%$, while CT and LMS are more strongly affected. CT has $\operatorname{PRMSE}=70.87\%$ and $\operatorname{PMAE}=78.17\%$, and LMS has $\operatorname{PRMSE}=71.67\%$ and $\operatorname{PMAE}=79.12\%$. Thus, under mean-shift contamination, MAD shows the smallest relative deterioration, followed by MSD.

Under variance contamination, the ranking is similar. At $p_c=0.35$, MAD again has the smallest relative increase, with $\operatorname{PRMSE}=53.44\%$ and $\operatorname{PMAE}=59.66\%$. The corresponding values for CT are $\operatorname{PRMSE}=72.09\%$ and $\operatorname{PMAE}=79.96\%$. Hence, in terms of percentage changes relative to the clean baseline, MAD is the most stable rule in S4.  However, this relative stability is attributable to a less favorable clean baseline. MAD starts with $\operatorname{RMSE}=4.03$, whereas CT and LMS start with $3.60$ and $3.58$, respectively. Thus, although MAD is less affected in relative terms, it also exhibits inferior performance prior to contamination. Consequently, S4 demonstrates the trade-off between CATE estimation accuracy and sensitivity to contamination. 

\FloatBarrier
\FloatBarrier

\section{Empirical Robustness Analysis}
\label{sec:empirical_contamination}

We consider three datasets that differ in sample size, application domain, and covariate structure. First,  we use the Lalonde dataset \citep{lalonde_evaluating_1986}, a causal inference benchmark with randomized treatment assignment.  Second, we use the Breast Cancer dataset, from the \texttt{ems} package in \texttt{R} \citep{ems_package}, which provides a smaller biomedical application with clinically meaningful covariates and a treatment indicator.  Third, we use the ACIC dataset, based on the \texttt{acic18} data from the \texttt{grf} replication files \citep{grf_package}, which provides a larger semi-synthetic benchmark with richer covariate information.
We consider the same two types of outcome irregularities described in Section~\ref{sec:contamination}, heavy-tailed outcome noise and artificial outcome contamination. 
\paragraph{Lalonde}
\label{sec:lalonde}
The \texttt{Lalonde} dataset, also known as the National Supported Work (\texttt{NSW}) Demonstration Program, is based on a job training program conducted in the United States in the 1970s. The dataset became a standard benchmark for treatment effect estimation after its use in \citet{lalonde_evaluating_1986} and has subsequently been used in related research \citep{parikh_validating_2022}. The NSW program was designed to assist low-income individuals facing barriers to employment, including limited education, work experience, or job skills.
The dataset contains 722 individuals, of whom 297 received the NSW job training program and 425 were assigned to the control group. The outcome variable is real earnings in 1978, and the treatment indicator records participation in the training program. The dataset also includes several pre-treatment covariates: age, years of education, race indicators, marital status, high-school diploma status, and real earnings in 1975. Summary statistics are reported in Table~\ref{tab:summary_Lalonde} in Appendix~\ref{App:Empirical Results}. Since this dataset is derived from a randomized controlled trial, treatment assignment is randomized, ensuring the absence of unobserved confounding. The unbiased estimate of the ATE for the NSW sample, calculated by a simple difference-in-means, is $\$886.30$.

\paragraph{Breast Cancer Data}
\label{sec:Breast_Cancer}

The \texttt{breastCancer} dataset from the \texttt{ems} package in \texttt{R} contains clinical and pathological information for 686 Breast Cancer patients from Germany, based on the studies of \citet{Schumacher1994,Schmoor2000}. It is commonly used for prognostic modeling and survival analysis. The dataset includes patient age, menopausal status, tumor size, tumor grade indicators, the number of positive lymph nodes, hormone receptor levels, and an indicator for tamoxifen treatment. The variable \texttt{rectime} measures the time in days to death or cancer recurrence, while \texttt{censrec} indicates whether this time is censored. Since we use \texttt{rectime} as the outcome variable, we restrict the sample to observations with observed event times. Censored observations do not reveal the actual time to death or recurrence. They only indicate that the event had not occurred up to the last follow-up time. We therefore remove the 387 censored observations and retain the 299 observations with observed event times. We use this dataset as an empirical robustness benchmark rather than to make any substantive causal claim about the effect of tamoxifen as the treatment assignment is observational and restricting the analysis to observed event times does not address potentially informative censoring. Table~\ref{tab:summary_breast_cancer} in Appendix~\ref{App:Empirical Results} reports descriptive statistics for the analysis sample.

\paragraph{ACIC}
\label{sec:ACIC}

The \texttt{acic18} data are obtained from the \texttt{grf} replication files \citep{grf_package}. The dataset is derived from the National Study of Learning Mindsets\footnote{See \url{https://studentexperiencenetwork.org/national-mindset-study/} for further information.} and was analyzed during the 2018 Atlantic Causal Inference Conference as part of a shared challenge to compare methods for estimating heterogeneous treatment effects. For the challenge, the data were modified to emulate an observational study while preserving key features of the original study, including covariate distributions, data structure, and effect sizes. The resulting dataset contains about 10,000 students from 76 schools. The outcome is a simulated continuous measure of student achievement, the treatment indicator records receipt of the growth-mindset intervention, and the covariates include student-level and school-level characteristics. Summary statistics for the continuous and categorical variables are reported in Tables~\ref{tab:summary_cont_ACIC} and~\ref{tab:summary_cat_ACIC} in Appendix~\ref{App:Empirical Results}.

\subsection{Evaluation Metrics}

We first establish baseline performance in the unperturbed setting. These baseline metrics provide the reference point for evaluating subsequent changes under outcome irregularities. We report three baseline quantities: the transformed-outcome criterion $\operatorname{RMSE}^{TOT}$, the empirical standard deviation of the estimated CATEs, $SD_{\hat{\tau}}$, and the estimated ATE. 
 
Since the true CATEs are unavailable in the empirical applications, we cannot directly compute the infeasible $\operatorname{RMSE}_{\tau} $, defined as \begin{align*}
\operatorname{RMSE}_{\tau} &=\sqrt{\frac{1}{N^{te}}\sum_{i=1}^{N^{te}}\left(\tau(X_i)-\hat{\tau}(X_i)\right)^2}.
\end{align*}
Instead, we use the transformed-outcome criterion
\begin{align*}
    \operatorname{RMSE}^{TOT}
&=
\sqrt{
\frac{1}{N^{te}}
\sum_{i=1}^{N^{te}}
\left(
Y_i^*-\hat{\tau}(X_i)
\right)^2
},
\\Y_i^*
&=
Y_i
\frac{D_i-\hat p(X_i)}
{\hat p(X_i)(1-\hat p(X_i))},
\end{align*} 
where $\hat p(X_i)$ denotes the estimated propensity score.
Under Assumptions \ref{assump:SUTVA}--\ref{assump:overlap} and using the true propensity score it follows that $\mathbb{E}[Y_i^* \mid X_i]=\tau(X_i)$. In practice, we replace the propensity score by its estimate $\hat p(X_i)$, so that $\operatorname{RMSE}^{TOT}$ provides a feasible, but noisy, proxy for the infeasible CATE prediction error. As discussed by \citet{wager_estimation_2018}, the transformed-outcome criterion often ranks estimators similarly to the infeasible $\operatorname{RMSE}_{\tau}$. 
Next, we report the empirical standard deviation of the estimated CATEs,
\begin{align*}
SD_{\hat{\tau}}=\sqrt{\frac{1}{N^{te}-1}\sum_{i=1}^{N^{te}}\left(\hat{\tau}(X_i)-\hat{\bar{\tau}}\right)^2}.\end{align*}
Here, $\hat{\bar{\tau}}$ denotes the average estimated CATE in the test sample.
Finally, to estimate the ATE, we use the doubly robust augmented inverse-propensity-weighted  ($\operatorname{AIPW}$) estimator \citep{chernozhukov_doubledebiased_2018}, which is defined as

\begin{align*}
\widehat{\tau}_{\operatorname{AIPW}} &= \frac{1}{N^{te}}\sum_{i=1}^{N^{te}} \left(\hat{\mu}_1(X_i) - \hat{\mu}_0(X_i) + \frac{D_i(Y_i - \hat{\mu}_1(X_i))}{\hat{p}(X_i)} - \frac{(1-D_i)(Y_i - \hat{\mu}_0(X_i))}{1-\hat{p}(X_i)} \right).
\end{align*} 
\subsubsection{Robustness metrics}
\label{sec:robustness_metrics_empirical}

To quantify robustness under artificial outcome contamination, we compare treatment effect estimates obtained from contaminated training data, $\hat{\tau}^c(X_i)$, with the corresponding estimates obtained from clean training data, $\hat{\tau}^{\mathrm{clean}}(X_i)$. Here, ``clean'' refers to the original, uncontaminated training sample. These metrics measure sensitivity relative to the uncontaminated-data estimates. We report the RMSE deviation,
\begin{align} 
\operatorname{RMSE}_{\Delta}
&=
\sqrt{
\frac{1}{N^{te}}
\sum_{i=1}^{N^{te}}
\left(
\hat{\tau}^c(X_i) - \hat{\tau}^{\mathrm{clean}}(X_i)
\right)^2
},
\intertext{the MAE deviation,}
\operatorname{MAE}_{\Delta}
&=
\frac{1}{N^{te}}
\sum_{i=1}^{N^{te}}
\left|
\hat{\tau}^c(X_i) - \hat{\tau}^{\mathrm{clean}}(X_i)
\right|,
\intertext{and the MAPE deviation,}
\operatorname{MAPE}_{\Delta} &= \frac{100}{N^{te}} \sum_{i=1}^{N^{te}}
 \left| \frac{ \hat{\tau}^c(X_i) - \hat{\tau}^{\mathrm{clean}}(X_i) }{ \left| \hat{\tau}^{\mathrm{clean}}(X_i) \right|  } \right|. \label{eq:accuracy2}
\end{align}

\subsection{Results}
All reported results are averaged over 100 independent repetitions. In each repetition, we randomly split the dataset into training and test samples, and evaluate the resulting CATE or ATE estimates on the corresponding test sample. The train/test split ratio is adapted to the size of each dataset: 75\%/25\% for Lalonde, 90\%/10\% for Breast Cancer because of its smaller sample size, and 60\%/40\% for the larger ACIC dataset.

\subsubsection{Baseline performance}
Table~\ref{tab:clean_baseline_empirical} reports the results before artificial irregularities are introduced. Overall, $\operatorname{RMSE}^{TOT}$ is similar across splitting rules within each dataset.  In the Lalonde application, for example, $\operatorname{RMSE}^{TOT}$ ranges from 17563.38 for CT to 17568.18 for MAD. In the Breast Cancer application, the corresponding values range from 2377.05 for MAD to 2377.94 for MSD, and in ACIC, all methods yield essentially the same value of 1.366. These findings indicate that median-based splitting criteria do not materially change transformed-outcome performance relative to the CT rule. In the Lalonde application, LMS and MAD generate substantially more variable CATE estimates than CT, with $SD_{\hat{\tau}}=294.82$ and $275.18$, respectively, compared with $128.54$ for CT. In the Breast Cancer application, MSD achieves the lowest dispersion, $SD_{\hat{\tau}}=48.04$, while LMS results in the highest, $SD_{\hat{\tau}}=88.26$. For ACIC, all methods yield highly similar ATE estimates, with only minor differences in $SD_{\hat{\tau}}$.

These results serve as a reference point for the subsequent robustness analysis. Since $\operatorname{RMSE}^{TOT}$ is similar across the different splitting rules, their baseline performance is broadly comparable. The differences that are observed in the robustness analysis that follows can therefore be viewed as being due primarily to differing sensitivities to the introduced outcome irregularities rather than to any significant pre-existing differences in baseline performance.

\FloatBarrier
\begin{table}[h]
\caption{Baseline performance across empirical applications.}
\label{tab:clean_baseline_empirical}
\centering
\setlength{\tabcolsep}{8pt}
\resizebox{.95\linewidth}{!}{
\begin{tabular}{l rrr @{\hskip 1.2cm} rrr @{\hskip 1.2cm} rrr}
\toprule
& \multicolumn{9}{c}{Dataset} \\
\cmidrule(l{3pt}r{3pt}){2-10}
& \multicolumn{3}{c}{Lalonde}
& \multicolumn{3}{c}{Breast Cancer}
& \multicolumn{3}{c}{ACIC} \\
\cmidrule(l{3pt}r{3pt}){2-4}
\cmidrule(l{3pt}r{3pt}){5-7}
\cmidrule(l{3pt}r{3pt}){8-10}
Method
& $\widehat{\tau}_{\text{AIPW}}$ & $\operatorname{RMSE}^{TOT}$ & $SD_{\hat{\tau}}$
& $\widehat{\tau}_{\text{AIPW}}$ & $\operatorname{RMSE}^{TOT}$ & $SD_{\hat{\tau}}$
& $\widehat{\tau}_{\text{AIPW}}$ & $\operatorname{RMSE}^{TOT}$ & $SD_{\hat{\tau}}$ \\
\midrule
CT
& 869.34 & 17563.38 & 128.54
& 103.81 & 2377.56 & 59.12
& 0.25 & 1.366 & 0.025 \\
LMS
& 875.16 & 17568.13 & 294.82
& 103.12 & 2377.41 & 88.26
& 0.25 & 1.366 & 0.034 \\
MAD
& 867.58 & 17568.18 & 275.18
& 103.99 & 2377.05 & 63.24
& 0.25 & 1.366 & 0.037 \\
MSD
& 873.08 & 17565.89 & 227.30
& 102.42 & 2377.94 & 48.04
& 0.25 & 1.366 & 0.041 \\
\bottomrule
\end{tabular}
}
\begin{notes} Entries report baseline metrics averaged over 100 train/test splits. The ATE estimates correspond to the unperturbed outcome setting.
\end{notes}
\end{table}
\FloatBarrier

\subsubsection{Heavy-tailed outcome noise}

Figure~\ref{fig:application_ate} and Table~\ref{tab:applications_ate} in Appendix \ref{App:Empirical Results} summarize the ATE estimation results, while Figures~\ref{fig:acic_cate}--\ref{fig:lalonde_cate}
 and Table~\ref{tab:apl_cate_metric} in Appendix \ref{App:Empirical Results} report the corresponding CATE estimation results across splitting rules and outcome-noise settings. Since the true treatment effects are unknown in the empirical applications, we focus on changes in the estimated ATE and in the distribution of estimated CATEs across noise settings.

\begin{figure}[!thb]
\caption{Estimated ATEs across splitting rules and outcome-noise settings in the three empirical applications.}
\centering
    \label{fig:application_ate}
    \resizebox{0.75\linewidth}{!}{%
   \input{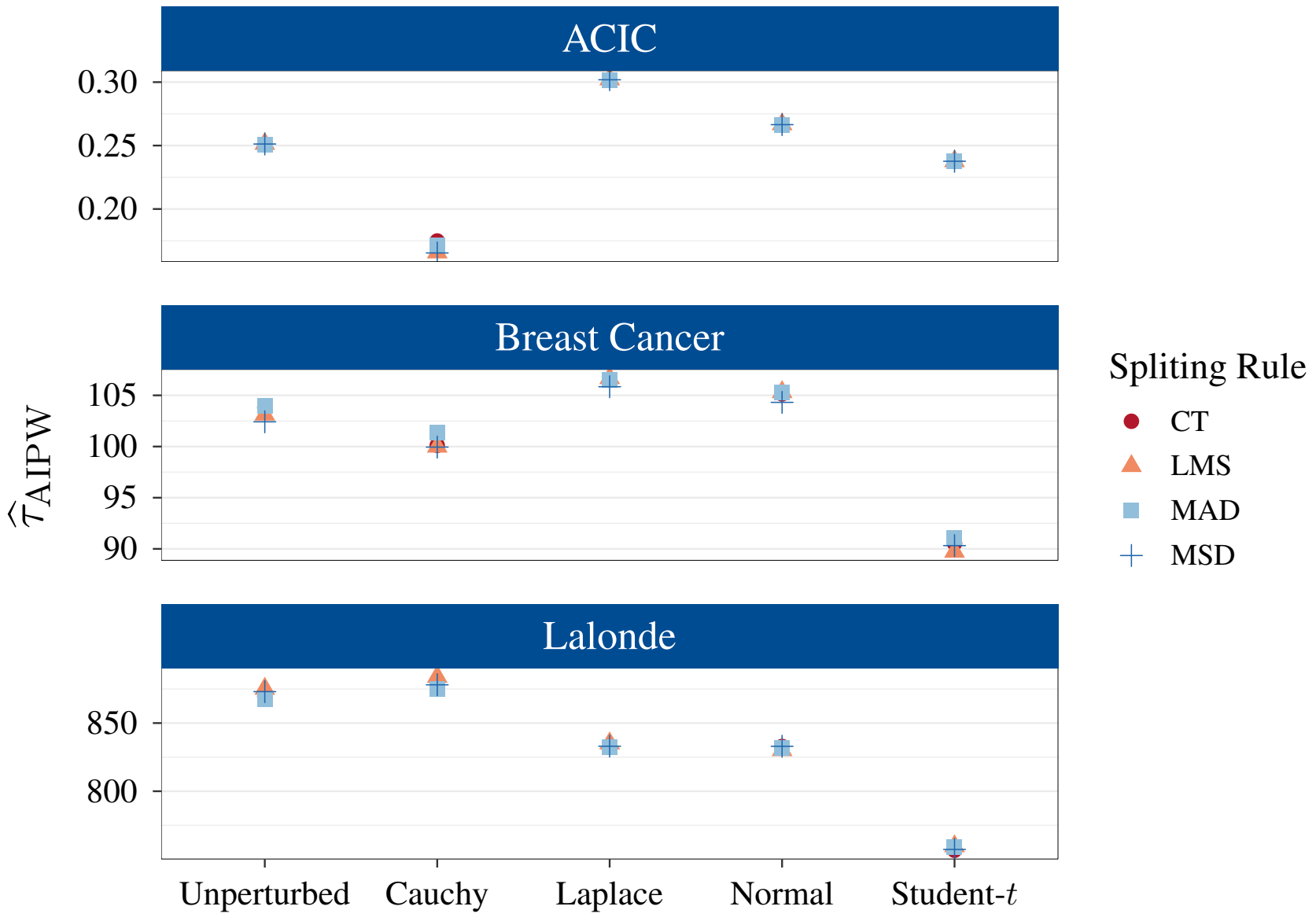}
    }
\begin{notes}
Panels correspond to the ACIC, Breast Cancer, and Lalonde datasets. The $y$-axis scales differ across applications to accommodate differences in outcome magnitudes.    \end{notes}
\end{figure}

For the unperturbed Lalonde application, the estimated ATEs are similar across splitting rules, ranging from $867.58$ for MAD to $875.16$ for LMS. In contrast, the average estimated CATEs vary more substantially across splitting rules, ranging from $804.53$ for MSD to $882.36$ for MAD. LMS falls in between, with an average estimated CATE of $852.37$. Under Cauchy noise, the estimated ATEs increase slightly across all splitting rules, reaching $879.22$ for CT, $883.97$ for LMS, $874.75$ for MAD and $878.07$ for MSD. Under Normal and Laplace noise, the estimated ATEs are lower than in the unperturbed case (ranging from $830.12$ to $832.96$ and from $832.17$ to $835.09$, respectively). The largest drop is under Student-$t$ noise (ranging from $756.64$ to $759.60$) across the four splitting rules. A similar pattern is observed for the average estimated CATEs. Under Normal and Student-$t$ noise, they decrease relative to the unperturbed setting across all splitting rules.  Under Laplace noise, the same holds for most methods, although the differences are less uniform. Under Cauchy noise, the results are mixed. The average estimated CATE decreases from $881.60$ to $849.42$ for CT, while it increases from $804.53$ to $856.77$ for MSD. LMS also increases slightly from $852.37$ to $858.60$, whereas MAD remains close to its unperturbed value, increasing from $882.36$ to $883.81$.

For the Breast Cancer application, the estimated ATE in the unperturbed setting is between $102.42$ and $103.99$ across the four splitting rules. Student-$t$ noise leads to the largest drop, with estimated ATEs of around $90$--$91$ for all methods. Normal and Laplace noise give rise to slightly higher ATE estimates, whereas Cauchy noise gives lower estimates.  Overall, no substantial differences between the splitting rules were observed in any of the four noise settings. The same general pattern is seen for the average estimated CATE, although the differences across the splitting rules are somewhat more pronounced. The Cauchy setting again produces much greater dispersion in the estimated CATEs than the other noise distributions.

For the ACIC application, the estimated ATE is approximately $0.25$ in the unperturbed case across all splitting rules.  Under Cauchy noise, the estimated ATE drops to around $0.17$ for all methods. Normal and Student-$t$ noise lead to only minor changes in the ATE estimate (around $0.27$ and $0.24$, respectively), whereas Laplace noise results in an increase in the estimated ATE to approximately $0.30$. 
The average estimated CATE follows a similar pattern. The one deviation is under Cauchy noise, where CT gives an average estimated CATE of around $0.07$ as opposed to the range of around $0.23$ to $0.25$ for the median-based splitting rules. This discrepancy seems reasonable, as the moments of the Cauchy distribution are undefined, which likely affects mean-based splitting rules such as CT more severely than median-based ones. The standard deviation under Cauchy noise is also substantially higher than in the unperturbed or any other noise setting.

\subsubsection{Mean-shift and variance contamination}
\begin{figure}[ht]
\caption{Percentage changes in sensitivity metrics under mean-shift and variance contamination in the Lalonde application.}
\centering
\includegraphics[width=.80\linewidth]{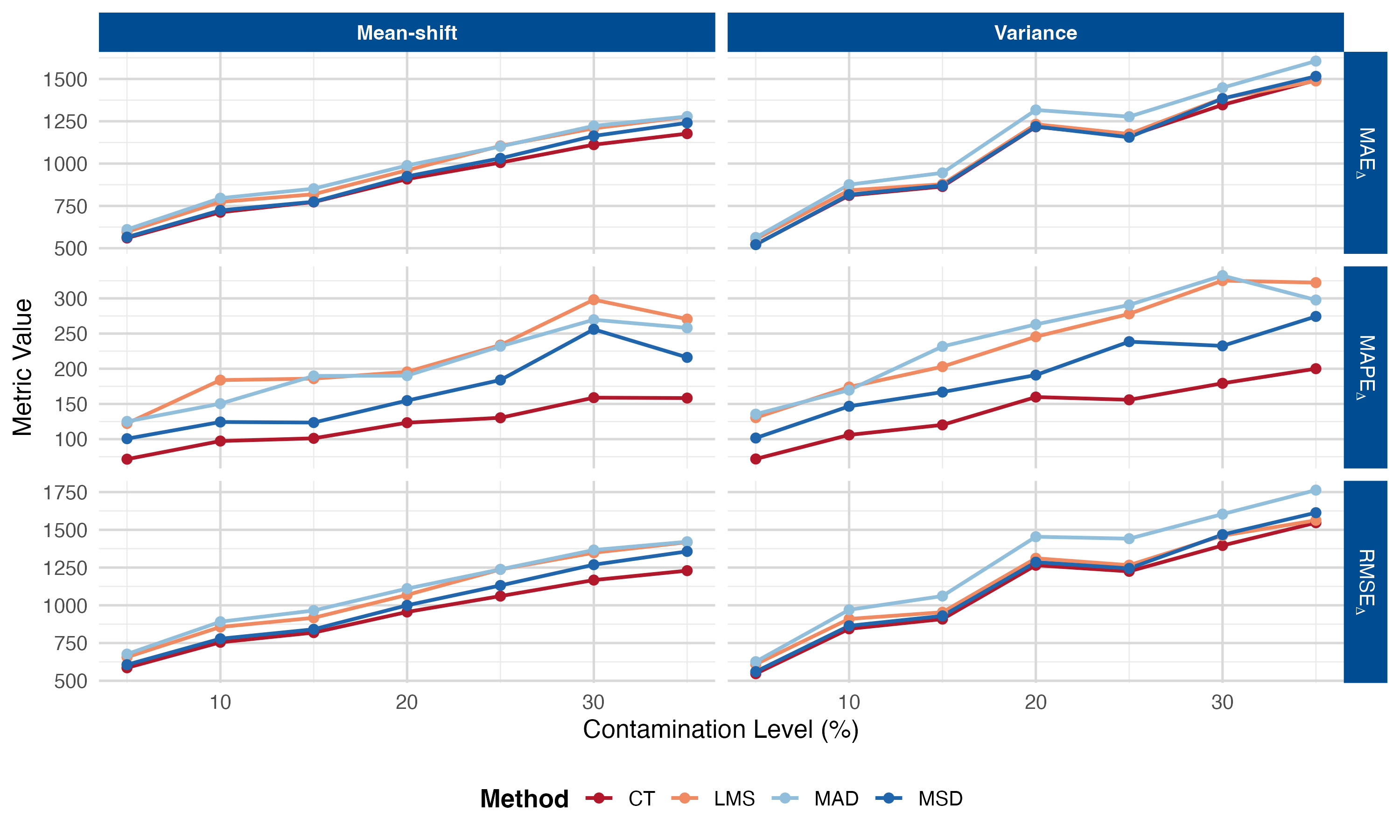}
\begin{notes} Columns distinguish mean-shift and variance contamination. Rows report percentage changes in $\operatorname{MAE}_{\Delta}$, $\operatorname{MAPE}_{\Delta}$, and $\operatorname{RMSE}_{\Delta}$.
\end{notes}
\label{fig:lalonde_pct_increase}
\end{figure}

\begin{figure}[ht]
\caption{Percentage changes in sensitivity metrics under mean-shift and variance contamination in the Breast Cancer application.}
\centering
\includegraphics[width=.80\linewidth]{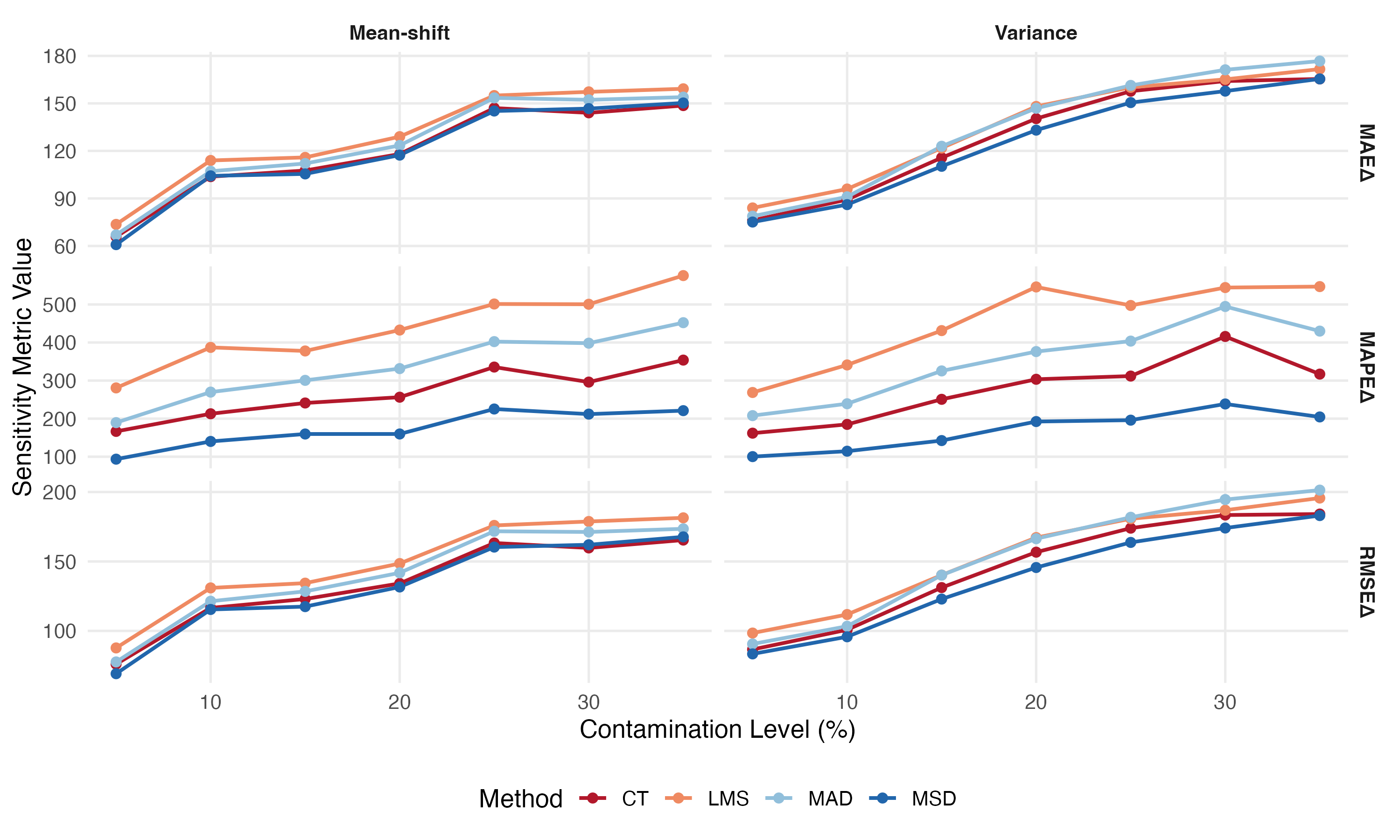}
\begin{notes} Columns distinguish mean-shift and variance contamination. Rows report percentage changes in $\operatorname{MAE}_{\Delta}$, $\operatorname{MAPE}_{\Delta}$, and $\operatorname{RMSE}_{\Delta}$.
\end{notes}
\label{fig:breast_pct_increase}
\end{figure}

\begin{figure}[ht]
\caption{Percentage changes in sensitivity metrics under mean-shift and variance contamination in the ACIC application.}
\centering
\includegraphics[width=.80\linewidth]{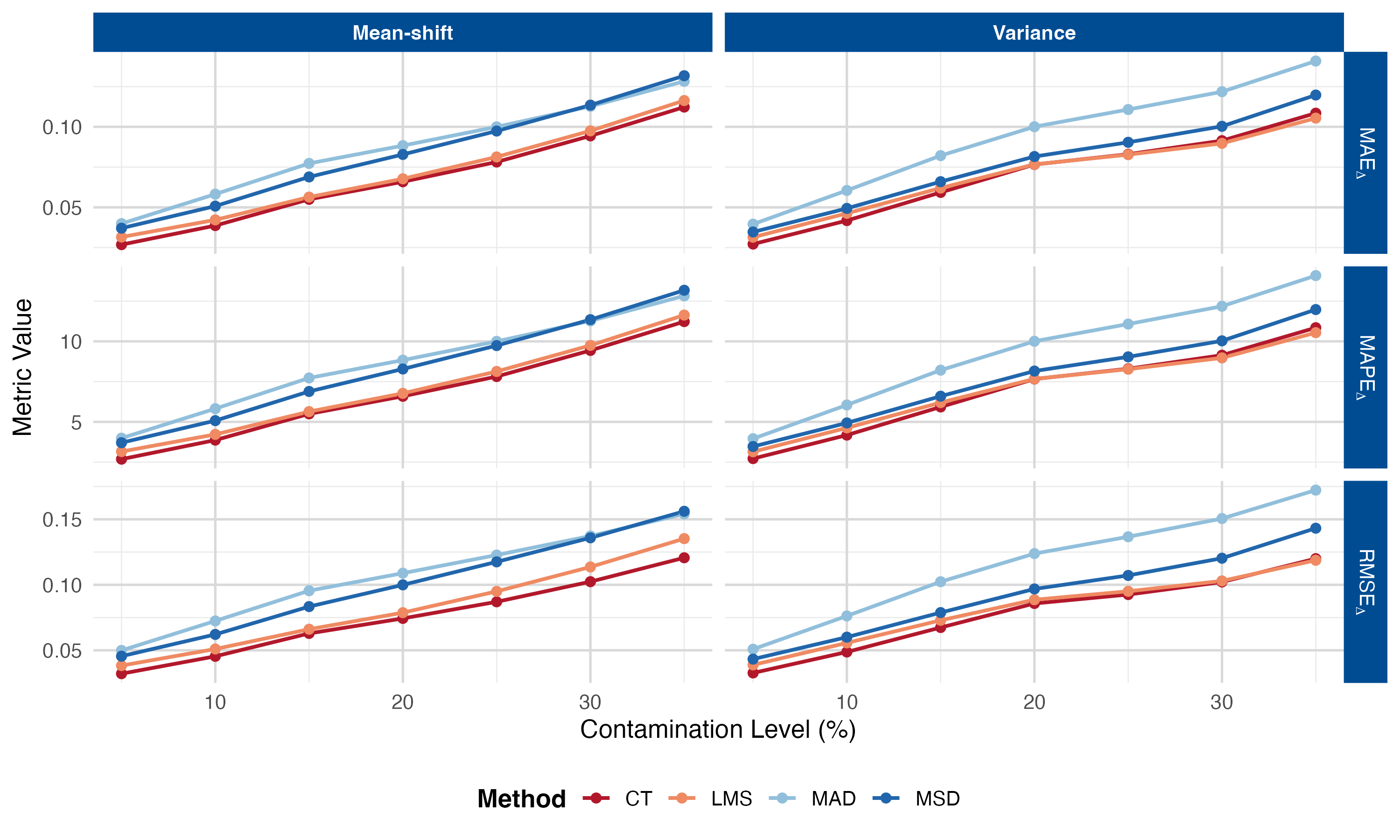}
\begin{notes} Columns distinguish mean-shift and variance contamination. Rows report percentage changes in $\operatorname{MAE}_{\Delta}$, $\operatorname{MAPE}_{\Delta}$, and $\operatorname{RMSE}_{\Delta}$.
\end{notes}
\label{fig:acic_pct_increase}
\end{figure}

Figures~\ref{fig:lalonde_pct_increase} to \ref{fig:acic_pct_increase} illustrate the sensitivity of the different splitting rules to mean-shift and variance contamination.
The corresponding Tables~\ref{tab:mean_contamination_empirical} and~\ref{tab:variance_contamination_empirical} are provided in the appendix. We also report percentage increases relative to the best-performing method in each setting in the appendix. Overall, the results do not indicate uniform dominance of the median-based splitting rules.

In the Lalonde dataset, CT performs best in most cases. Under mean-shift contamination, CT gives $\operatorname{RMSE}_{\Delta}=1229.45$ and $\operatorname{MAE}_{\Delta}=1176.90$ at $p_c=0.35$, which is roughly $13\%$ lower in $\operatorname{RMSE}_{\Delta}$ and $8\%$ lower in $\operatorname{MAE}_{\Delta}$ than LMS and MAD. MSD is closer to CT, but still has higher deviations. At the same contamination level, CT has about $9.4\%$ lower $\operatorname{RMSE}_{\Delta}$ and $5.1\%$ lower $\operatorname{MAE}_{\Delta}$ than MSD. The variance-contamination results are similar. At $p_c=0.35$, CT has about $12\%$ lower $\operatorname{RMSE}_{\Delta}$ than MAD and about $4\%$ lower than MSD. Thus, for Lalonde, the median-based rules do not improve on CT.

\FloatBarrier
\FloatBarrier

In the Breast Cancer application, the ranking of the methods differs. In this case, MSD is often the best or nearly the best method, especially when contamination is low or moderate. For example, under mean-shift contamination with $p_c=0.05$, MSD gives $\operatorname{RMSE}_{\Delta}=69.17$ and $\operatorname{MAE}_{\Delta}=60.83$,  compared with $76.06$ and $65.84$ for CT. This corresponds to roughly $9\%$ lower $\operatorname{RMSE}_{\Delta}$ and $8\%$ lower $\operatorname{MAE}_{\Delta}$. At higher contamination levels, CT closes the gap and performs on par with MSD. Under variance contamination, MSD remains competitive, but the differences become smaller at higher contamination levels. At $p_c=0.35$, MSD gives $\operatorname{RMSE}_{\Delta}=183.04$ and $\operatorname{MAE}_{\Delta}=165.46$, while CT gives $184.13$ and $165.43$. Thus, by the highest contamination level, both methods perform almost identically. The Breast Cancer application is therefore the clearest case where a median-based split rule reduces sensitivity relative to the CT benchmark. 
\FloatBarrier

\FloatBarrier
For ACIC, the values are generally much smaller because the estimated effects are on a different scale. The curves start close together at low contamination levels and increase only gradually. Under mean-shift contamination, CT remains slightly lower than the other methods across most of the contamination range. At $p_c=0.35$, CT gives $\operatorname{RMSE}_{\Delta}=0.12$ and $\operatorname{MAE}_{\Delta}=0.11$. The corresponding $\operatorname{RMSE}_{\Delta}$ values are $0.14$ for LMS, $0.15$ for MAD, and $0.16$ for MSD. Under variance contamination, CT and LMS remain close throughout the contamination range and have $\operatorname{RMSE}_{\Delta}=0.12$ and $\operatorname{MAE}_{\Delta}=0.11$ at $p_c=0.35$. MAD and MSD increase more strongly, especially at higher contamination levels, with $\operatorname{RMSE}_{\Delta}=0.17$ for MAD and $0.14$ for MSD at $p_c=0.35$. Hence, for ACIC, the benchmark CT remains competitive, and the median-based criteria do not provide a systematic improvement.
\FloatBarrier
\FloatBarrier

\section{Discussion}
\label{sec:Robust_discussion}
Building on the robustness analysis of \citet{roy_robustness_2012} for regression forests, we studied whether median-based splitting criteria can improve the robustness of causal trees and causal forests to outcome irregularities. By modifying only the splitting step of the causal tree algorithm, we keep the estimand fixed as the CATE and maintain mean-based leaf-level treatment effect estimation. The proposed criteria modify the standard MSE-based splitting rule using robust alternatives, including the Hodges–Lehmann estimator and least median squares.

Our Monte Carlo experiments, based on four data-generating processes, show that median-based splitting rules can reduce sensitivity to contamination in selected settings, but they generally do not dominate the standard causal tree splitting rule. Their advantages are most visible under more extreme outcome irregularities and in some nonlinear treatment effect designs, whereas CT remains competitive in many other settings. The results therefore point to a clear trade-off, more robust split selection can reduce sensitivity to outcome irregularities, but this does not necessarily translate into uniformly better CATE estimation performance. With respect to inference, the IJ generally achieves slightly higher empirical coverage than the BLB, although neither variance estimator is uniformly well calibrated across all designs. Overall, the results suggest that median-based alternatives  may be useful when the treatment effect surface is expected to be highly nonlinear, when the dataset contains extreme outcome outliers, or both.

The empirical robustness analysis leads to a similar conclusion. In the heavy-tailed-noise analysis, the relative behavior of the splitting rules depends on the application and noise distribution, with the clearest differences emerging under Cauchy noise. Under artificial contamination, CT remains competitive and often performs best in terms of sensitivity to contamination in the Lalonde and ACIC data. In the Breast Cancer application, MSD is frequently the most stable or nearly the most stable method, especially at low and moderate contamination levels. Thus, the empirical results reinforce the finding that no single splitting rule dominates across settings.  

These findings are consistent with the main conclusion of \citet{roy_robustness_2012}, who showed that robust forest variants can provide additional protection against contamination, while robustifying the splitting rule addresses only one part of the robustness problem in general. A similar problem arises in the causal tree setting studied here. Median-based splitting can make the selected partitions less sensitive to extreme outliers, but the final CATE estimates are still computed using mean-based treatment-control differences within leaves and are then averaged across trees in the forest. Consequently, even if partitions are selected more robustly, outlying observations within terminal leaves can still affect the final treatment effect estimates.

This findings provides several natural extensions. First, future work could study more robust leaf-level CATE estimators, for example by replacing mean-based leaf estimates with median-based or trimmed estimators. Second, future work could investigate contamination in the covariate space. Although tree-based methods are invariant to monotone transformations of individual covariates, perturbations that change the ordering of covariate values can affect split locations and therefore the estimated treatment effect surface. Third, the contamination mechanisms considered here are outcome-side and largely random. More structured forms of contamination, depending on the outcome, treatment status, or covariates, may generate different robustness patterns. Finally, data-driven procedures, such as cross-validation or sample splitting, could be used to choose among CT, LMS, MAD, and MSD in practice, since no single rule dominates across all settings.

\acks{We would like to thank Christoph Hanck and Lennard Maßmann for valuable feedback on earlier drafts of this manuscript and we are grateful to seminar participants at the RuhrMetrics Research Seminar for helpful comments and discussions. The authors acknowledge partial financial support from TRR 391 Spatio-temporal Statistics for the Transition of Energy and Transport (520388526) by the Deutsche Forschungsgemeinschaft (DFG, German Research Foundation) and from the Rhine-Ruhr Center for Scientific Data Literacy (DKZ.2R) by the German Federal Ministry of Education and Research (BMBF). During the preparation of this manuscript, we used OpenAI's large language models (up to GPT-5.5) to assist with proofreading and editing. We reviewed and verified all resulting content and are solely responsible for any remaining errors.
}

\newpage
\bibliography{citation}
\appendix
\newpage
\section{Additional Theory} \label{appendix:additional_theory}

\subsection{Alternative Splitting Rules}

For completeness, we report two additional splitting criteria. First, we give the adaptive counterpart of the honest causal tree criterion in ~\eqref{eq:robust_MSE_CT_H}. Second, we present the honest version of the MSD criterion in~\eqref{eq:mse_tau_med_short}.

\subsubsection{Adaptive Causal Tree Criterion}
\label{appendix:Robust_Alternative_Estimators}
 In contrast to honest estimation, the adaptive causal tree uses the same sample $\mathcal S^{tr}$ both to determine the partition and to estimate treatment effects within the resulting leaves. Accordingly, its risk criterion evaluates
$\operatorname{MSE}_{\tau}
(\mathcal S^{tr},\mathcal S^{tr},\Pi).$
The corresponding adaptive splitting objective is then
\begin{align}
\widehat Q_\tau^{\mathrm{MSE,A}}
(\mathcal S^{tr},\Pi)
&=
\frac{1}{N^{tr}}
\sum_{i\in\mathcal S^{tr}}
\hat\tau^2(X_i;\mathcal S^{tr},\Pi).
\end{align}

Thus, the adaptive criterion rewards partitions that generate large differences in estimated treatment effects across leaves, without the variance correction appearing in the honest criterion in Equation~\eqref{eq:robust_MSE_CT_H}. 

\subsubsection{Honest Median Squared Deviation Criterion}
\label{appendix:Honest_Splitting_MSD}

The MSD criterion introduced in
Section~\ref{sec:MSD_short} can be related directly to the honest causal-tree
criterion in ~\eqref{eq:robust_MSE_CT_H}. The derivation follows the same
expected-MSE argument as for the mean-based honest splitting rule in \cite{athey_recursive_2016}, but replaces
the difference-in-means estimator by the leafwise Hodges--Lehmann estimator
$\hat{\tau}_{\mathrm{HL}}(l)$. \cite{maßmann2026medianbasedsplittingrulescausal} show that
under the leafwise location-shift conditions the Hodges--Lehmann estimator is consistent and
asymptotically normal for the leafwise treatment effect. Its asymptotic variance can be written as
\begin{align}
    \Var\!\left(
        \hat{\tau}_{\mathrm{HL}}(l;\mathcal{S}^{est})
    \right)
    \approx
    \frac{1}{N_l^{est}}
    V_l,
    \qquad
    V_l
    =
    \frac{1}{
        12\,p(1-p)
        \left(\int f_l^2(u)\,du\right)^2
    },
\end{align}
where $f_l$ denotes the common within-leaf outcome density up to the treatment
shift and $p$ is the treatment probability.

Using the same bias correction as in the honest mean-based criterion therefore
gives the honest MSD splitting objective
\begin{align}
\label{eq:Q_MSD_H}
    \widehat Q_{\tau}^{\mathrm{MSD,H}}
    (\mathcal{S}^{tr},\Pi)
    =
    \sum_{l\in\Pi}
    \frac{N_l^{tr}}{N^{tr}}
    \left[
        \hat{\tau}_{\mathrm{HL}}^2(l;\mathcal{S}^{tr})
        -
        \left(
            \frac{1}{N_l^{tr}}
            +
            \frac{1}{N_l^{est}}
        \right)
        \widehat V_l
    \right],
\end{align}
where $\widehat V_l$ is an estimate of the Hodges--Lehmann asymptotic variance
factor $V_l$.
Equation \eqref{eq:Q_MSD_H} is the direct median-based analog of the honest causal-tree splitting criterion. The first term rewards treatment effect heterogeneity, whereas the
second penalizes leaves in which the Hodges--Lehmann effect estimate is imprecise. In practice, however, evaluating $\widehat V_l$ requires estimating
the density functional $\int f_l^2(u)\,du$ repeatedly for every candidate
leaf and split. We therefore use the density-free MSD approximation described
in Section~\ref{sec:MSD_short}.

\subsection{Variance Estimation}
\label{appendix:Robust_variance_discussion}

The confidence interval construction of \citet{athey_generalized_2019} is derived within the generalized random forest (GRF) framework, which differs from the original causal forest algorithm considered in this work. We briefly summarize the GRF formulation.

In the GRF framework, $\tau(x)$ is characterized as the solution to the local moment condition\footnote{More generally, the GRF framework estimates parameters of the form $\theta(x)=\xi^\top\beta(x)$, where $\xi$ is a contrast vector. Since our interest is restricted to the CATE, we present the special case corresponding to $\tau(x)$.}
\begin{align}
\mathbb{E}\!\left[
\left(Y_i-\tau(x)D_i-c(x)\right)
(1,D_i)^\top
\,\middle|\,
X_i=x
\right]
=
0,
\end{align}
where $c(x)$ denotes a local intercept. Rather than averaging the treatment effect estimates produced by the individual trees, generalized random forests first construct adaptive similarity weights $\alpha$. Let
\[
\alpha_{bi}(x)
=
\frac{\mathds 1\{X_i\in L_b(x)\}}
{|L_b(x)|}
\]
denote the weight assigned to unit $i$ by tree $b$, where $L_b(x)$ is the leaf containing the test point $x$. Averaging over the $B$ trees yields the forest weights
\[
\alpha_i(x)
=
\frac 1B
\sum_{b=1}^B
\alpha_{bi}(x).
\]
Using these weights, the local treatment effect estimator is
\begin{align} \label{eq:Robust_tau_hat_grf}
\hat{\tau}(x)
=
\left(
\sum_{i=1}^N
\alpha_i(x)
(D_i-D_\alpha)^2
\right)^{-1}
\sum_{i=1}^N
\alpha_i(x)
(D_i-D_\alpha)
(Y_i-Y_\alpha),
\end{align}
where
\begin{align*}
D_\alpha
=
\sum_i
\alpha_i(x)D_i,
\qquad
Y_\alpha
=
\sum_i
\alpha_i(x)Y_i.
\end{align*}
The variance theory developed by \citet{athey_generalized_2019} is based on \eqref{eq:Robust_tau_hat_grf}. Specifically, the estimator is linearized around the true parameter, and the stochastic component is represented by the weighted estimating equation
\[
\Psi(\tau,c)
=
\sum_{i=1}^N
\alpha_i(x)
\psi_{\tau,c}(O_i),
\]
where $O_i=(Y_i,D_i,X_i)$ denotes the observed data for unit $i$, and
\[
\psi_{\tau,c}(O_i)
=
\left(Y_i-\tau D_i-c\right)
(1,D_i)^\top .
\]
The corresponding covariance matrix is
\[H_N^{\mathrm{GRF,full}}(x)
=
\operatorname{Var}
\!\left[
\Psi(\tau,c)
\right]
=
\operatorname{Var}
\!\left[
\sum_{i=1}^n
\alpha_i(x)
\psi_{\tau,c}(O_i)
\right].
\]
Since our interest is restricted to the treatment effect $\tau(x)$, we consider the corresponding scalar variance component, 
\[
H_N^{\mathrm{GRF}}(x)=
\operatorname{Var}\left[
\sum_{i=1}^N
\alpha_i(x)
(D_i-p(x))
\left(Y_i-\tau(x)D_i-c(x)\right)
\right].
\]
Following \cite{athey_generalized_2019} this yields the asymptotic variance
\begin{align} \label{eq:appendix_grf_variance}
\operatorname{Var}\left(\hat\tau(x)\right)
=
V(x)^{-2}
H_N^{\mathrm{GRF}}(x)
\end{align}
where $V(x)$ is the curvature term of the estimating equation. Hence, $H_N^{\mathrm{GRF}}(x)$ cannot itself be identified with the variance of the completed GRF CATE estimator. To see this, suppose that
\[H_N^{\mathrm{GRF}}(x)
=
\operatorname{Var}\!\left(\hat\tau_{\mathrm{GRF}}(x)\right).\]
Substitution into Equation~\eqref{eq:appendix_grf_variance} would imply
\[\operatorname{Var}\!\left(\hat\tau_{\mathrm{GRF}}(x)\right)
=
V(x)^{-2}
\operatorname{Var}\!\left(\hat\tau_{\mathrm{GRF}}(x)\right).\]
For a nondegenerate estimator with positive variance, this equality requires $V(x)^2=1$. In the binary-treatment setting,
$$V(x)
=
\mathbb E\!\left[
\left(D_i-p(X_i)\right)^2
\mid X_i=x
\right]
=
p(x)\bigl(1-p(x)\bigr)
\leq \frac{1}{4}$$
so $V(x)^2=1$ is not possible. Therefore, $ H_N^{\mathrm{GRF}}(x)$ is not the variance of the completed CATE estimator, but an intermediate variance component that must be transformed by $V(x)^{-2}$ within the GRF framework.

Our implementation follows the original causal forest construction, in which each tree produces a completed treatment effect prediction $\hat\tau_b(x)$ and the forest estimate is
$$\hat\tau(x)
=
\frac{1}{B}
\sum_{b=1}^{B}\hat\tau_b(x).$$
Both the Bootstrap of Little Bags and the Infinitesimal Jackknife are applied directly to these completed tree-level predictions. The resulting empirical forest-variance estimators $\widehat H_N^{\mathrm{BLB}}(x)$ and 
$\widehat H_N^{\mathrm{IJ}}(x)$ therefore quantify the variability of the completed causal forest predictions rather than the intermediate variance component $H_N^{\mathrm{GRF}}(x)$.  Nevertheless, for comparison with the GRF formulation of \citet{athey_generalized_2019}, we additionally report curvature-adjusted versions obtained by multiplying the empirical variance estimates by $ V(x)^{-2}$.
Although this adjustment is not theoretically required for variance estimators applied directly to completed tree-level predictions, it produces wider confidence intervals and improve empirical coverage when the unadjusted intervals undercover, bu at the cost of overcoverage.

\section{Additional Monte Carlo Results}

All simulations are conducted in \texttt{R} using RStudio as the
development environment \citep{r_2025, rstudio_2025}. The code for reproducibility is available at
\url{https://github.com/karo93/RobustMedianCausalForest}.  

\label{appendix:Robust_ct_monte_additional_results}

\subsection{Heavy-tailed Outcome Noise}

\begin{landscape}
    \begin{table}[!h]
\centering
\caption{\label{tab:ci_metric_ct_lms} Coverage and width of 95\% confidence intervals for BLB and IJ across effect scenarios and noise distributions. Results for splitting rules CT and LMS.}
\centering
\resizebox{\ifdim\width>\linewidth\linewidth\else\width\fi}{!}{
\fontsize{10}{12}\selectfont
\begin{tabular}[t]{ll>{\raggedright\arraybackslash}p{0.25cm}rrrrrrrr>{\raggedleft\arraybackslash}p{0.25cm}rrrrrrrr}
\toprule
\multicolumn{1}{c}{ } & \multicolumn{1}{c}{ } & \multicolumn{1}{c}{ } & \multicolumn{8}{c}{CT} & \multicolumn{1}{c}{ } & \multicolumn{8}{c}{LMS} \\
\cmidrule(l{3pt}r{3pt}){4-11} \cmidrule(l{3pt}r{3pt}){13-20}
\multicolumn{1}{c}{ } & \multicolumn{1}{c}{ } & \multicolumn{1}{c}{ } & \multicolumn{2}{c}{BLB} & \multicolumn{2}{c}{$\text{BLB}^{V}$} & \multicolumn{2}{c}{IJ} & \multicolumn{2}{c}{$\text{IJ}^{V}$} & \multicolumn{1}{c}{ } & \multicolumn{2}{c}{BLB} & \multicolumn{2}{c}{$\text{BLB}^{V}$} & \multicolumn{2}{c}{IJ} & \multicolumn{2}{c}{$\text{IJ}^{V}$} \\
\cmidrule(l{3pt}r{3pt}){4-5} \cmidrule(l{3pt}r{3pt}){6-7} \cmidrule(l{3pt}r{3pt}){8-9} \cmidrule(l{3pt}r{3pt}){10-11} \cmidrule(l{3pt}r{3pt}){13-14} \cmidrule(l{3pt}r{3pt}){15-16} \cmidrule(l{3pt}r{3pt}){17-18} \cmidrule(l{3pt}r{3pt}){19-20}
Effect & Noise &   & Coverage & Width & Coverage & Width & Coverage & Width & Coverage & Width &   & Coverage & Width & Coverage & Width & Coverage & Width & Coverage & Width\\
\midrule
 &  &  & \phantom{(}0.20\phantom{)} & \phantom{(}35.84\phantom{)} & \phantom{(}0.50\phantom{)} & \phantom{(}143.18\phantom{)} & \phantom{(}0.21\phantom{)} & \phantom{(}38.11\phantom{)} & \phantom{(}0.60\phantom{)} & \phantom{(}152.24\phantom{)} &  & \phantom{(}0.19\phantom{)} & \phantom{(}36.44\phantom{)} & \phantom{(}0.47\phantom{)} & \phantom{(}145.57\phantom{)} & \phantom{(}0.20\phantom{)} & \phantom{(}37.06\phantom{)} & \phantom{(}0.57\phantom{)} & \phantom{(}148.03\phantom{)}\\

 & \multirow{-2}{*}{\raggedright\arraybackslash Cauchy} &  & (0.14) & (271.95) & (0.17) & (1088.27) & (0.11) & (380.43) & (0.14) & (1521.94) &  & (0.14) & (339.65) & (0.18) & (1359.02) & (0.12) & (366.10) & (0.16) & (1464.20)\\
\addlinespace[4pt]

 &  &  & \phantom{(}0.06\phantom{)} & \phantom{(}4.18\phantom{)} & \phantom{(}0.23\phantom{)} & \phantom{(}16.66\phantom{)} & \phantom{(}0.09\phantom{)} & \phantom{(}6.23\phantom{)} & \phantom{(}0.34\phantom{)} & \phantom{(}24.86\phantom{)} &  & \phantom{(}0.05\phantom{)} & \phantom{(}3.83\phantom{)} & \phantom{(}0.20\phantom{)} & \phantom{(}15.26\phantom{)} & \phantom{(}0.07\phantom{)} & \phantom{(}5.21\phantom{)} & \phantom{(}0.28\phantom{)} & \phantom{(}20.80\phantom{)}\\

 & \multirow{-2}{*}{\raggedright\arraybackslash Laplace} &  & (0.01) & (1.02) & (0.03) & (4.06) & (0.01) & (1.36) & (0.01) & (5.44) &  & (0.01) & (0.96) & (0.03) & (3.83) & (0.01) & (1.13) & (0.01) & (4.49)\\
\addlinespace[4pt]

 &  &  & \phantom{(}0.06\phantom{)} & \phantom{(}4.17\phantom{)} & \phantom{(}0.23\phantom{)} & \phantom{(}16.64\phantom{)} & \phantom{(}0.09\phantom{)} & \phantom{(}6.20\phantom{)} & \phantom{(}0.34\phantom{)} & \phantom{(}24.73\phantom{)} &  & \phantom{(}0.05\phantom{)} & \phantom{(}3.84\phantom{)} & \phantom{(}0.20\phantom{)} & \phantom{(}15.30\phantom{)} & \phantom{(}0.07\phantom{)} & \phantom{(}5.21\phantom{)} & \phantom{(}0.28\phantom{)} & \phantom{(}20.80\phantom{)}\\

 & \multirow{-2}{*}{\raggedright\arraybackslash Normal} &  & (0.01) & (1.07) & (0.03) & (4.25) & (0.01) & (1.34) & (0.01) & (5.32) &  & (0.01) & (0.96) & (0.03) & (3.81) & (0.01) & (1.11) & (0.01) & (4.44)\\
\addlinespace[4pt]

 &  &  & \phantom{(}0.06\phantom{)} & \phantom{(}4.23\phantom{)} & \phantom{(}0.23\phantom{)} & \phantom{(}16.86\phantom{)} & \phantom{(}0.09\phantom{)} & \phantom{(}6.24\phantom{)} & \phantom{(}0.34\phantom{)} & \phantom{(}24.88\phantom{)} &  & \phantom{(}0.05\phantom{)} & \phantom{(}3.80\phantom{)} & \phantom{(}0.20\phantom{)} & \phantom{(}15.16\phantom{)} & \phantom{(}0.07\phantom{)} & \phantom{(}5.22\phantom{)} & \phantom{(}0.28\phantom{)} & \phantom{(}20.84\phantom{)}\\

\multirow{-8}{*}{\raggedright\arraybackslash S1} & \multirow{-2}{*}{\raggedright\arraybackslash Student-$t$} &  & (0.01) & (1.04) & (0.03) & (4.16) & (0.01) & (1.32) & (0.01) & (5.27) &  & (0.01) & (0.92) & (0.03) & (3.66) & (0.01) & (1.11) & (0.01) & (4.42)\\
\addlinespace[4pt]

\cmidrule{1-20}
 &  &  & \phantom{(}0.73\phantom{)} & \phantom{(}25.10\phantom{)} & \phantom{(}0.85\phantom{)} & \phantom{(}100.20\phantom{)} & \phantom{(}0.86\phantom{)} & \phantom{(}22.71\phantom{)} & \phantom{(}1.00\phantom{)} & \phantom{(}90.65\phantom{)} &  & \phantom{(}0.73\phantom{)} & \phantom{(}24.14\phantom{)} & \phantom{(}0.86\phantom{)} & \phantom{(}96.35\phantom{)} & \phantom{(}0.87\phantom{)} & \phantom{(}23.27\phantom{)} & \phantom{(}1.00\phantom{)} & \phantom{(}92.89\phantom{)}\\

 & \multirow{-2}{*}{\raggedright\arraybackslash Cauchy} &  & (0.11) & (77.83) & (0.07) & (310.91) & (0.15) & (65.11) & (0.00) & (260.05) &  & (0.11) & (65.04) & (0.07) & (259.87) & (0.14) & (69.33) & (0.00) & (276.97)\\
\addlinespace[4pt]

 &  &  & \phantom{(}0.11\phantom{)} & \phantom{(}0.29\phantom{)} & \phantom{(}0.70\phantom{)} & \phantom{(}1.16\phantom{)} & \phantom{(}0.12\phantom{)} & \phantom{(}0.37\phantom{)} & \phantom{(}0.89\phantom{)} & \phantom{(}1.47\phantom{)} &  & \phantom{(}0.06\phantom{)} & \phantom{(}0.29\phantom{)} & \phantom{(}0.69\phantom{)} & \phantom{(}1.15\phantom{)} & \phantom{(}0.06\phantom{)} & \phantom{(}0.35\phantom{)} & \phantom{(}0.89\phantom{)} & \phantom{(}1.39\phantom{)}\\

 & \multirow{-2}{*}{\raggedright\arraybackslash Laplace} &  & (0.10) & (0.04) & (0.08) & (0.16) & (0.12) & (0.01) & (0.01) & (0.05) &  & (0.07) & (0.04) & (0.09) & (0.16) & (0.08) & (0.01) & (0.01) & (0.04)\\
\addlinespace[4pt]

 &  &  & \phantom{(}0.05\phantom{)} & \phantom{(}0.22\phantom{)} & \phantom{(}0.62\phantom{)} & \phantom{(}0.88\phantom{)} & \phantom{(}0.04\phantom{)} & \phantom{(}0.28\phantom{)} & \phantom{(}0.88\phantom{)} & \phantom{(}1.11\phantom{)} &  & \phantom{(}0.02\phantom{)} & \phantom{(}0.22\phantom{)} & \phantom{(}0.60\phantom{)} & \phantom{(}0.86\phantom{)} & \phantom{(}0.02\phantom{)} & \phantom{(}0.26\phantom{)} & \phantom{(}0.86\phantom{)} & \phantom{(}1.05\phantom{)}\\

 & \multirow{-2}{*}{\raggedright\arraybackslash Normal} &  & (0.05) & (0.03) & (0.09) & (0.12) & (0.06) & (0.01) & (0.03) & (0.03) &  & (0.02) & (0.03) & (0.10) & (0.11) & (0.02) & (0.01) & (0.05) & (0.02)\\
\addlinespace[4pt]

 &  &  & \phantom{(}0.06\phantom{)} & \phantom{(}0.22\phantom{)} & \phantom{(}0.61\phantom{)} & \phantom{(}0.86\phantom{)} & \phantom{(}0.05\phantom{)} & \phantom{(}0.28\phantom{)} & \phantom{(}0.87\phantom{)} & \phantom{(}1.10\phantom{)} &  & \phantom{(}0.02\phantom{)} & \phantom{(}0.21\phantom{)} & \phantom{(}0.58\phantom{)} & \phantom{(}0.84\phantom{)} & \phantom{(}0.02\phantom{)} & \phantom{(}0.25\phantom{)} & \phantom{(}0.82\phantom{)} & \phantom{(}1.00\phantom{)}\\

\multirow{-8}{*}{\raggedright\arraybackslash S2} & \multirow{-2}{*}{\raggedright\arraybackslash Student-$t$} &  & (0.06) & (0.03) & (0.09) & (0.13) & (0.06) & (0.02) & (0.04) & (0.07) &  & (0.01) & (0.03) & (0.11) & (0.13) & (0.01) & (0.02) & (0.10) & (0.06)\\
\addlinespace[4pt]

\cmidrule{1-20}
 &  &  & \phantom{(}0.75\phantom{)} & \phantom{(}35.12\phantom{)} & \phantom{(}0.85\phantom{)} & \phantom{(}140.07\phantom{)} & \phantom{(}0.89\phantom{)} & \phantom{(}31.09\phantom{)} & \phantom{(}1.00\phantom{)} & \phantom{(}124.02\phantom{)} &  & \phantom{(}0.75\phantom{)} & \phantom{(}31.05\phantom{)} & \phantom{(}0.85\phantom{)} & \phantom{(}123.87\phantom{)} & \phantom{(}0.90\phantom{)} & \phantom{(}31.34\phantom{)} & \phantom{(}1.00\phantom{)} & \phantom{(}125.03\phantom{)}\\

 & \multirow{-2}{*}{\raggedright\arraybackslash Cauchy} &  & (0.11) & (252.59) & (0.07) & (1006.90) & (0.16) & (238.77) & (0.00) & (951.89) &  & (0.12) & (148.54) & (0.08) & (592.40) & (0.15) & (184.19) & (0.00) & (734.46)\\
\addlinespace[4pt]

 &  &  & \phantom{(}0.78\phantom{)} & \phantom{(}0.27\phantom{)} & \phantom{(}0.88\phantom{)} & \phantom{(}1.07\phantom{)} & \phantom{(}0.96\phantom{)} & \phantom{(}0.33\phantom{)} & \phantom{(}1.00\phantom{)} & \phantom{(}1.32\phantom{)} &  & \phantom{(}0.78\phantom{)} & \phantom{(}0.27\phantom{)} & \phantom{(}0.88\phantom{)} & \phantom{(}1.07\phantom{)} & \phantom{(}0.96\phantom{)} & \phantom{(}0.33\phantom{)} & \phantom{(}1.00\phantom{)} & \phantom{(}1.30\phantom{)}\\

 & \multirow{-2}{*}{\raggedright\arraybackslash Laplace} &  & (0.13) & (0.04) & (0.06) & (0.15) & (0.08) & (0.01) & (0.00) & (0.04) &  & (0.12) & (0.04) & (0.06) & (0.15) & (0.09) & (0.01) & (0.00) & (0.04)\\
\addlinespace[4pt]

 &  &  & \phantom{(}0.78\phantom{)} & \phantom{(}0.19\phantom{)} & \phantom{(}0.88\phantom{)} & \phantom{(}0.76\phantom{)} & \phantom{(}0.95\phantom{)} & \phantom{(}0.23\phantom{)} & \phantom{(}1.00\phantom{)} & \phantom{(}0.93\phantom{)} &  & \phantom{(}0.77\phantom{)} & \phantom{(}0.19\phantom{)} & \phantom{(}0.88\phantom{)} & \phantom{(}0.75\phantom{)} & \phantom{(}0.95\phantom{)} & \phantom{(}0.23\phantom{)} & \phantom{(}1.00\phantom{)} & \phantom{(}0.93\phantom{)}\\

 & \multirow{-2}{*}{\raggedright\arraybackslash Normal} &  & (0.13) & (0.03) & (0.06) & (0.11) & (0.09) & (0.01) & (0.00) & (0.02) &  & (0.12) & (0.03) & (0.06) & (0.10) & (0.09) & (0.01) & (0.00) & (0.02)\\
\addlinespace[4pt]

 &  &  & \phantom{(}0.78\phantom{)} & \phantom{(}0.19\phantom{)} & \phantom{(}0.88\phantom{)} & \phantom{(}0.75\phantom{)} & \phantom{(}0.95\phantom{)} & \phantom{(}0.23\phantom{)} & \phantom{(}1.00\phantom{)} & \phantom{(}0.91\phantom{)} &  & \phantom{(}0.78\phantom{)} & \phantom{(}0.19\phantom{)} & \phantom{(}0.88\phantom{)} & \phantom{(}0.75\phantom{)} & \phantom{(}0.95\phantom{)} & \phantom{(}0.22\phantom{)} & \phantom{(}1.00\phantom{)} & \phantom{(}0.90\phantom{)}\\

\multirow{-8}{*}{\raggedright\arraybackslash S3} & \multirow{-2}{*}{\raggedright\arraybackslash Student-$t$} &  & (0.12) & (0.04) & (0.06) & (0.16) & (0.10) & (0.03) & (0.00) & (0.13) &  & (0.12) & (0.04) & (0.06) & (0.16) & (0.10) & (0.03) & (0.00) & (0.13)\\
\addlinespace[4pt]

\cmidrule{1-20}
 &  &  & \phantom{(}0.37\phantom{)} & \phantom{(}32.50\phantom{)} & \phantom{(}0.71\phantom{)} & \phantom{(}129.69\phantom{)} & \phantom{(}0.40\phantom{)} & \phantom{(}33.51\phantom{)} & \phantom{(}0.87\phantom{)} & \phantom{(}133.72\phantom{)} &  & \phantom{(}0.36\phantom{)} & \phantom{(}33.02\phantom{)} & \phantom{(}0.70\phantom{)} & \phantom{(}131.73\phantom{)} & \phantom{(}0.39\phantom{)} & \phantom{(}32.88\phantom{)} & \phantom{(}0.85\phantom{)} & \phantom{(}131.18\phantom{)}\\

 & \multirow{-2}{*}{\raggedright\arraybackslash Cauchy} &  & (0.17) & (199.87) & (0.12) & (798.45) & (0.13) & (299.71) & (0.10) & (1196.94) &  & (0.17) & (216.65) & (0.13) & (865.35) & (0.15) & (258.37) & (0.12) & (1031.75)\\
\addlinespace[4pt]

 &  &  & \phantom{(}0.06\phantom{)} & \phantom{(}1.48\phantom{)} & \phantom{(}0.24\phantom{)} & \phantom{(}5.91\phantom{)} & \phantom{(}0.09\phantom{)} & \phantom{(}2.17\phantom{)} & \phantom{(}0.35\phantom{)} & \phantom{(}8.65\phantom{)} &  & \phantom{(}0.06\phantom{)} & \phantom{(}1.38\phantom{)} & \phantom{(}0.21\phantom{)} & \phantom{(}5.49\phantom{)} & \phantom{(}0.07\phantom{)} & \phantom{(}1.81\phantom{)} & \phantom{(}0.29\phantom{)} & \phantom{(}7.22\phantom{)}\\

 & \multirow{-2}{*}{\raggedright\arraybackslash Laplace} &  & (0.01) & (0.23) & (0.04) & (0.91) & (0.01) & (0.07) & (0.01) & (0.27) &  & (0.01) & (0.23) & (0.03) & (0.90) & (0.01) & (0.07) & (0.01) & (0.28)\\
\addlinespace[4pt]

 &  &  & \phantom{(}0.06\phantom{)} & \phantom{(}1.46\phantom{)} & \phantom{(}0.23\phantom{)} & \phantom{(}5.82\phantom{)} & \phantom{(}0.09\phantom{)} & \phantom{(}2.14\phantom{)} & \phantom{(}0.34\phantom{)} & \phantom{(}8.54\phantom{)} &  & \phantom{(}0.06\phantom{)} & \phantom{(}1.36\phantom{)} & \phantom{(}0.21\phantom{)} & \phantom{(}5.43\phantom{)} & \phantom{(}0.07\phantom{)} & \phantom{(}1.77\phantom{)} & \phantom{(}0.28\phantom{)} & \phantom{(}7.08\phantom{)}\\

 & \multirow{-2}{*}{\raggedright\arraybackslash Normal} &  & (0.01) & (0.23) & (0.04) & (0.92) & (0.01) & (0.07) & (0.01) & (0.26) &  & (0.01) & (0.22) & (0.03) & (0.88) & (0.01) & (0.07) & (0.01) & (0.28)\\
\addlinespace[4pt]

 &  &  & \phantom{(}0.06\phantom{)} & \phantom{(}1.46\phantom{)} & \phantom{(}0.23\phantom{)} & \phantom{(}5.83\phantom{)} & \phantom{(}0.09\phantom{)} & \phantom{(}2.14\phantom{)} & \phantom{(}0.35\phantom{)} & \phantom{(}8.55\phantom{)} &  & \phantom{(}0.05\phantom{)} & \phantom{(}1.35\phantom{)} & \phantom{(}0.21\phantom{)} & \phantom{(}5.40\phantom{)} & \phantom{(}0.07\phantom{)} & \phantom{(}1.77\phantom{)} & \phantom{(}0.28\phantom{)} & \phantom{(}7.08\phantom{)}\\

\multirow{-8}{*}{\raggedright\arraybackslash S4} & \multirow{-2}{*}{\raggedright\arraybackslash Student-$t$} &  & (0.01) & (0.23) & (0.04) & (0.93) & (0.01) & (0.07) & (0.01) & (0.27) &  & (0.01) & (0.23) & (0.04) & (0.93) & (0.01) & (0.07) & (0.01) & (0.29)\\
\addlinespace[4pt]
\bottomrule
\end{tabular}}%
\vspace{0.5em}
\begin{minipage}{0.95\linewidth}
\footnotesize
\textit{Notes:} Entries report Monte Carlo averages across $MC=1000$ simulation runs, with standard deviations in parentheses. Coverage refers to the proportion of confidence intervals containing the true CATE. Superscript $V$ denotes variance estimates additionally scaled by $V(x)^{-2}$.
\end{minipage}

\end{table}

\end{landscape}

\begin{landscape}
    \begin{table}[!h]
\centering
\caption{\label{tab:ci_metric_mad_msd} Coverage and width of 95\% bootstrap confidence intervals for BLB and IJ across effect scenarios and noise distributions. Results for splitting rules MAD and MSD.}
\centering
\resizebox{\ifdim\width>\linewidth\linewidth\else\width\fi}{!}{
\fontsize{10}{12}\selectfont
\begin{tabular}[t]{ll>{\raggedright\arraybackslash}p{0.25cm}rrrrrrrr>{\raggedleft\arraybackslash}p{0.25cm}rrrrrrrr}
\toprule
\multicolumn{1}{c}{ } & \multicolumn{1}{c}{ } & \multicolumn{1}{c}{ } & \multicolumn{8}{c}{MAD} & \multicolumn{1}{c}{ } & \multicolumn{8}{c}{MSD} \\
\cmidrule(l{3pt}r{3pt}){4-11} \cmidrule(l{3pt}r{3pt}){13-20}
\multicolumn{1}{c}{ } & \multicolumn{1}{c}{ } & \multicolumn{1}{c}{ } & \multicolumn{2}{c}{BLB} & \multicolumn{2}{c}{$\text{BLB}^{V}$} & \multicolumn{2}{c}{IJ} & \multicolumn{2}{c}{$\text{IJ}^{V}$} & \multicolumn{1}{c}{ } & \multicolumn{2}{c}{BLB} & \multicolumn{2}{c}{$\text{BLB}^{V}$} & \multicolumn{2}{c}{IJ} & \multicolumn{2}{c}{$\text{IJ}^{V}$} \\
\cmidrule(l{3pt}r{3pt}){4-5} \cmidrule(l{3pt}r{3pt}){6-7} \cmidrule(l{3pt}r{3pt}){8-9} \cmidrule(l{3pt}r{3pt}){10-11} \cmidrule(l{3pt}r{3pt}){13-14} \cmidrule(l{3pt}r{3pt}){15-16} \cmidrule(l{3pt}r{3pt}){17-18} \cmidrule(l{3pt}r{3pt}){19-20}
Effect & Noise &   & Coverage & Width & Coverage & Width & Coverage & Width & Coverage & Width &   & Coverage & Width & Coverage & Width & Coverage & Width & Coverage & Width\\
\midrule
 &  &  & \phantom{(}0.19\phantom{)} & \phantom{(}37.43\phantom{)} & \phantom{(}0.49\phantom{)} & \phantom{(}149.51\phantom{)} & \phantom{(}0.21\phantom{)} & \phantom{(}37.62\phantom{)} & \phantom{(}0.59\phantom{)} & \phantom{(}150.27\phantom{)} &  & \phantom{(}0.18\phantom{)} & \phantom{(}36.76\phantom{)} & \phantom{(}0.47\phantom{)} & \phantom{(}146.88\phantom{)} & \phantom{(}0.19\phantom{)} & \phantom{(}37.97\phantom{)} & \phantom{(}0.53\phantom{)} & \phantom{(}151.69\phantom{)}\\

 & \multirow{-2}{*}{\raggedright\arraybackslash Cauchy} &  & (0.14) & (336.90) & (0.17) & (1347.90) & (0.11) & (377.82) & (0.15) & (1511.03) &  & (0.15) & (329.10) & (0.20) & (1317.18) & (0.15) & (337.45) & (0.20) & (1350.50)\\
\addlinespace[4pt]

 &  &  & \phantom{(}0.06\phantom{)} & \phantom{(}4.20\phantom{)} & \phantom{(}0.23\phantom{)} & \phantom{(}16.77\phantom{)} & \phantom{(}0.08\phantom{)} & \phantom{(}5.93\phantom{)} & \phantom{(}0.32\phantom{)} & \phantom{(}23.65\phantom{)} &  & \phantom{(}0.05\phantom{)} & \phantom{(}3.78\phantom{)} & \phantom{(}0.18\phantom{)} & \phantom{(}15.09\phantom{)} & \phantom{(}0.06\phantom{)} & \phantom{(}4.60\phantom{)} & \phantom{(}0.22\phantom{)} & \phantom{(}18.33\phantom{)}\\

 & \multirow{-2}{*}{\raggedright\arraybackslash Laplace} &  & (0.01) & (1.04) & (0.03) & (4.14) & (0.01) & (1.32) & (0.01) & (5.25) &  & (0.01) & (0.95) & (0.03) & (3.78) & (0.01) & (0.99) & (0.01) & (3.95)\\
\addlinespace[4pt]

 &  &  & \phantom{(}0.06\phantom{)} & \phantom{(}4.20\phantom{)} & \phantom{(}0.23\phantom{)} & \phantom{(}16.74\phantom{)} & \phantom{(}0.08\phantom{)} & \phantom{(}5.90\phantom{)} & \phantom{(}0.32\phantom{)} & \phantom{(}23.54\phantom{)} &  & \phantom{(}0.05\phantom{)} & \phantom{(}3.74\phantom{)} & \phantom{(}0.18\phantom{)} & \phantom{(}14.94\phantom{)} & \phantom{(}0.06\phantom{)} & \phantom{(}4.58\phantom{)} & \phantom{(}0.22\phantom{)} & \phantom{(}18.27\phantom{)}\\

 & \multirow{-2}{*}{\raggedright\arraybackslash Normal} &  & (0.01) & (1.05) & (0.03) & (4.17) & (0.01) & (1.30) & (0.01) & (5.16) &  & (0.01) & (0.92) & (0.03) & (3.67) & (0.01) & (0.97) & (0.01) & (3.85)\\
\addlinespace[4pt]

 &  &  & \phantom{(}0.06\phantom{)} & \phantom{(}4.19\phantom{)} & \phantom{(}0.22\phantom{)} & \phantom{(}16.71\phantom{)} & \phantom{(}0.08\phantom{)} & \phantom{(}5.92\phantom{)} & \phantom{(}0.32\phantom{)} & \phantom{(}23.61\phantom{)} &  & \phantom{(}0.05\phantom{)} & \phantom{(}3.78\phantom{)} & \phantom{(}0.18\phantom{)} & \phantom{(}15.08\phantom{)} & \phantom{(}0.06\phantom{)} & \phantom{(}4.59\phantom{)} & \phantom{(}0.22\phantom{)} & \phantom{(}18.31\phantom{)}\\

\multirow{-8}{*}{\raggedright\arraybackslash S1} & \multirow{-2}{*}{\raggedright\arraybackslash Student-$t$} &  & (0.01) & (1.03) & (0.03) & (4.10) & (0.01) & (1.27) & (0.01) & (5.06) &  & (0.01) & (0.91) & (0.03) & (3.65) & (0.01) & (0.97) & (0.01) & (3.87)\\
\addlinespace[4pt]

\cmidrule{1-20}
 &  &  & \phantom{(}0.72\phantom{)} & \phantom{(}24.24\phantom{)} & \phantom{(}0.84\phantom{)} & \phantom{(}96.74\phantom{)} & \phantom{(}0.86\phantom{)} & \phantom{(}22.57\phantom{)} & \phantom{(}1.00\phantom{)} & \phantom{(}90.07\phantom{)} &  & \phantom{(}0.77\phantom{)} & \phantom{(}25.07\phantom{)} & \phantom{(}0.90\phantom{)} & \phantom{(}100.07\phantom{)} & \phantom{(}0.89\phantom{)} & \phantom{(}24.63\phantom{)} & \phantom{(}1.00\phantom{)} & \phantom{(}98.32\phantom{)}\\

 & \multirow{-2}{*}{\raggedright\arraybackslash Cauchy} &  & (0.11) & (66.84) & (0.07) & (267.09) & (0.15) & (61.39) & (0.00) & (245.19) &  & (0.13) & (77.84) & (0.07) & (311.01) & (0.12) & (76.68) & (0.00) & (306.44)\\
\addlinespace[4pt]

 &  &  & \phantom{(}0.09\phantom{)} & \phantom{(}0.29\phantom{)} & \phantom{(}0.70\phantom{)} & \phantom{(}1.16\phantom{)} & \phantom{(}0.11\phantom{)} & \phantom{(}0.37\phantom{)} & \phantom{(}0.89\phantom{)} & \phantom{(}1.48\phantom{)} &  & \phantom{(}0.04\phantom{)} & \phantom{(}0.29\phantom{)} & \phantom{(}0.71\phantom{)} & \phantom{(}1.15\phantom{)} & \phantom{(}0.04\phantom{)} & \phantom{(}0.33\phantom{)} & \phantom{(}0.89\phantom{)} & \phantom{(}1.33\phantom{)}\\

 & \multirow{-2}{*}{\raggedright\arraybackslash Laplace} &  & (0.08) & (0.04) & (0.08) & (0.16) & (0.11) & (0.01) & (0.01) & (0.05) &  & (0.06) & (0.04) & (0.09) & (0.16) & (0.06) & (0.01) & (0.01) & (0.04)\\
\addlinespace[4pt]

 &  &  & \phantom{(}0.03\phantom{)} & \phantom{(}0.22\phantom{)} & \phantom{(}0.61\phantom{)} & \phantom{(}0.87\phantom{)} & \phantom{(}0.04\phantom{)} & \phantom{(}0.28\phantom{)} & \phantom{(}0.87\phantom{)} & \phantom{(}1.13\phantom{)} &  & \phantom{(}0.01\phantom{)} & \phantom{(}0.21\phantom{)} & \phantom{(}0.60\phantom{)} & \phantom{(}0.85\phantom{)} & \phantom{(}0.02\phantom{)} & \phantom{(}0.25\phantom{)} & \phantom{(}0.83\phantom{)} & \phantom{(}0.98\phantom{)}\\

 & \multirow{-2}{*}{\raggedright\arraybackslash Normal} &  & (0.04) & (0.03) & (0.09) & (0.12) & (0.05) & (0.01) & (0.03) & (0.03) &  & (0.01) & (0.03) & (0.11) & (0.12) & (0.01) & (0.01) & (0.08) & (0.02)\\
\addlinespace[4pt]

 &  &  & \phantom{(}0.03\phantom{)} & \phantom{(}0.22\phantom{)} & \phantom{(}0.60\phantom{)} & \phantom{(}0.86\phantom{)} & \phantom{(}0.04\phantom{)} & \phantom{(}0.27\phantom{)} & \phantom{(}0.85\phantom{)} & \phantom{(}1.10\phantom{)} &  & \phantom{(}0.01\phantom{)} & \phantom{(}0.21\phantom{)} & \phantom{(}0.60\phantom{)} & \phantom{(}0.84\phantom{)} & \phantom{(}0.02\phantom{)} & \phantom{(}0.24\phantom{)} & \phantom{(}0.80\phantom{)} & \phantom{(}0.96\phantom{)}\\

\multirow{-8}{*}{\raggedright\arraybackslash S2} & \multirow{-2}{*}{\raggedright\arraybackslash Student-$t$} &  & (0.04) & (0.03) & (0.10) & (0.13) & (0.05) & (0.02) & (0.05) & (0.06) &  & (0.01) & (0.03) & (0.11) & (0.13) & (0.01) & (0.02) & (0.10) & (0.07)\\
\addlinespace[4pt]

\cmidrule{1-20}
 &  &  & \phantom{(}0.73\phantom{)} & \phantom{(}31.68\phantom{)} & \phantom{(}0.84\phantom{)} & \phantom{(}126.38\phantom{)} & \phantom{(}0.89\phantom{)} & \phantom{(}32.03\phantom{)} & \phantom{(}1.00\phantom{)} & \phantom{(}127.77\phantom{)} &  & \phantom{(}0.80\phantom{)} & \phantom{(}33.23\phantom{)} & \phantom{(}0.90\phantom{)} & \phantom{(}132.59\phantom{)} & \phantom{(}0.93\phantom{)} & \phantom{(}34.20\phantom{)} & \phantom{(}1.00\phantom{)} & \phantom{(}136.45\phantom{)}\\

 & \multirow{-2}{*}{\raggedright\arraybackslash Cauchy} &  & (0.12) & (185.34) & (0.07) & (738.95) & (0.16) & (247.87) & (0.00) & (987.97) &  & (0.12) & (166.56) & (0.07) & (664.36) & (0.13) & (230.32) & (0.00) & (918.46)\\
\addlinespace[4pt]

 &  &  & \phantom{(}0.78\phantom{)} & \phantom{(}0.27\phantom{)} & \phantom{(}0.88\phantom{)} & \phantom{(}1.07\phantom{)} & \phantom{(}0.96\phantom{)} & \phantom{(}0.33\phantom{)} & \phantom{(}1.00\phantom{)} & \phantom{(}1.32\phantom{)} &  & \phantom{(}0.81\phantom{)} & \phantom{(}0.26\phantom{)} & \phantom{(}0.90\phantom{)} & \phantom{(}1.06\phantom{)} & \phantom{(}0.96\phantom{)} & \phantom{(}0.31\phantom{)} & \phantom{(}1.00\phantom{)} & \phantom{(}1.23\phantom{)}\\

 & \multirow{-2}{*}{\raggedright\arraybackslash Laplace} &  & (0.12) & (0.04) & (0.06) & (0.15) & (0.09) & (0.01) & (0.00) & (0.04) &  & (0.13) & (0.04) & (0.06) & (0.15) & (0.10) & (0.01) & (0.00) & (0.04)\\
\addlinespace[4pt]

 &  &  & \phantom{(}0.78\phantom{)} & \phantom{(}0.19\phantom{)} & \phantom{(}0.88\phantom{)} & \phantom{(}0.76\phantom{)} & \phantom{(}0.95\phantom{)} & \phantom{(}0.23\phantom{)} & \phantom{(}1.00\phantom{)} & \phantom{(}0.93\phantom{)} &  & \phantom{(}0.82\phantom{)} & \phantom{(}0.19\phantom{)} & \phantom{(}0.91\phantom{)} & \phantom{(}0.76\phantom{)} & \phantom{(}0.95\phantom{)} & \phantom{(}0.22\phantom{)} & \phantom{(}1.00\phantom{)} & \phantom{(}0.87\phantom{)}\\

 & \multirow{-2}{*}{\raggedright\arraybackslash Normal} &  & (0.12) & (0.03) & (0.06) & (0.11) & (0.10) & (0.01) & (0.00) & (0.02) &  & (0.13) & (0.03) & (0.06) & (0.11) & (0.11) & (0.01) & (0.00) & (0.02)\\
\addlinespace[4pt]

 &  &  & \phantom{(}0.77\phantom{)} & \phantom{(}0.19\phantom{)} & \phantom{(}0.87\phantom{)} & \phantom{(}0.75\phantom{)} & \phantom{(}0.95\phantom{)} & \phantom{(}0.23\phantom{)} & \phantom{(}1.00\phantom{)} & \phantom{(}0.91\phantom{)} &  & \phantom{(}0.81\phantom{)} & \phantom{(}0.19\phantom{)} & \phantom{(}0.90\phantom{)} & \phantom{(}0.75\phantom{)} & \phantom{(}0.95\phantom{)} & \phantom{(}0.21\phantom{)} & \phantom{(}1.00\phantom{)} & \phantom{(}0.85\phantom{)}\\

\multirow{-8}{*}{\raggedright\arraybackslash S3} & \multirow{-2}{*}{\raggedright\arraybackslash Student-$t$} &  & (0.12) & (0.04) & (0.06) & (0.15) & (0.10) & (0.03) & (0.00) & (0.13) &  & (0.13) & (0.04) & (0.06) & (0.14) & (0.11) & (0.03) & (0.00) & (0.13)\\
\addlinespace[4pt]

\cmidrule{1-20}
 &  &  & \phantom{(}0.36\phantom{)} & \phantom{(}33.22\phantom{)} & \phantom{(}0.70\phantom{)} & \phantom{(}132.55\phantom{)} & \phantom{(}0.39\phantom{)} & \phantom{(}34.05\phantom{)} & \phantom{(}0.86\phantom{)} & \phantom{(}135.88\phantom{)} &  & \phantom{(}0.35\phantom{)} & \phantom{(}34.77\phantom{)} & \phantom{(}0.71\phantom{)} & \phantom{(}138.70\phantom{)} & \phantom{(}0.38\phantom{)} & \phantom{(}36.24\phantom{)} & \phantom{(}0.83\phantom{)} & \phantom{(}144.57\phantom{)}\\

 & \multirow{-2}{*}{\raggedright\arraybackslash Cauchy} &  & (0.17) & (224.31) & (0.12) & (896.29) & (0.14) & (313.98) & (0.11) & (1254.03) &  & (0.19) & (253.64) & (0.15) & (1012.35) & (0.19) & (292.36) & (0.14) & (1166.98)\\
\addlinespace[4pt]

 &  &  & \phantom{(}0.06\phantom{)} & \phantom{(}1.47\phantom{)} & \phantom{(}0.23\phantom{)} & \phantom{(}5.87\phantom{)} & \phantom{(}0.08\phantom{)} & \phantom{(}1.97\phantom{)} & \phantom{(}0.32\phantom{)} & \phantom{(}7.87\phantom{)} &  & \phantom{(}0.05\phantom{)} & \phantom{(}1.36\phantom{)} & \phantom{(}0.19\phantom{)} & \phantom{(}5.44\phantom{)} & \phantom{(}0.06\phantom{)} & \phantom{(}1.68\phantom{)} & \phantom{(}0.24\phantom{)} & \phantom{(}6.71\phantom{)}\\

 & \multirow{-2}{*}{\raggedright\arraybackslash Laplace} &  & (0.01) & (0.24) & (0.04) & (0.95) & (0.01) & (0.07) & (0.01) & (0.26) &  & (0.01) & (0.20) & (0.03) & (0.82) & (0.01) & (0.07) & (0.01) & (0.28)\\
\addlinespace[4pt]

 &  &  & \phantom{(}0.06\phantom{)} & \phantom{(}1.46\phantom{)} & \phantom{(}0.23\phantom{)} & \phantom{(}5.82\phantom{)} & \phantom{(}0.08\phantom{)} & \phantom{(}1.95\phantom{)} & \phantom{(}0.31\phantom{)} & \phantom{(}7.79\phantom{)} &  & \phantom{(}0.05\phantom{)} & \phantom{(}1.34\phantom{)} & \phantom{(}0.19\phantom{)} & \phantom{(}5.36\phantom{)} & \phantom{(}0.06\phantom{)} & \phantom{(}1.66\phantom{)} & \phantom{(}0.23\phantom{)} & \phantom{(}6.61\phantom{)}\\

 & \multirow{-2}{*}{\raggedright\arraybackslash Normal} &  & (0.01) & (0.23) & (0.04) & (0.92) & (0.01) & (0.07) & (0.01) & (0.28) &  & (0.01) & (0.20) & (0.03) & (0.80) & (0.01) & (0.07) & (0.01) & (0.28)\\
\addlinespace[4pt]

 &  &  & \phantom{(}0.06\phantom{)} & \phantom{(}1.46\phantom{)} & \phantom{(}0.23\phantom{)} & \phantom{(}5.83\phantom{)} & \phantom{(}0.08\phantom{)} & \phantom{(}1.95\phantom{)} & \phantom{(}0.31\phantom{)} & \phantom{(}7.79\phantom{)} &  & \phantom{(}0.05\phantom{)} & \phantom{(}1.35\phantom{)} & \phantom{(}0.19\phantom{)} & \phantom{(}5.37\phantom{)} & \phantom{(}0.06\phantom{)} & \phantom{(}1.66\phantom{)} & \phantom{(}0.23\phantom{)} & \phantom{(}6.61\phantom{)}\\

\multirow{-8}{*}{\raggedright\arraybackslash S4} & \multirow{-2}{*}{\raggedright\arraybackslash Student-$t$} &  & (0.01) & (0.24) & (0.04) & (0.95) & (0.01) & (0.07) & (0.01) & (0.27) &  & (0.01) & (0.20) & (0.03) & (0.81) & (0.01) & (0.07) & (0.01) & (0.28)\\
\addlinespace[4pt]
\bottomrule
\end{tabular}}%
\vspace{0.5em}
\begin{minipage}{0.95\linewidth}
\footnotesize
\textit{Notes:} Entries report Monte Carlo averages across $MC=1000$ simulation runs, with standard deviations in parentheses. Coverage refers to the proportion of confidence intervals containing the true CATE. Superscript $V$ denotes variance estimates additionally scaled by $V(x)^{-2}$.
\end{minipage}

\end{table}

\end{landscape}

\FloatBarrier  
\begin{figure}
    \centering
    \caption{Coverage and width of $95\%$ confidence intervals for CATE estimates for scenario S1 across different splitting rules and outcome noise settings.}
    \label{fig:ci_s1}
    \resizebox{1\linewidth}{!}{%
        \input{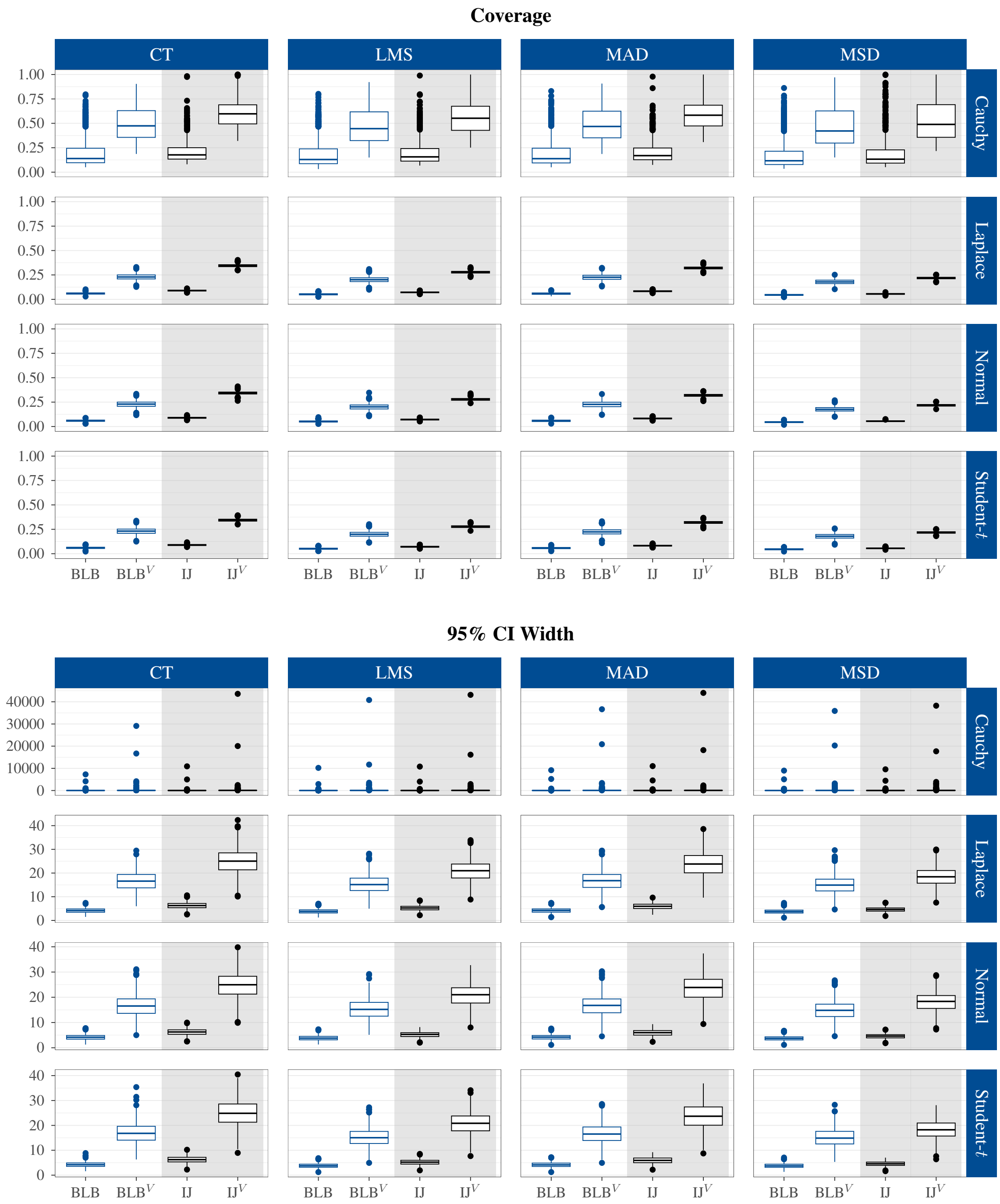}}
        \begin{notes}
            The boxplots of the confidence intervals do not share a common $y$-axis range across the different noise settings. As the Cauchy noise produces substantially wider intervals, applying a common scale would make the boxplots for the other settings difficult to read.  
        \end{notes}
\end{figure}
\FloatBarrier

\FloatBarrier  
\begin{figure}
    \centering
    \caption{Coverage and width of $95\%$ confidence intervals for CATE estimates for scenario S2 across different splitting rules and outcome noise settings.}
    \label{fig:ci_s2}
    \resizebox{1\linewidth}{!}{%
        \input{ci_S2}}
        \begin{notes}
            The boxplots of the confidence intervals do not share a common $y$-axis range across the different noise settings. As the Cauchy noise produces substantially wider intervals, applying a common scale would make the boxplots for the other settings difficult to read.  
        \end{notes}
\end{figure}
\FloatBarrier

\FloatBarrier  
\begin{figure}
    \centering
    \caption{Coverage and width of $95\%$ confidence intervals for CATE estimates for scenario S3 across different splitting rules and outcome noise settings.}
    \label{fig:ci_s3}
    \resizebox{1\linewidth}{!}{%
        \input{ci_S3}}
    \begin{notes}
            The boxplots of the confidence intervals do not share a common $y$-axis range across the different noise settings. As the Cauchy noise produces substantially wider intervals, applying a common scale would make the boxplots for the other settings difficult to read.  
        \end{notes}    
\end{figure}
\FloatBarrier

\FloatBarrier  
\begin{figure}
    \centering
    \caption{
    Coverage and width of $95\%$ confidence intervals for CATE estimates for scenario S4 across different splitting rules and outcome noise settings.}
    \label{fig:ci_s4}
    \resizebox{1\linewidth}{!}{%
        \input{ci_S4}}
        \begin{notes}
            The boxplots of the confidence intervals do not share a common $y$-axis range across the different noise settings. As the Cauchy noise produces substantially wider intervals, applying a common scale would make the boxplots for the other settings difficult to read.  
        \end{notes}
\end{figure}
\FloatBarrier

\subsection{Mean-shift \& Variance Contamination}

\FloatBarrier
\begin{table}[h]
\caption{Simulation performance and percentage changes relative to the uncontaminated baseline under both contamination levels for scenario S1.}
\label{tab:sim_raw_pct_s1}
\centering
\scriptsize
\setlength{\tabcolsep}{3pt}
\renewcommand{\arraystretch}{0.9}
\resizebox{0.98\linewidth}{!}{%
\begin{tabular}{lr rrr @{\hskip 0.45cm} rrr}
\\[-1.8ex]\hline
\hline \\[-1.8ex]
 & & \multicolumn{3}{c}{Mean-shift contamination}
 & \multicolumn{3}{c}{Variance contamination} \\
\cmidrule(l{3pt}r{3pt}){3-5}
\cmidrule(l{3pt}r{3pt}){6-8}
Method & \(p_c\)
& \({\operatorname{RMSE}}\) \((\operatorname{PRMSE})\)
& \({\operatorname{MAE}}\) \((\operatorname{PMAE})\)
& \({\operatorname{MAPE}}\) \((\operatorname{PMAPE})\)
& \({\operatorname{RMSE}}\) \((\operatorname{PRMSE})\)
& \({\operatorname{MAE}}\) \((\operatorname{PMAE})\)
& \({\operatorname{MAPE}}\) \((\operatorname{PMAPE})\) \\
\midrule

\multirow{8}{*}{CT}
& 0  & 33.165 (0.000) & 26.485 (0.000) & 117.146 (0.000)
     & 33.165 (0.000) & 26.485 (0.000) & 117.146 (0.000) \\
& 5  & 33.365 (0.603) & 26.649 (0.619) & 123.208 (5.175)
     & 33.314 (0.449) & 26.599 (0.432) & 121.811 (3.982) \\
& 10 & 33.491 (0.984) & 26.753 (1.014) & 127.700 (9.009)
     & 33.506 (1.028) & 26.757 (1.028) & 128.013 (9.277) \\
& 15 & 33.729 (1.701) & 26.944 (1.736) & 133.457 (13.924)
     & 33.672 (1.527) & 26.904 (1.583) & 132.106 (12.770) \\
& 20 & 33.819 (1.972) & 27.023 (2.032) & 136.117 (16.195)
     & 33.899 (2.214) & 27.091 (2.290) & 138.093 (17.881) \\
& 25 & 33.839 (2.031) & 27.030 (2.060) & 136.306 (16.356)
     & 34.147 (2.960) & 27.295 (3.061) & 144.105 (23.013) \\
& 30 & 33.910 (2.246) & 27.101 (2.328) & 139.096 (18.737)
     & 34.334 (3.523) & 27.450 (3.647) & 148.829 (27.046) \\
& 35 & 34.060 (2.700) & 27.217 (2.767) & 142.204 (21.391)
     & 34.444 (3.857) & 27.544 (4.000) & 152.194 (29.919) \\
\addlinespace

\multirow{8}{*}{LMS}
& 0  & 33.117 (0.000) & 26.444 (0.000) & 115.203 (0.000)
     & 33.117 (0.000) & 26.444 (0.000) & 115.203 (0.000) \\
& 5  & 33.317 (0.604) & 26.608 (0.619) & 121.979 (5.882)
     & 33.117 (0.002) & 26.434 (-0.040) & 114.571 (-0.549) \\
& 10 & 33.475 (1.082) & 26.739 (1.114) & 126.918 (10.169)
     & 33.328 (0.638) & 26.607 (0.616) & 121.947 (5.854) \\
& 15 & 33.718 (1.815) & 26.936 (1.861) & 133.221 (15.640)
     & 33.453 (1.015) & 26.719 (1.039) & 126.089 (9.449) \\
& 20 & 33.834 (2.165) & 27.035 (2.232) & 136.025 (18.074)
     & 33.683 (1.710) & 26.910 (1.760) & 132.343 (14.877) \\
& 25 & 33.841 (2.187) & 27.032 (2.221) & 136.563 (18.541)
     & 33.906 (2.383) & 27.094 (2.459) & 138.498 (20.220) \\
& 30 & 33.904 (2.375) & 27.095 (2.462) & 139.422 (21.023)
     & 34.106 (2.986) & 27.262 (3.091) & 143.733 (24.765) \\
& 35 & 34.088 (2.933) & 27.239 (3.004) & 143.209 (24.309)
     & 34.204 (3.284) & 27.346 (3.408) & 146.447 (27.121) \\
\addlinespace

\multirow{8}{*}{MAD}
& 0  & 33.157 (0.000) & 26.479 (0.000) & 117.186 (0.000)
     & 33.157 (0.000) & 26.479 (0.000) & 117.186 (0.000) \\
& 5  & 33.366 (0.631) & 26.650 (0.645) & 123.530 (5.414)
     & 33.337 (0.543) & 26.621 (0.534) & 122.366 (4.420) \\
& 10 & 33.515 (1.079) & 26.774 (1.114) & 128.187 (9.387)
     & 33.517 (1.085) & 26.766 (1.085) & 128.466 (9.626) \\
& 15 & 33.720 (1.697) & 26.937 (1.728) & 133.788 (14.167)
     & 33.709 (1.665) & 26.937 (1.728) & 132.995 (13.490) \\
& 20 & 33.853 (2.098) & 27.049 (2.154) & 136.520 (16.498)
     & 33.925 (2.316) & 27.114 (2.398) & 138.933 (18.558) \\
& 25 & 33.884 (2.193) & 27.069 (2.230) & 137.307 (17.170)
     & 34.199 (3.142) & 27.335 (3.233) & 145.306 (23.995) \\
& 30 & 33.941 (2.365) & 27.126 (2.445) & 139.795 (19.293)
     & 34.361 (3.630) & 27.474 (3.758) & 149.779 (27.813) \\
& 35 & 34.066 (2.741) & 27.221 (2.803) & 142.359 (21.481)
     & 34.469 (3.957) & 27.567 (4.108) & 153.428 (30.926) \\
\addlinespace

\multirow{8}{*}{MSD}
& 0  & 33.181 (0.000) & 26.498 (0.000) & 117.415 (0.000)
     & 33.181 (0.000) & 26.498 (0.000) & 117.415 (0.000) \\
& 5  & 33.387 (0.621) & 26.667 (0.636) & 123.725 (5.374)
     & 33.324 (0.431) & 26.610 (0.419) & 121.815 (3.747) \\
& 10 & 33.510 (0.990) & 26.766 (1.011) & 128.176 (9.164)
     & 33.500 (0.962) & 26.752 (0.958) & 128.184 (9.171) \\
& 15 & 33.755 (1.729) & 26.968 (1.770) & 134.362 (14.433)
     & 33.696 (1.552) & 26.924 (1.605) & 132.343 (12.714) \\
& 20 & 33.852 (2.022) & 27.049 (2.076) & 136.497 (16.252)
     & 33.900 (2.165) & 27.089 (2.228) & 138.117 (17.631) \\
& 25 & 33.901 (2.170) & 27.080 (2.194) & 137.785 (17.348)
     & 34.144 (2.902) & 27.292 (2.994) & 144.207 (22.818) \\
& 30 & 33.972 (2.384) & 27.153 (2.470) & 140.571 (19.721)
     & 34.370 (3.583) & 27.480 (3.706) & 149.732 (27.523) \\
& 35 & 34.073 (2.688) & 27.227 (2.748) & 142.793 (21.613)
     & 34.446 (3.812) & 27.547 (3.956) & 152.774 (30.114) \\
\bottomrule
\end{tabular}
}
\caption*{\scriptsize\textit{Notes:} Entries report RMSE, MAE, and MAPE, with percentage changes relative to the uncontaminated baseline of the corresponding splitting rule in parentheses. The row \(p_c=0\) has percentage increase zero by construction. Negative values indicate that the average metric is lower than in the clean baseline.}
\end{table}
\FloatBarrier
\FloatBarrier
\begin{table}[h]
\caption{Simulation performance and percentage changes relative to the uncontaminated baseline under both contamination levels for scenario S2.}
\label{tab:sim_raw_pct_s2}
\centering
\scriptsize
\setlength{\tabcolsep}{2.8pt}
\renewcommand{\arraystretch}{0.88}
\resizebox{0.98\linewidth}{!}{%
\begin{tabular}{lr rrr @{\hskip 0.35cm} rrr}
\\[-1.8ex]\hline
\hline \\[-1.8ex]
 & & \multicolumn{3}{c}{Mean-shift contamination}
 & \multicolumn{3}{c}{Variance contamination} \\
\cmidrule(l{3pt}r{3pt}){3-5}
\cmidrule(l{3pt}r{3pt}){6-8}
Method & \(p_c\)
& \({\operatorname{RMSE}}\) \((\operatorname{PRMSE})\)
& \({\operatorname{MAE}}\) \((\operatorname{PMAE})\)
& \({\operatorname{MAPE}}\) \((\operatorname{PMAPE})\)
& \({\operatorname{RMSE}}\) \((\operatorname{PRMSE})\)
& \({\operatorname{MAE}}\) \((\operatorname{PMAE})\)
& \({\operatorname{MAPE}}\) \((\operatorname{PMAPE})\) \\
\midrule

\multirow{8}{*}{CT}
& 0  & 0.991 (0.000) & 0.601 (0.000) & 39.557 (0.000)
     & 0.991 (0.000) & 0.601 (0.000) & 39.557 (0.000) \\
& 5  & 1.008 (1.799) & 0.597 (-0.643) & 38.737 (-2.073)
     & 1.006 (1.581) & 0.595 (-1.006) & 38.550 (-2.546) \\
& 10 & 1.020 (3.021) & 0.591 (-1.629) & 37.833 (-4.357)
     & 1.020 (2.982) & 0.591 (-1.671) & 37.814 (-4.405) \\
& 15 & 1.029 (3.851) & 0.591 (-1.668) & 37.637 (-4.854)
     & 1.030 (3.945) & 0.583 (-3.002) & 36.772 (-7.040) \\
& 20 & 1.036 (4.573) & 0.577 (-3.923) & 35.993 (-9.010)
     & 1.037 (4.670) & 0.576 (-4.122) & 35.869 (-9.323) \\
& 25 & 1.043 (5.306) & 0.576 (-4.121) & 35.708 (-9.729)
     & 1.048 (5.764) & 0.577 (-3.853) & 35.822 (-9.441) \\
& 30 & 1.050 (6.045) & 0.563 (-6.307) & 34.124 (-13.733)
     & 1.056 (6.577) & 0.571 (-4.894) & 34.969 (-11.598) \\
& 35 & 1.055 (6.499) & 0.552 (-8.005) & 32.907 (-16.811)
     & 1.062 (7.205) & 0.575 (-4.197) & 35.271 (-10.834) \\
\addlinespace

\multirow{8}{*}{LMS}
& 0  & 1.013 (0.000) & 0.617 (0.000) & 40.691 (0.000)
     & 1.013 (0.000) & 0.617 (0.000) & 40.691 (0.000) \\
& 5  & 1.021 (0.800) & 0.606 (-1.784) & 39.367 (-3.254)
     & 1.021 (0.772) & 0.605 (-1.959) & 39.244 (-3.555) \\
& 10 & 1.028 (1.428) & 0.596 (-3.374) & 38.192 (-6.140)
     & 1.029 (1.605) & 0.596 (-3.358) & 38.160 (-6.219) \\
& 15 & 1.033 (1.944) & 0.592 (-4.074) & 37.635 (-7.509)
     & 1.036 (2.263) & 0.587 (-4.900) & 37.025 (-9.009) \\
& 20 & 1.038 (2.436) & 0.578 (-6.318) & 36.025 (-11.467)
     & 1.042 (2.859) & 0.579 (-6.185) & 36.031 (-11.451) \\
& 25 & 1.044 (3.021) & 0.575 (-6.825) & 35.576 (-12.570)
     & 1.051 (3.738) & 0.579 (-6.060) & 35.943 (-11.668) \\
& 30 & 1.050 (3.608) & 0.562 (-8.839) & 34.105 (-16.184)
     & 1.058 (4.425) & 0.573 (-7.130) & 35.084 (-13.780) \\
& 35 & 1.054 (4.029) & 0.553 (-10.309) & 33.013 (-18.867)
     & 1.063 (4.908) & 0.576 (-6.550) & 35.343 (-13.143) \\
\addlinespace

\multirow{8}{*}{MAD}
& 0  & 1.003 (0.000) & 0.609 (0.000) & 40.146 (0.000)
     & 1.003 (0.000) & 0.609 (0.000) & 40.146 (0.000) \\
& 5  & 1.013 (0.996) & 0.600 (-1.526) & 38.954 (-2.967)
     & 1.012 (0.921) & 0.598 (-1.847) & 38.750 (-3.477) \\
& 10 & 1.022 (1.846) & 0.592 (-2.841) & 37.917 (-5.550)
     & 1.022 (1.885) & 0.590 (-3.184) & 37.684 (-6.132) \\
& 15 & 1.029 (2.542) & 0.588 (-3.399) & 37.415 (-6.801)
     & 1.030 (2.670) & 0.583 (-4.276) & 36.839 (-8.237) \\
& 20 & 1.035 (3.201) & 0.576 (-5.444) & 35.906 (-10.560)
     & 1.036 (3.274) & 0.576 (-5.480) & 35.899 (-10.579) \\
& 25 & 1.042 (3.874) & 0.575 (-5.629) & 35.641 (-11.221)
     & 1.046 (4.253) & 0.577 (-5.282) & 35.826 (-10.761) \\
& 30 & 1.049 (4.582) & 0.564 (-7.477) & 34.269 (-14.637)
     & 1.054 (5.077) & 0.572 (-6.136) & 35.087 (-12.602) \\
& 35 & 1.054 (5.028) & 0.554 (-9.140) & 33.053 (-17.667)
     & 1.060 (5.614) & 0.575 (-5.624) & 35.292 (-12.091) \\
\addlinespace

\multirow{8}{*}{MSD}
& 0  & 1.010 (0.000) & 0.613 (0.000) & 40.396 (0.000)
     & 1.010 (0.000) & 0.613 (0.000) & 40.396 (0.000) \\
& 5  & 1.020 (0.979) & 0.605 (-1.370) & 39.303 (-2.705)
     & 1.020 (0.985) & 0.603 (-1.669) & 39.090 (-3.231) \\
& 10 & 1.028 (1.758) & 0.595 (-2.967) & 38.095 (-5.695)
     & 1.030 (1.966) & 0.595 (-2.937) & 38.054 (-5.797) \\
& 15 & 1.033 (2.320) & 0.592 (-3.440) & 37.683 (-6.716)
     & 1.037 (2.698) & 0.589 (-3.951) & 37.270 (-7.738) \\
& 20 & 1.039 (2.911) & 0.578 (-5.804) & 35.974 (-10.945)
     & 1.043 (3.215) & 0.580 (-5.339) & 36.221 (-10.335) \\
& 25 & 1.045 (3.499) & 0.577 (-5.882) & 35.796 (-11.387)
     & 1.051 (4.074) & 0.579 (-5.498) & 35.954 (-10.996) \\
& 30 & 1.052 (4.188) & 0.565 (-7.876) & 34.324 (-15.031)
     & 1.058 (4.767) & 0.573 (-6.489) & 35.145 (-12.997) \\
& 35 & 1.056 (4.519) & 0.555 (-9.477) & 33.167 (-17.894)
     & 1.064 (5.334) & 0.580 (-5.462) & 35.693 (-11.642) \\
\bottomrule
\end{tabular}%
}
\caption*{\scriptsize\textit{Notes:} Entries report RMSE, MAE, and MAPE, with percentage changes relative to the uncontaminated baseline of the corresponding splitting rule in parentheses. The row \(p_c=0\) has percentage increase zero by construction. Negative values indicate that the average metric is lower than in the clean baseline.}
\end{table}
\FloatBarrier
\FloatBarrier
\begin{table}[h]
\caption{Simulation performance and percentage changes relative to the uncontaminated baseline under both contamination levels for scenario S3.}
\label{tab:sim_raw_pct_s3}
\centering
\scriptsize
\setlength{\tabcolsep}{2.8pt}
\renewcommand{\arraystretch}{0.88}
\resizebox{0.98\linewidth}{!}{%
\begin{tabular}{lr rrr @{\hskip 0.35cm} rrr}
\\[-1.8ex]\hline
\hline \\[-1.8ex]
 & & \multicolumn{3}{c}{Mean-shift contamination}
 & \multicolumn{3}{c}{Variance contamination} \\
\cmidrule(l{3pt}r{3pt}){3-5}
\cmidrule(l{3pt}r{3pt}){6-8}
Method & \(p_c\)
& \({\operatorname{RMSE}}\) \((\operatorname{PRMSE})\)
& \({\operatorname{MAE}}\) \((\operatorname{PMAE})\)
& \({\operatorname{MAPE}}\) \((\operatorname{PMAPE})\)
& \({\operatorname{RMSE}}\) \((\operatorname{PRMSE})\)
& \({\operatorname{MAE}}\) \((\operatorname{PMAE})\)
& \({\operatorname{MAPE}}\) \((\operatorname{PMAPE})\) \\
\midrule

\multirow{8}{*}{CT}
& 0  & 0.911 (0.000) & 0.830 (0.000) & 83.038 (0.000)
     & 0.911 (0.000) & 0.830 (0.000) & 83.038 (0.000) \\
& 5  & 1.450 (59.147) & 1.327 (59.791) & 132.687 (59.791)
     & 1.441 (58.177) & 1.318 (58.706) & 131.786 (58.706) \\
& 10 & 1.774 (94.784) & 1.626 (95.871) & 162.646 (95.871)
     & 1.855 (103.653) & 1.709 (105.801) & 170.892 (105.801) \\
& 15 & 2.080 (128.362) & 1.910 (130.035) & 191.016 (130.035)
     & 2.087 (129.132) & 1.918 (130.938) & 191.765 (130.938) \\
& 20 & 2.285 (150.840) & 2.098 (152.691) & 209.829 (152.691)
     & 2.327 (155.489) & 2.142 (157.895) & 214.150 (157.895) \\
& 25 & 2.485 (172.762) & 2.285 (175.166) & 228.491 (175.166)
     & 2.529 (177.602) & 2.330 (180.549) & 232.962 (180.549) \\
& 30 & 2.526 (177.294) & 2.314 (178.648) & 231.383 (178.648)
     & 2.634 (189.177) & 2.424 (191.912) & 242.397 (191.912) \\
& 35 & 2.685 (194.759) & 2.463 (196.607) & 246.295 (196.607)
     & 2.731 (199.755) & 2.512 (202.527) & 251.211 (202.527) \\
\addlinespace

\multirow{8}{*}{LMS}
& 0  & 0.899 (0.000) & 0.817 (0.000) & 81.651 (0.000)
     & 0.899 (0.000) & 0.817 (0.000) & 81.651 (0.000) \\
& 5  & 1.460 (62.402) & 1.338 (63.837) & 133.774 (63.837)
     & 1.433 (59.428) & 1.312 (60.648) & 131.170 (60.648) \\
& 10 & 1.764 (96.207) & 1.621 (98.473) & 162.055 (98.473)
     & 1.848 (105.605) & 1.705 (108.865) & 170.540 (108.865) \\
& 15 & 2.085 (131.901) & 1.919 (135.085) & 191.949 (135.085)
     & 2.094 (132.994) & 1.930 (136.351) & 192.983 (136.351) \\
& 20 & 2.285 (154.173) & 2.099 (157.089) & 209.915 (157.089)
     & 2.319 (157.913) & 2.133 (161.182) & 213.258 (161.182) \\
& 25 & 2.517 (179.995) & 2.311 (182.976) & 231.052 (182.976)
     & 2.565 (185.329) & 2.359 (188.855) & 235.852 (188.855) \\
& 30 & 2.558 (184.605) & 2.329 (185.294) & 232.945 (185.294)
     & 2.643 (194.011) & 2.417 (196.029) & 241.710 (196.029) \\
& 35 & 2.729 (203.597) & 2.486 (204.489) & 248.618 (204.489)
     & 2.769 (208.040) & 2.530 (209.819) & 252.969 (209.819) \\
\addlinespace

\multirow{8}{*}{MAD}
& 0  & 0.914 (0.000) & 0.830 (0.000) & 83.043 (0.000)
     & 0.914 (0.000) & 0.830 (0.000) & 83.043 (0.000) \\
& 5  & 1.468 (60.657) & 1.337 (61.036) & 133.730 (61.036)
     & 1.462 (59.977) & 1.330 (60.213) & 133.046 (60.213) \\
& 10 & 1.812 (98.268) & 1.652 (98.951) & 165.215 (98.951)
     & 1.885 (106.251) & 1.726 (107.824) & 172.584 (107.824) \\
& 15 & 2.106 (130.367) & 1.923 (131.571) & 192.304 (131.571)
     & 2.129 (132.980) & 1.946 (134.314) & 194.582 (134.314) \\
& 20 & 2.318 (153.597) & 2.116 (154.866) & 211.649 (154.866)
     & 2.360 (158.159) & 2.162 (160.393) & 216.238 (160.393) \\
& 25 & 2.527 (176.442) & 2.312 (178.374) & 231.171 (178.374)
     & 2.574 (181.640) & 2.358 (183.968) & 235.816 (183.968) \\
& 30 & 2.539 (177.750) & 2.310 (178.159) & 230.992 (178.159)
     & 2.647 (189.599) & 2.421 (191.545) & 242.108 (191.545) \\
& 35 & 2.726 (198.222) & 2.491 (199.943) & 249.083 (199.943)
     & 2.745 (200.272) & 2.513 (202.671) & 251.348 (202.671) \\
\addlinespace

\multirow{8}{*}{MSD}
& 0  & 0.942 (0.000) & 0.850 (0.000) & 85.020 (0.000)
     & 0.942 (0.000) & 0.850 (0.000) & 85.020 (0.000) \\
& 5  & 1.483 (57.431) & 1.349 (58.690) & 134.918 (58.690)
     & 1.471 (56.156) & 1.337 (57.308) & 133.742 (57.308) \\
& 10 & 1.804 (91.590) & 1.646 (93.658) & 164.647 (93.658)
     & 1.882 (99.854) & 1.726 (103.030) & 172.616 (103.030) \\
& 15 & 2.103 (123.270) & 1.922 (126.016) & 192.158 (126.016)
     & 2.132 (126.423) & 1.951 (129.453) & 195.080 (129.453) \\
& 20 & 2.325 (146.863) & 2.124 (149.799) & 212.378 (149.799)
     & 2.358 (150.435) & 2.158 (153.816) & 215.793 (153.816) \\
& 25 & 2.543 (169.991) & 2.323 (173.282) & 232.343 (173.282)
     & 2.574 (173.371) & 2.356 (177.109) & 235.597 (177.109) \\
& 30 & 2.574 (173.321) & 2.334 (174.497) & 233.376 (174.497)
     & 2.669 (183.372) & 2.435 (186.363) & 243.464 (186.363) \\
& 35 & 2.743 (191.307) & 2.495 (193.485) & 249.520 (193.485)
     & 2.813 (198.727) & 2.567 (201.972) & 256.735 (201.972) \\
\bottomrule
\end{tabular}%
}
\caption*{\scriptsize\textit{Notes:} Entries report RMSE, MAE, and MAPE, with percentage changes relative to the uncontaminated baseline of the corresponding splitting rule in parentheses. The row \(p_c=0\) has percentage increase zero by construction. Negative values indicate that the average metric is lower than in the clean baseline. In S3, \(\tau(x)=0\), so the stabilized MAPE reduces to \(100\cdot\operatorname{MAE}\). Consequently, PMAPE and PMAE are identical up to rounding.}
\end{table}
\FloatBarrier
\FloatBarrier
\begin{table}[h]
\caption{Simulation performance and percentage changes relative to the uncontaminated baseline under both contamination levels for scenario S4.}
\label{tab:sim_raw_pct_s4}
\centering
\scriptsize
\setlength{\tabcolsep}{2.8pt}
\renewcommand{\arraystretch}{0.88}
\resizebox{0.98\linewidth}{!}{%
\begin{tabular}{lr rrr @{\hskip 0.35cm} rrr}
\\[-1.8ex]\hline
\hline \\[-1.8ex]
 & & \multicolumn{3}{c}{Mean-shift contamination}
 & \multicolumn{3}{c}{Variance contamination} \\
\cmidrule(l{3pt}r{3pt}){3-5}
\cmidrule(l{3pt}r{3pt}){6-8}
Method & \(p_c\)
& \({\operatorname{RMSE}}\) \((\operatorname{PRMSE})\)
& \({\operatorname{MAE}}\) \((\operatorname{PMAE})\)
& \({\operatorname{MAPE}}\) \((\operatorname{PMAPE})\)
& \({\operatorname{RMSE}}\) \((\operatorname{PRMSE})\)
& \({\operatorname{MAE}}\) \((\operatorname{PMAE})\)
& \({\operatorname{MAPE}}\) \((\operatorname{PMAPE})\) \\
\midrule

\multirow{8}{*}{CT}
& 0  & 3.599 (0.000) & 2.907 (0.000) & 30.530 (0.000)
     & 3.599 (0.000) & 2.907 (0.000) & 30.530 (0.000) \\
& 5  & 3.838 (6.631) & 3.106 (6.862) & 30.601 (0.233)
     & 3.826 (6.289) & 3.096 (6.502) & 30.528 (-0.008) \\
& 10 & 4.104 (14.028) & 3.327 (14.444) & 30.737 (0.679)
     & 4.095 (13.762) & 3.317 (14.130) & 30.615 (0.277) \\
& 15 & 4.416 (22.682) & 3.588 (23.428) & 31.222 (2.266)
     & 4.427 (22.999) & 3.599 (23.808) & 31.225 (2.274) \\
& 20 & 4.780 (32.820) & 3.905 (34.358) & 32.185 (5.420)
     & 4.809 (33.608) & 3.933 (35.318) & 32.421 (6.193) \\
& 25 & 5.191 (44.227) & 4.273 (46.992) & 33.705 (10.398)
     & 5.225 (45.177) & 4.308 (48.220) & 33.962 (11.241) \\
& 30 & 5.617 (56.064) & 4.667 (60.575) & 35.608 (16.630)
     & 5.694 (58.196) & 4.745 (63.255) & 36.170 (18.472) \\
& 35 & 6.150 (70.870) & 5.179 (78.166) & 38.414 (25.824)
     & 6.194 (72.086) & 5.231 (79.961) & 38.978 (27.669) \\
\addlinespace

\multirow{8}{*}{LMS}
& 0  & 3.580 (0.000) & 2.891 (0.000) & 30.379 (0.000)
     & 3.580 (0.000) & 2.891 (0.000) & 30.379 (0.000) \\
& 5  & 3.871 (8.123) & 3.134 (8.407) & 30.888 (1.677)
     & 3.826 (6.857) & 3.095 (7.072) & 30.527 (0.489) \\
& 10 & 4.127 (15.270) & 3.345 (15.706) & 30.929 (1.813)
     & 4.102 (14.571) & 3.323 (14.951) & 30.701 (1.063) \\
& 15 & 4.428 (23.688) & 3.597 (24.431) & 31.268 (2.929)
     & 4.442 (24.063) & 3.610 (24.866) & 31.321 (3.102) \\
& 20 & 4.778 (33.446) & 3.902 (34.978) & 32.129 (5.762)
     & 4.830 (34.903) & 3.949 (36.595) & 32.572 (7.219) \\
& 25 & 5.174 (44.517) & 4.259 (47.309) & 33.594 (10.585)
     & 5.243 (46.447) & 4.321 (49.481) & 34.093 (12.228) \\
& 30 & 5.603 (56.496) & 4.656 (61.049) & 35.506 (16.878)
     & 5.702 (59.274) & 4.749 (64.273) & 36.212 (19.201) \\
& 35 & 6.146 (71.673) & 5.178 (79.124) & 38.383 (26.350)
     & 6.207 (73.375) & 5.239 (81.227) & 39.049 (28.542) \\
\addlinespace

\multirow{8}{*}{MAD}
& 0  & 4.029 (0.000) & 3.269 (0.000) & 34.177 (0.000)
     & 4.029 (0.000) & 3.269 (0.000) & 34.177 (0.000) \\
& 5  & 4.045 (0.392) & 3.279 (0.296) & 32.325 (-5.418)
     & 4.058 (0.723) & 3.289 (0.626) & 32.459 (-5.028) \\
& 10 & 4.199 (4.204) & 3.402 (4.073) & 31.509 (-7.807)
     & 4.220 (4.724) & 3.418 (4.571) & 31.618 (-7.487) \\
& 15 & 4.463 (10.769) & 3.623 (10.825) & 31.547 (-7.697)
     & 4.492 (11.495) & 3.648 (11.607) & 31.678 (-7.313) \\
& 20 & 4.802 (19.184) & 3.920 (19.915) & 32.315 (-5.450)
     & 4.834 (19.986) & 3.950 (20.824) & 32.568 (-4.707) \\
& 25 & 5.190 (28.819) & 4.269 (30.583) & 33.717 (-1.348)
     & 5.243 (30.123) & 4.320 (32.150) & 34.046 (-0.384) \\
& 30 & 5.615 (39.363) & 4.663 (42.647) & 35.589 (4.131)
     & 5.695 (41.344) & 4.744 (45.129) & 36.134 (5.725) \\
& 35 & 6.143 (52.466) & 5.170 (58.169) & 38.376 (12.287)
     & 6.182 (53.443) & 5.219 (59.660) & 38.889 (13.785) \\
\addlinespace

\multirow{8}{*}{MSD}
& 0  & 3.738 (0.000) & 3.022 (0.000) & 31.832 (0.000)
     & 3.738 (0.000) & 3.022 (0.000) & 31.832 (0.000) \\
& 5  & 4.024 (7.657) & 3.263 (7.962) & 32.297 (1.460)
     & 4.009 (7.233) & 3.250 (7.529) & 32.208 (1.179) \\
& 10 & 4.280 (14.480) & 3.473 (14.906) & 32.281 (1.410)
     & 4.277 (14.423) & 3.471 (14.852) & 32.253 (1.320) \\
& 15 & 4.550 (21.713) & 3.700 (22.405) & 32.420 (1.847)
     & 4.582 (22.575) & 3.727 (23.317) & 32.610 (2.442) \\
& 20 & 4.869 (30.258) & 3.978 (31.635) & 33.011 (3.702)
     & 4.936 (32.045) & 4.036 (33.548) & 33.537 (5.354) \\
& 25 & 5.254 (40.538) & 4.324 (43.075) & 34.268 (7.650)
     & 5.323 (42.385) & 4.384 (45.065) & 34.787 (9.282) \\
& 30 & 5.648 (51.088) & 4.691 (55.208) & 35.926 (12.861)
     & 5.758 (54.021) & 4.792 (58.554) & 36.666 (15.183) \\
& 35 & 6.164 (64.895) & 5.191 (71.746) & 38.585 (21.211)
     & 6.238 (66.857) & 5.261 (74.073) & 39.291 (23.430) \\
\bottomrule
\end{tabular}%
}
\caption*{\scriptsize\textit{Notes:} Entries report RMSE, MAE, and MAPE, with percentage changes relative to the uncontaminated baseline of the corresponding splitting rule in parentheses. The row \(p_c=0\) has percentage increase zero by construction. Negative values indicate that the average metric is lower than in the clean baseline.}
\end{table}
\FloatBarrier



\section{Additional Empirical Figures and Results} \label{App:Empirical Results}
\begin{table}[!htbp]
\centering
\caption{Summary statistics for the Lalonde application.}
\label{tab:summary_Lalonde}
\begin{tabular}{lccccc}
\\[-1.8ex]\hline
\hline \\[-1.8ex]
Variable & N & Mean & St. Dev. & Min & Max \\
\hline \\[-1.8ex]
Treatment indicator & 722 & 0.41 & 0.49 & 0 & 1 \\
Age & 722 & 24.52 & 6.63 & 17 & 55 \\
Years of education & 722 & 10.27 & 1.70 & 3 & 16 \\
Black indicator & 722 & 0.80 & 0.40 & 0 & 1 \\
Hispanic indicator & 722 & 0.11 & 0.31 & 0 & 1 \\
Married indicator & 722 & 0.16 & 0.37 & 0 & 1 \\
No high-school degree & 722 & 0.78 & 0.41 & 0 & 1 \\
Real earnings in 1975 & 722 & 3,042.90 & 5,066.14 & 0 & 37,431.66 \\
Real earnings in 1978 & 722 & 5,454.64 & 6,252.94 & 0 & 60,307.93 \\
\hline \\[-1.8ex]
\end{tabular}
\begin{notes}
Entries report summary statistics for the Lalonde application. The outcome variable is real earnings in 1978. The treatment indicator equals one for individuals assigned to the NSW job training program and zero for controls. Real earnings are measured in US dollars.
\end{notes}
\end{table}

\begin{table}[!htbp]
\centering
\caption{Summary statistics for the Breast Cancer application.}
\label{tab:summary_breast_cancer}
\begin{tabular}{lccccc}
\\[-1.8ex]\hline
\hline \\[-1.8ex]
Statistic & N & Mean & St. Dev. & Min & Max \\
\hline \\[-1.8ex]
age     & 299 & 53.003 & 10.862 & 21 & 80 \\
size    & 299 & 31.458 & 15.752 & 5 & 120 \\
grade   & 299 & 2.204 & 0.533 & 1 & 3 \\
nodes   & 299 & 6.518 & 6.136 & 1 & 36 \\
enodes  & 299 & 0.551 & 0.261 & 0.013 & 0.887 \\
pgr     & 299 & 70.525 & 122.206 & 0 & 912 \\
er      & 299 & 85.538 & 152.682 & 0 & 1,091 \\
rectime & 299 & 797.766 & 500.660 & 72 & 2,456 \\
\hline \\[-1.8ex]
\end{tabular}
\begin{notes}
Summary statistics are reported for the analysis sample after removing censored observations. The sample contains 299 patients with observed death or recurrence times.
\end{notes}
\end{table}

\FloatBarrier
\begin{table}[!htbp]
\centering
\caption{Summary statistics for the ACIC application.}
\label{tab:summary_cont_ACIC}
\begin{tabular}{@{\extracolsep{5pt}}lccccc}
\\[-1.8ex]\hline
\hline \\[-1.8ex]
Variable & N & Mean & St. Dev. & Min & Max \\
\hline \\[-1.8ex]
Student achievement outcome & 10,391 & -0.10 & 0.64 & -2.10 & 2.19 \\
Student success expectations & 10,391 & 5.27 & 1.12 & 1.00 & 7.00 \\
School-level mean fixed mindset & 10,391 & -0.04 & 0.97 & -3.09 & 2.83 \\
School achievement level & 10,391 & 0.05 & 0.94 & -3.35 & 2.17 \\
School racial/ethnic minority composition & 10,391 & -0.09 & 0.96 & -1.58 & 2.36 \\
School poverty concentration & 10,391 & -0.05 & 0.97 & -1.92 & 2.82 \\
School size & 10,391 & -0.03 & 1.01 & -1.81 & 1.89 \\
\hline \\[-1.8ex]
\end{tabular}
\begin{notes}
Entries report summary statistics for the continuous variables used in the ACIC application. Several school-level variables are standardized, so their means are close to zero and their standard deviations are close to one.
\end{notes}
\end{table}
\FloatBarrier

\FloatBarrier
\begin{table}[!htbp]
\centering
\caption{Counts for categorical variables in the ACIC application.}
\label{tab:summary_cat_ACIC}
\small
\setlength{\tabcolsep}{7pt}
\begin{tabular}{@{\extracolsep{5pt}}lc}
Variable/category & Count \\
\hline \\[-1.8ex]
Student's identified gender: 1 & 5,300 \\
Student's identified gender: 2 & 5,091 \\
\hline
Student's first-generation status: 0 & 3,835 \\
Student's first-generation status: 1 & 6,556 \\
\hline
Student race/ethnicity: 1 & 983 \\
Student race/ethnicity: 2 & 1,578 \\
Student race/ethnicity: 3 & 113 \\
Student race/ethnicity: 4 & 5,030 \\
Student race/ethnicity: 5 & 420 \\
Student race/ethnicity: 6 & 40 \\
Student race/ethnicity: 7 & 43 \\
Student race/ethnicity: 8 & 195 \\
Student race/ethnicity: 9 & 135 \\
Student race/ethnicity: 10 & 176 \\
Student race/ethnicity: 11 & 170 \\
Student race/ethnicity: 12 & 325 \\
Student race/ethnicity: 13 & 182 \\
Student race/ethnicity: 14 & 652 \\
Student race/ethnicity: 15 & 349 \\
\hline
Urbanicity of the school: 0 & 845 \\
Urbanicity of the school: 1 & 2,425 \\
Urbanicity of the school: 2 & 1,979 \\
Urbanicity of the school: 3 & 1,516 \\
Urbanicity of the school: 4 & 3,626 \\
\hline \\[-1.8ex]
\end{tabular}
\begin{notes}
Entries report counts for the categorical variables used in the ACIC application. Category labels follow the coding of the original dataset. 
\end{notes}
\end{table}
\FloatBarrier

\FloatBarrier
\begin{table}[h]
\caption{$\operatorname{RMSE}_{\Delta}$ and $\operatorname{MAE}_{\Delta}$ results under mean-shift contamination across the three empirical applications.}
\label{tab:mean_contamination_empirical}
\centering
\setlength{\tabcolsep}{7pt}
\resizebox{.8\linewidth}{!}{
\begin{tabular}{ll rr @{\hskip 1.2cm} rr @{\hskip 1.2cm} rr}
\\[-1.8ex]\hline
\hline \\[-1.8ex]
 &  & \multicolumn{6}{c}{Dataset} \\
\cmidrule(l{3pt}r{3pt}){3-8}
 &  & \multicolumn{2}{c}{Lalonde}
 & \multicolumn{2}{c}{Breast Cancer}
 & \multicolumn{2}{c}{ACIC} \\
\cmidrule(l{3pt}r{3pt}){3-4}
\cmidrule(l{3pt}r{3pt}){5-6}
\cmidrule(l{3pt}r{3pt}){7-8}
Method & $p_c$
& $\operatorname{RMSE}_{\Delta}$ & $\operatorname{MAE}_{\Delta}$
& $\operatorname{RMSE}_{\Delta}$ & $\operatorname{MAE}_{\Delta}$
& $\operatorname{RMSE}_{\Delta}$ & $\operatorname{MAE}_{\Delta}$ \\
\midrule
\multirow{7}{*}{CT}
 & 0.05 & 586.01 & 560.41 & 76.06 & 65.84 & 0.03 & 0.03 \\
 & 0.10 & 754.95 & 712.51 & 116.55 & 103.76 & 0.05 & 0.04 \\
 & 0.15 & 819.07 & 772.91 & 122.95 & 107.63 & 0.06 & 0.05 \\
 & 0.20 & 955.82 & 908.64 & 134.16 & 118.04 & 0.07 & 0.07 \\
 & 0.25 & 1060.98 & 1005.87 & 163.28 & 147.05 & 0.09 & 0.08 \\
 & 0.30 & 1167.04 & 1111.58 & 159.74 & 144.06 & 0.10 & 0.09 \\
 & 0.35 & 1229.45 & 1176.90 & 165.40 & 148.63 & 0.12 & 0.11 \\
\addlinespace
\multirow{7}{*}{LMS}
 & 0.05 & 656.44 & 597.01 & 87.72 & 73.62 & 0.04 & 0.03 \\
 & 0.10 & 856.28 & 773.48 & 130.95 & 113.97 & 0.05 & 0.04 \\
 & 0.15 & 917.11 & 818.75 & 134.37 & 115.90 & 0.07 & 0.06 \\
 & 0.20 & 1068.89 & 960.73 & 148.50 & 129.01 & 0.08 & 0.07 \\
 & 0.25 & 1237.64 & 1104.38 & 175.94 & 154.95 & 0.09 & 0.08 \\
 & 0.30 & 1348.29 & 1208.49 & 178.77 & 157.29 & 0.11 & 0.10 \\
 & 0.35 & 1417.98 & 1276.10 & 181.41 & 159.23 & 0.14 & 0.12 \\
\addlinespace
\multirow{7}{*}{MAD}
 & 0.05 & 676.69 & 610.38 & 77.65 & 66.96 & 0.05 & 0.04 \\
 & 0.10 & 891.08 & 795.60 & 121.34 & 107.17 & 0.07 & 0.06 \\
 & 0.15 & 965.32 & 851.77 & 128.44 & 112.05 & 0.10 & 0.08 \\
 & 0.20 & 1110.55 & 988.82 & 141.81 & 123.46 & 0.11 & 0.09 \\
 & 0.25 & 1238.19 & 1101.12 & 171.73 & 153.47 & 0.12 & 0.10 \\
 & 0.30 & 1366.23 & 1222.57 & 171.22 & 152.26 & 0.14 & 0.11 \\
 & 0.35 & 1420.96 & 1278.19 & 173.50 & 153.96 & 0.15 & 0.13 \\
\addlinespace
\multirow{7}{*}{MSD}
 & 0.05 & 606.45 & 565.63 & 69.17 & 60.83 & 0.05 & 0.04 \\
 & 0.10 & 778.48 & 723.77 & 115.42 & 104.25 & 0.06 & 0.05 \\
 & 0.15 & 841.00 & 774.22 & 117.47 & 105.48 & 0.08 & 0.07 \\
 & 0.20 & 999.91 & 924.38 & 131.61 & 117.35 & 0.10 & 0.08 \\
 & 0.25 & 1131.36 & 1031.06 & 160.34 & 145.20 & 0.12 & 0.10 \\
 & 0.30 & 1268.87 & 1163.36 & 162.00 & 146.72 & 0.14 & 0.11 \\
 & 0.35 & 1356.89 & 1240.45 & 167.69 & 150.28 & 0.16 & 0.13 \\
\bottomrule
\end{tabular}}
\begin{notes} Entries report robustness metrics averaged over 100 train/test splits. $\operatorname{RMSE}_{\Delta}$ and $\operatorname{MAE}_{\Delta}$ measure deviations between CATE estimates obtained from contaminated training data and the corresponding uncontaminated data estimates. Larger values indicate stronger sensitivity to contamination.
\end{notes} 
\end{table}
\FloatBarrier

\FloatBarrier
\begin{table}[h]
\caption{$\operatorname{RMSE}_{\Delta}$ and $\operatorname{MAE}_{\Delta}$ results under variance contamination across the three empirical applications.}
\label{tab:variance_contamination_empirical}
\centering
\setlength{\tabcolsep}{7pt}
\resizebox{.8\linewidth}{!}{
\begin{tabular}{ll rr @{\hskip 1.2cm} rr @{\hskip 1.2cm} rr}
\\[-1.8ex]\hline
\hline \\[-1.8ex]
 &  & \multicolumn{6}{c}{Dataset} \\
\cmidrule(l{3pt}r{3pt}){3-8}
 &  & \multicolumn{2}{c}{Lalonde}
 & \multicolumn{2}{c}{Breast Cancer}
 & \multicolumn{2}{c}{ACIC} \\
\cmidrule(l{3pt}r{3pt}){3-4}
\cmidrule(l{3pt}r{3pt}){5-6}
\cmidrule(l{3pt}r{3pt}){7-8}
Method & $p_c$
& $\operatorname{RMSE}_{\Delta}$ & $\operatorname{MAE}_{\Delta}$
& $\operatorname{RMSE}_{\Delta}$ & $\operatorname{MAE}_{\Delta}$
& $\operatorname{RMSE}_{\Delta}$ & $\operatorname{MAE}_{\Delta}$ \\
\midrule
\multirow{7}{*}{CT}
 & 0.05 & 546.84 & 522.88 & 86.62 & 76.15 & 0.03 & 0.03 \\
 & 0.10 & 844.20 & 811.90 & 100.79 & 89.29 & 0.05 & 0.04 \\
 & 0.15 & 908.45 & 864.50 & 131.18 & 115.70 & 0.07 & 0.06 \\
 & 0.20 & 1265.36 & 1217.21 & 156.68 & 140.39 & 0.09 & 0.08 \\
 & 0.25 & 1224.91 & 1162.94 & 173.97 & 157.71 & 0.09 & 0.08 \\
 & 0.30 & 1396.23 & 1346.94 & 183.42 & 164.11 & 0.10 & 0.09 \\
 & 0.35 & 1546.91 & 1492.48 & 184.13 & 165.43 & 0.12 & 0.11 \\
\addlinespace
\multirow{7}{*}{LMS}
 & 0.05 & 609.22 & 553.91 & 98.43 & 84.01 & 0.04 & 0.03 \\
 & 0.10 & 909.69 & 842.44 & 111.79 & 95.95 & 0.06 & 0.05 \\
 & 0.15 & 954.00 & 877.89 & 140.18 & 121.82 & 0.07 & 0.06 \\
 & 0.20 & 1311.37 & 1232.35 & 167.19 & 148.07 & 0.09 & 0.08 \\
 & 0.25 & 1265.92 & 1175.26 & 180.75 & 160.15 & 0.10 & 0.08 \\
 & 0.30 & 1461.34 & 1385.39 & 186.90 & 165.18 & 0.10 & 0.09 \\
 & 0.35 & 1562.91 & 1488.05 & 195.66 & 171.65 & 0.12 & 0.11 \\
\addlinespace
\multirow{7}{*}{MAD}
 & 0.05 & 626.33 & 563.44 & 90.57 & 78.77 & 0.05 & 0.04 \\
 & 0.10 & 970.27 & 875.54 & 103.32 & 91.01 & 0.08 & 0.06 \\
 & 0.15 & 1061.32 & 945.29 & 140.16 & 122.93 & 0.10 & 0.08 \\
 & 0.20 & 1454.15 & 1316.80 & 166.38 & 146.93 & 0.12 & 0.10 \\
 & 0.25 & 1441.32 & 1277.24 & 181.81 & 161.36 & 0.14 & 0.11 \\
 & 0.30 & 1604.13 & 1448.20 & 194.57 & 171.21 & 0.15 & 0.12 \\
 & 0.35 & 1763.04 & 1605.72 & 201.37 & 176.74 & 0.17 & 0.14 \\
\addlinespace
\multirow{7}{*}{MSD}
 & 0.05 & 560.51 & 521.16 & 83.42 & 75.10 & 0.04 & 0.03 \\
 & 0.10 & 864.27 & 816.83 & 95.76 & 86.10 & 0.06 & 0.05 \\
 & 0.15 & 929.71 & 870.42 & 122.91 & 110.29 & 0.08 & 0.07 \\
 & 0.20 & 1285.70 & 1219.06 & 145.63 & 133.12 & 0.10 & 0.08 \\
 & 0.25 & 1244.48 & 1155.49 & 163.70 & 150.51 & 0.11 & 0.09 \\
 & 0.30 & 1468.27 & 1383.96 & 174.07 & 157.76 & 0.12 & 0.10 \\
 & 0.35 & 1613.47 & 1515.46 & 183.04 & 165.46 & 0.14 & 0.12 \\
\bottomrule
\end{tabular}}
\begin{notes} Entries report robustness metrics averaged over 100 train/test splits. $\operatorname{RMSE}_{\Delta}$ and $\operatorname{MAE}_{\Delta}$ measure deviations between CATE estimates obtained from contaminated training data and the corresponding uncontaminated data estimates. Larger values indicate stronger sensitivity to contamination.
\end{notes} 
\end{table}
\FloatBarrier
\subsection*{Relative sensitivity summaries}
Following the relative-performance summary of \citet{roy_robustness_2012}, we report percentage increases relative to the best-performing method within each dataset, contamination type, contamination level, and metric. Let $M_{\Delta,m}(d,c,p_c)$ denote one of the sensitivity metrics for method $m$, dataset $d$, contamination type $c$, and contamination level $p_c$. We define
\[
\%M_{\Delta,m}(d,c,p_c)
=
100\cdot
\frac{
M_{\Delta,m}(d,c,p_c)
-
\min_{m'} M_{\Delta,m'}(d,c,p_c)
}{
\min_{m'} M_{\Delta,m'}(d,c,p_c)
}.
\]
A value of zero indicates the most stable method in that setting, while larger values indicate stronger sensitivity relative to the best-performing method.

\begin{figure}[ht]
\caption{Percentage increase in sensitivity metrics relative to the best-performing method in the Lalonde application.}
\centering
\includegraphics[width=.8\linewidth]{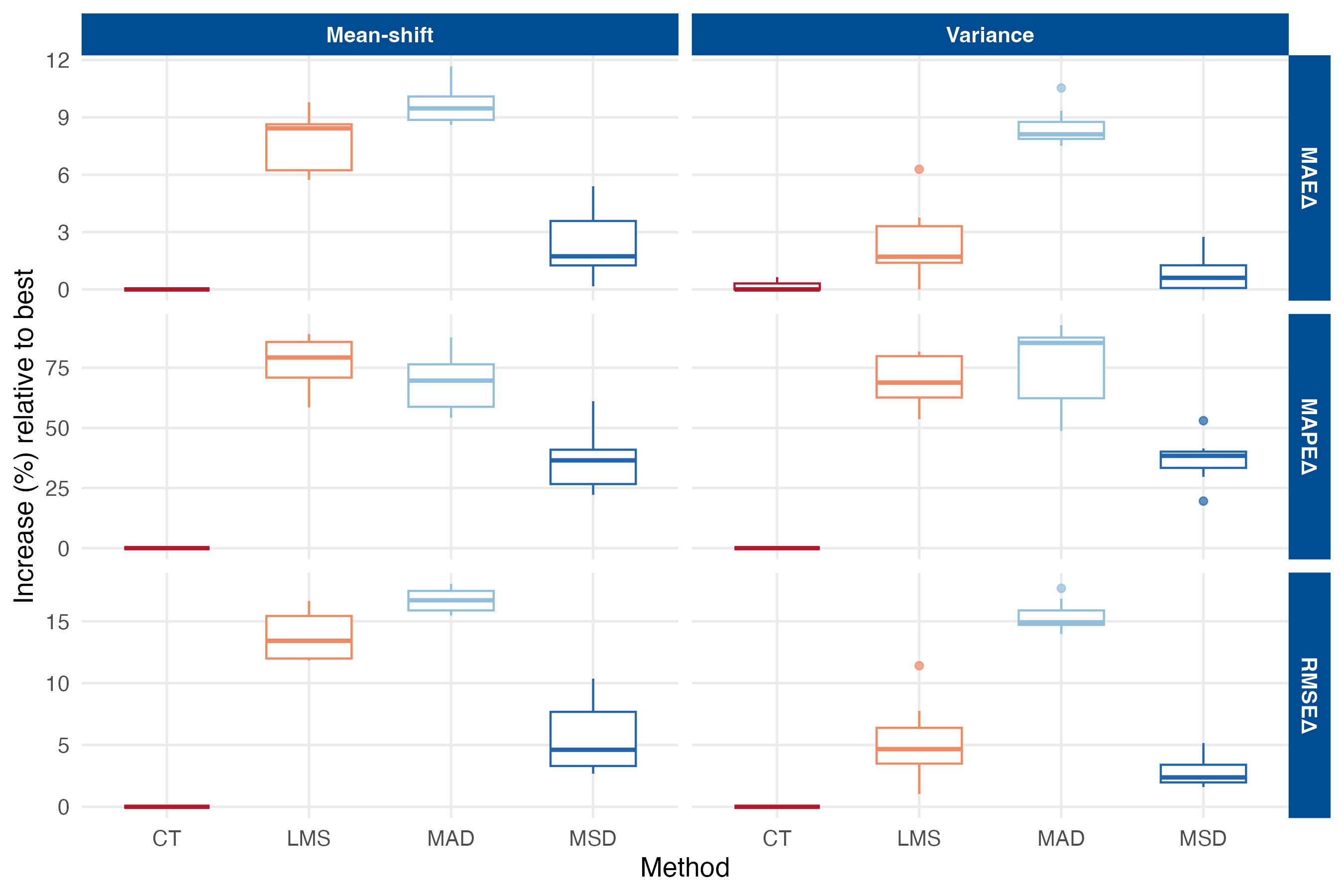}
\begin{notes} For each contamination type, contamination level, and sensitivity metric, the best-performing method is assigned a value of zero. Larger values indicate greater sensitivity relative to the best method in the same setting.
\end{notes}
\label{fig:relative_est_lalonde}

\end{figure}

\begin{figure}[ht]
\caption{Percentage increase in sensitivity metrics relative to the best-performing method in the Breast Cancer application.}
\centering
\includegraphics[width=.8\linewidth]{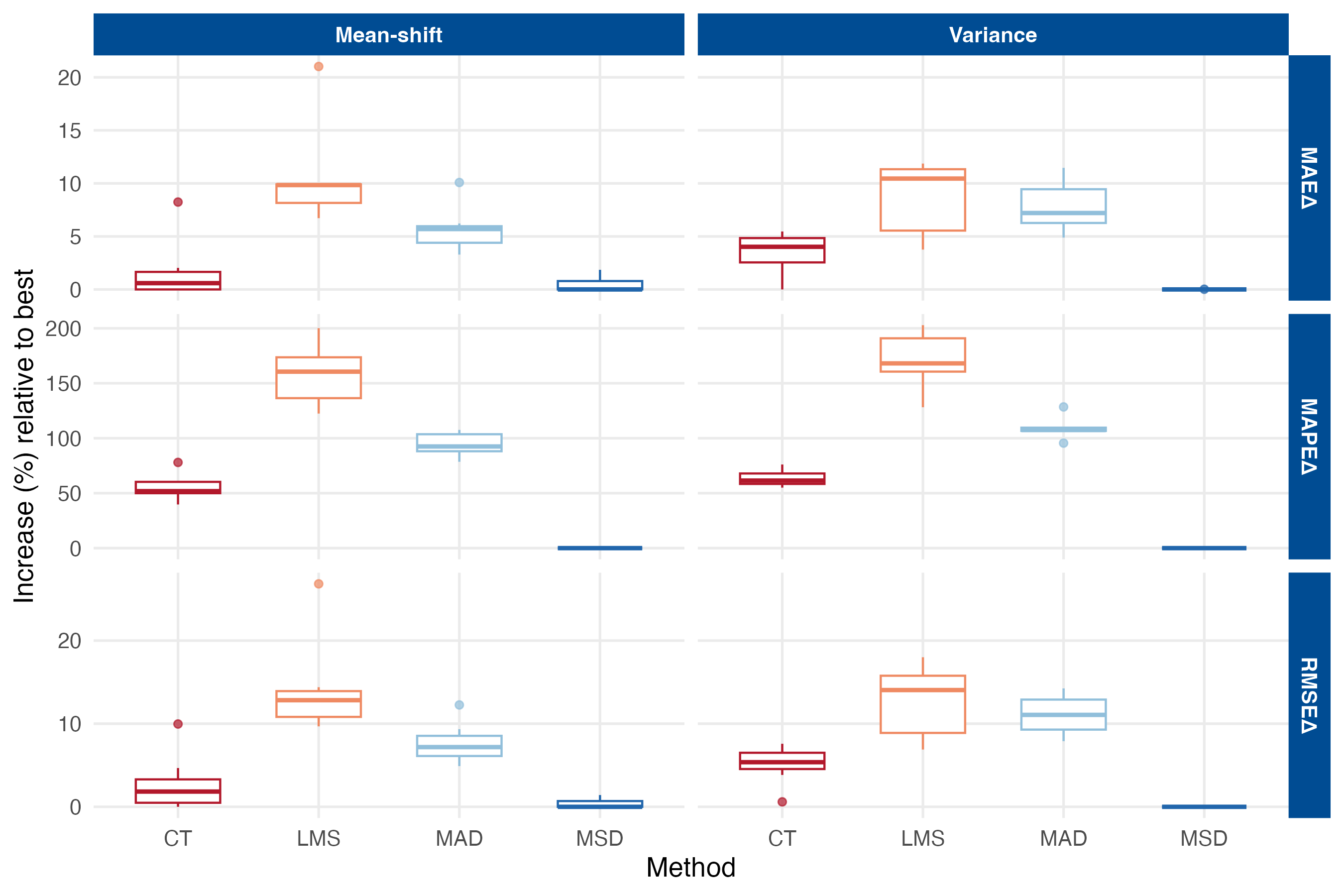}
\begin{notes} For each contamination type, contamination level, and sensitivity metric, the best-performing method is assigned a value of zero. Larger values indicate greater sensitivity relative to the best method in the same setting.
\end{notes}
\label{fig:relative_best_breast}

\end{figure}

\begin{figure}[ht]
\caption{Percentage increase in sensitivity metrics relative to the best-performing method in the ACIC application.}
\centering
\includegraphics[width=.8\linewidth]{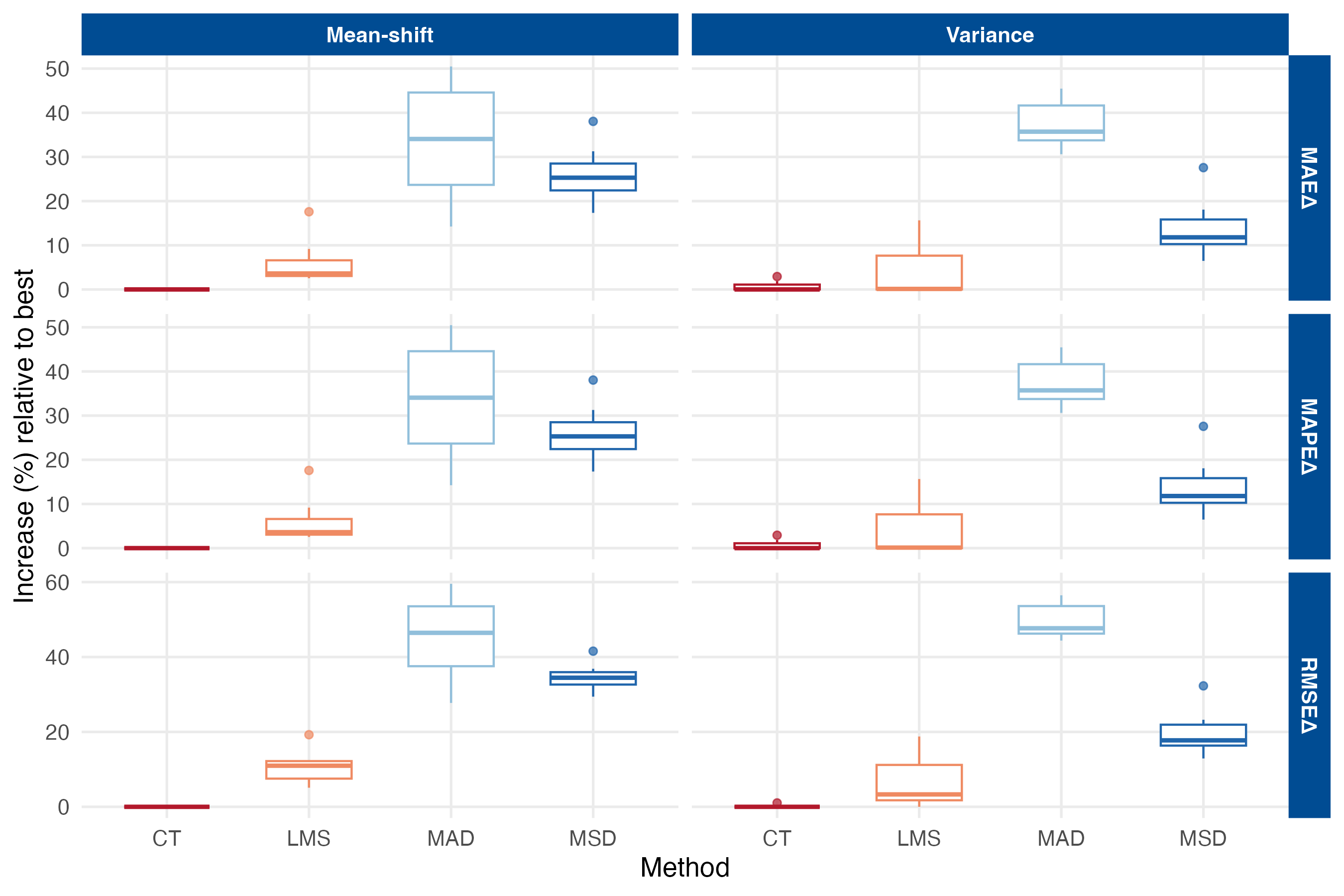}
\begin{notes} For each contamination type, contamination level, and sensitivity metric, the best-performing method is assigned a value of zero. Larger values indicate greater sensitivity relative to the best method in the same setting.
\end{notes} 
\label{fig:relative_best_acic}

\end{figure}

\FloatBarrier

\subsection{Heavy-tailed Outcome Noise}

\begin{table}[!h]
\centering
\caption{\label{tab:applications_ate} ATE estimates across splitting rules and outcome noise distributions for the three empirical applications.}
\centering
\resizebox{\ifdim\width>\linewidth\linewidth\else\width\fi}{!}{
\begin{tabular}[t]{ll>{\raggedright\arraybackslash}p{0.125cm}rrrr}
\toprule
\toprule
Dataset & Noise &   & CT & LMS & MAD & MSD\\
\midrule
 & Unperturbed &  & 0.25 & 0.25 & 0.25 & 0.25\\

 & Cauchy &  & 0.17 & 0.17 & 0.17 & 0.17\\

 & Laplace &  & 0.30 & 0.30 & 0.30 & 0.30\\

 & Normal &  & 0.27 & 0.27 & 0.27 & 0.27\\

\multirow{-5}{*}{\raggedright\arraybackslash ACIC} & Student-$t$ &  & 0.24 & 0.24 & 0.24 & 0.24\\
\cmidrule{1-7}
 & Unperturbed &  & 103.81 & 103.12 & 103.99 & 102.42\\

 & Cauchy &  & 100.11 & 99.99 & 101.35 & 99.95\\

 & Laplace &  & 106.49 & 106.69 & 106.52 & 105.84\\

 & Normal &  & 105.14 & 105.32 & 105.30 & 104.31\\

\multirow{-5}{*}{\raggedright\arraybackslash Breast Cancer} & Student-$t$ &  & 90.54 & 89.73 & 91.09 & 90.32\\
\cmidrule{1-7}
 & Unperturbed &  & 869.34 & 875.16 & 867.58 & 873.08\\

 & Cauchy &  & 879.22 & 883.97 & 874.75 & 878.07\\

 & Laplace &  & 833.30 & 835.09 & 832.17 & 833.02\\

 & Normal &  & 832.84 & 830.12 & 831.56 & 832.96\\

\multirow{-5}{*}{\raggedright\arraybackslash Lalonde} & Student-$t$ &  & 756.64 & 759.60 & 759.21 & 757.28\\
\bottomrule
\end{tabular}}
\begin{notes}
Entries report ATE estimates averaged over 100 train/test splits. The unperturbed setting uses the original outcomes the remaining settings add artificial noise drawn from the indicated distributions.
\end{notes}
\end{table}

\begin{table}[!h]
\centering
\caption{\label{tab:apl_cate_metric} 
Summary statistics of estimated CATEs across splitting rules and outcome noise distributions for the three empirical applications.}
\centering
\resizebox{.7\linewidth}{!}{
\begin{tabular}[t]{ll>{\raggedright\arraybackslash}p{0.25cm}rrrr}
\toprule
\toprule
Dataset & Noise &   & CT & LMS & MAD & MSD\\
\midrule
 &  &  & \phantom{(}0.25\phantom{)} & \phantom{(}0.25\phantom{)} & \phantom{(}0.25\phantom{)} & \phantom{(}0.25\phantom{)}\\

 & \multirow{-2}{*}{\raggedright\arraybackslash Unperturbed} &  & (0.02) & (0.01) & (0.02) & (0.01)\\
\addlinespace[4pt]

 &  &  & \phantom{(}0.07\phantom{)} & \phantom{(}0.23\phantom{)} & \phantom{(}0.25\phantom{)} & \phantom{(}0.24\phantom{)}\\

 & \multirow{-2}{*}{\raggedright\arraybackslash Cauchy} &  & (1.46) & (1.23) & (2.56) & (1.28)\\
\addlinespace[4pt]

 &  &  & \phantom{(}0.29\phantom{)} & \phantom{(}0.29\phantom{)} & \phantom{(}0.29\phantom{)} & \phantom{(}0.30\phantom{)}\\

 & \multirow{-2}{*}{\raggedright\arraybackslash Laplace} &  & (0.02) & (0.02) & (0.04) & (0.03)\\
\addlinespace[4pt]

 &  &  & \phantom{(}0.27\phantom{)} & \phantom{(}0.27\phantom{)} & \phantom{(}0.27\phantom{)} & \phantom{(}0.27\phantom{)}\\

 & \multirow{-2}{*}{\raggedright\arraybackslash Normal} &  & (0.03) & (0.01) & (0.05) & (0.03)\\
\addlinespace[4pt]

 &  &  & \phantom{(}0.24\phantom{)} & \phantom{(}0.24\phantom{)} & \phantom{(}0.24\phantom{)} & \phantom{(}0.24\phantom{)}\\

\multirow{-10}{*}{\raggedright\arraybackslash ACIC} & \multirow{-2}{*}{\raggedright\arraybackslash Student-$t$} &  & (0.02) & (0.01) & (0.04) & (0.03)\\
\addlinespace[4pt]

\cmidrule{1-7}
 &  &  & \phantom{(}106.92\phantom{)} & \phantom{(}107.59\phantom{)} & \phantom{(}106.25\phantom{)} & \phantom{(}105.47\phantom{)}\\

 & \multirow{-2}{*}{\raggedright\arraybackslash Unperturbed} &  & (30.13) & (23.01) & (34.27) & (38.45)\\
\addlinespace[4pt]

 &  &  & \phantom{(}99.91\phantom{)} & \phantom{(}102.36\phantom{)} & \phantom{(}104.92\phantom{)} & \phantom{(}102.67\phantom{)}\\

 & \multirow{-2}{*}{\raggedright\arraybackslash Cauchy} &  & (53.12) & (52.41) & (55.53) & (55.56)\\
\addlinespace[4pt]

 &  &  & \phantom{(}112.19\phantom{)} & \phantom{(}111.72\phantom{)} & \phantom{(}107.24\phantom{)} & \phantom{(}109.72\phantom{)}\\

 & \multirow{-2}{*}{\raggedright\arraybackslash Laplace} &  & (36.61) & (32.14) & (40.03) & (39.81)\\
\addlinespace[4pt]

 &  &  & \phantom{(}99.21\phantom{)} & \phantom{(}101.39\phantom{)} & \phantom{(}100.56\phantom{)} & \phantom{(}97.89\phantom{)}\\

 & \multirow{-2}{*}{\raggedright\arraybackslash Normal} &  & (34.20) & (25.94) & (35.91) & (36.51)\\
\addlinespace[4pt]

 &  &  & \phantom{(}90.11\phantom{)} & \phantom{(}93.86\phantom{)} & \phantom{(}93.08\phantom{)} & \phantom{(}96.92\phantom{)}\\

\multirow{-10}{*}{\raggedright\arraybackslash Breast Cancer} & \multirow{-2}{*}{\raggedright\arraybackslash Student-$t$} &  & (28.38) & (20.58) & (34.53) & (30.82)\\
\addlinespace[4pt]

\cmidrule{1-7}
 &  &  & \phantom{(}881.60\phantom{)} & \phantom{(}852.37\phantom{)} & \phantom{(}882.36\phantom{)} & \phantom{(}804.53\phantom{)}\\

 & \multirow{-2}{*}{\raggedright\arraybackslash Unperturbed} &  & (131.41) & (27.27) & (212.26) & (187.09)\\
\addlinespace[4pt]

 &  &  & \phantom{(}849.42\phantom{)} & \phantom{(}858.60\phantom{)} & \phantom{(}883.81\phantom{)} & \phantom{(}856.77\phantom{)}\\

 & \multirow{-2}{*}{\raggedright\arraybackslash Cauchy} &  & (328.69) & (344.29) & (403.53) & (336.49)\\
\addlinespace[4pt]

 &  &  & \phantom{(}845.71\phantom{)} & \phantom{(}810.06\phantom{)} & \phantom{(}811.97\phantom{)} & \phantom{(}828.32\phantom{)}\\

 & \multirow{-2}{*}{\raggedright\arraybackslash Laplace} &  & (226.87) & (235.27) & (273.76) & (249.81)\\
\addlinespace[4pt]

 &  &  & \phantom{(}810.29\phantom{)} & \phantom{(}847.75\phantom{)} & \phantom{(}843.02\phantom{)} & \phantom{(}792.85\phantom{)}\\

 & \multirow{-2}{*}{\raggedright\arraybackslash Normal} &  & (298.47) & (289.74) & (341.74) & (310.13)\\
\addlinespace[4pt]

 &  &  & \phantom{(}785.60\phantom{)} & \phantom{(}762.76\phantom{)} & \phantom{(}791.22\phantom{)} & \phantom{(}791.33\phantom{)}\\

\multirow{-10}{*}{\raggedright\arraybackslash Lalonde} & \multirow{-2}{*}{\raggedright\arraybackslash Student-$t$} &  & (348.23) & (312.65) & (339.68) & (358.36)\\
\addlinespace[4pt]
\bottomrule
\end{tabular}}
\begin{notes}
Entries report the average standard deviation of estimated CATEs across 100 train/test splits, with the standard deviation across splits in parentheses. The unperturbed setting uses the original outcomes, while the remaining settings add artificial noise drawn from the indicated distributions.
\end{notes}
\end{table}


\FloatBarrier  
\begin{figure}[!htbp]
    \caption{Boxplots of the CATE estimates for the ACIC dataset across different splitting rules and outcome noise settings.}
    \centering
    \label{fig:acic_cate}
    \resizebox{0.95\linewidth}{!}{%
        \input{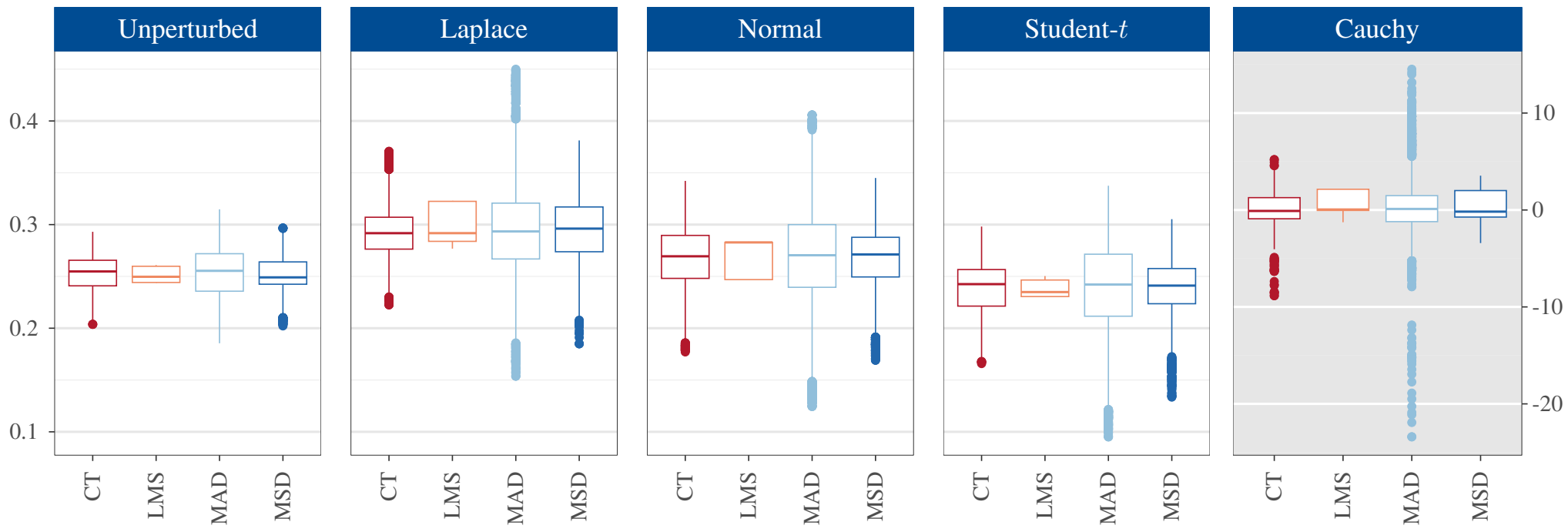}
    }
    \begin{notes}
        All noise settings share a common $y$-axis except for the Cauchy noise setting (highlighted in grey),  which uses a different scale to accommodate the greater dispersion of CATE estimates. Therefore, a shared axis would make the interpretation of the other settings impractical.
    \end{notes}
\end{figure}
\FloatBarrier  




\FloatBarrier  
\begin{figure}[!htbp]
    \centering
    \label{fig:breastcancer_cate}
    \caption{CATE estimates for the BreastCancer dataset across different splitting rules and outcome noise settings.}
    \resizebox{0.9\linewidth}{!}{%
    \input{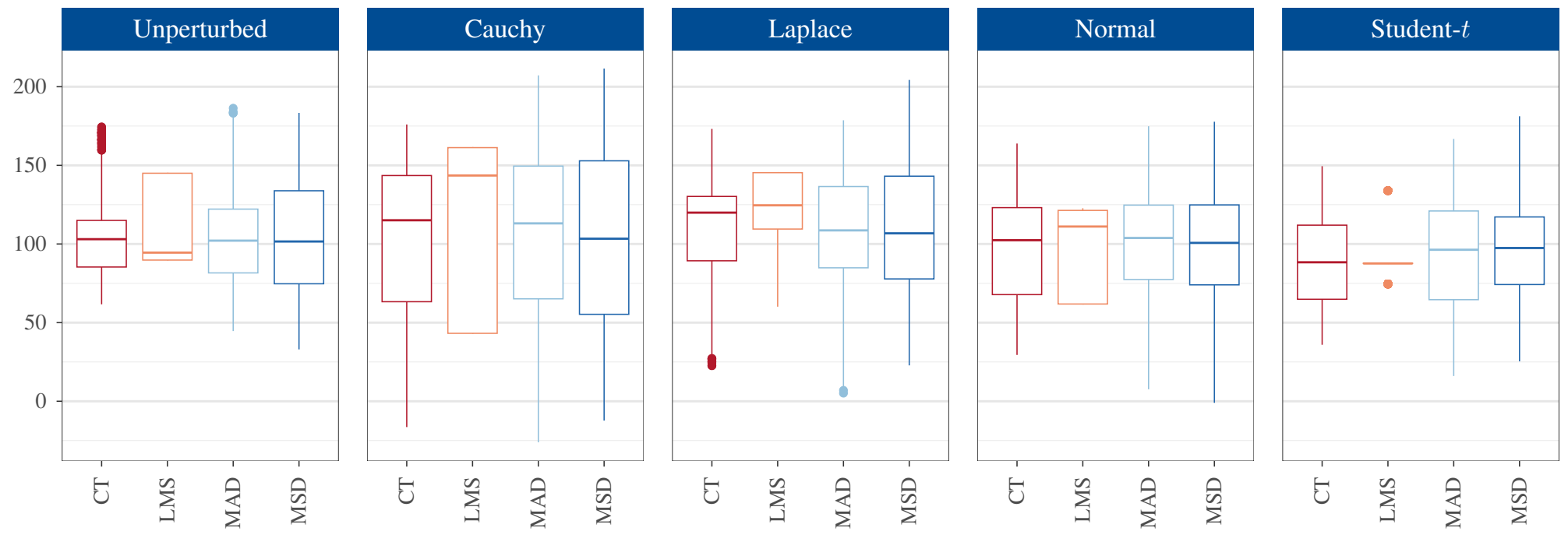}}
 \begin{notes}
Boxplots show the distribution of estimated CATEs across test observations for each splitting rule and outcome noise setting. All panels use a common y-axis scale.
 \end{notes}
\end{figure}
\FloatBarrier


    

\FloatBarrier  
\begin{figure}[!htbp]
    \centering
    \caption{CATE estimates for the Lalonde dataset across different splitting rules and outcome noise settings.}
    \label{fig:lalonde_cate}
    \resizebox{0.9\linewidth}{!}{%
    \input{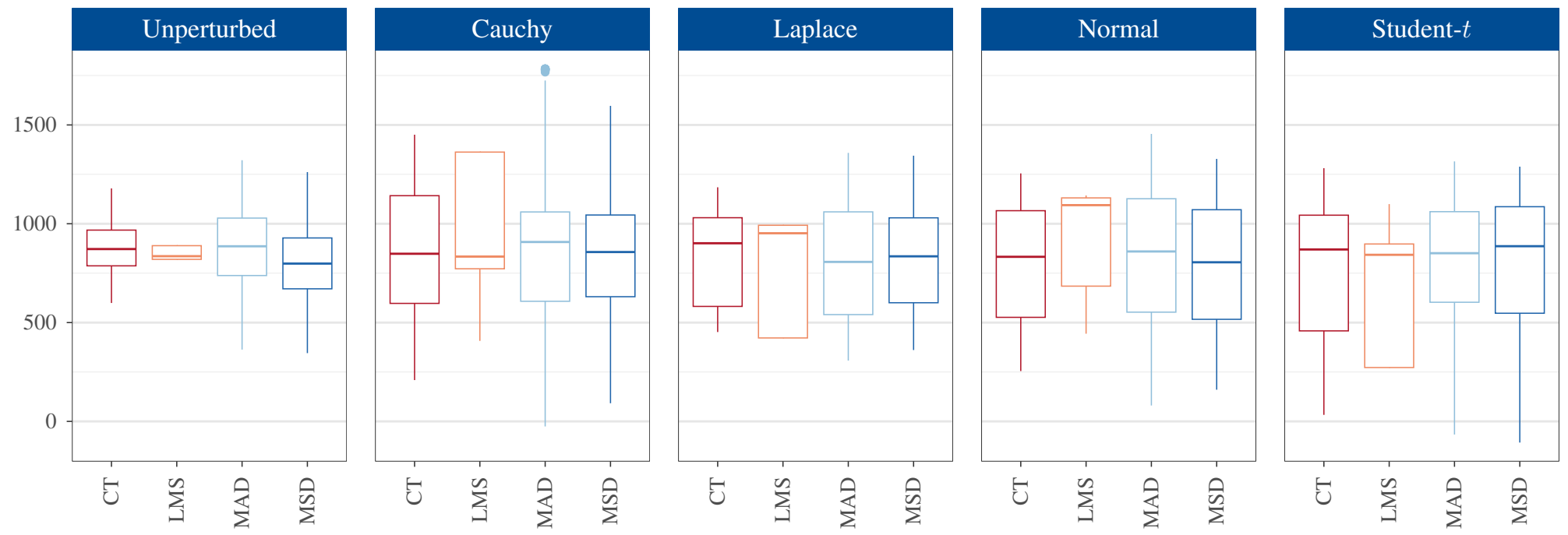}}
 \begin{notes}
Boxplots show the distribution of estimated CATEs across test observations for each splitting rule and outcome noise setting. All panels use a common y-axis scale.
 \end{notes}
\end{figure}
\FloatBarrier



\end{document}